\documentclass[fleqn,twoside]{article}
\usepackage{amsmath,amssymb,espcrc2,color,cite}

\usepackage{graphicx,soul}
\usepackage[figuresright]{rotating}

\newcommand{\url}[1]{{\tt #1}}
\newcommand{\lsim}{\;\raisebox{-.3em}{$\stackrel{\displaystyle <}{\sim}$}\;}
\newcommand{\gsim}{\;\raisebox{-.3em}{$\stackrel{\displaystyle >}{\sim}$}\;}

\newcommand{\gmt}{\ensuremath{(g-2)_\mu}}
\newcommand{\br}{{\rm BR}}
\newcommand{\bsg}{\ensuremath{\br(b \to s \gamma)}}
\newcommand{\btn}{\ensuremath{\br(B_u \to \tau \nu_\tau)}}

\newcommand{\bsmm}{\ensuremath{\br(B_s \to \mu^+\mu^-)}}
\newcommand{\bdmm}{\ensuremath{\br(B_d \to \mu^+\mu^-)}}
\newcommand{\bsdmm}{\ensuremath{\br(B_{s, d} \to \mu^+\mu^-)}}
\newcommand{\bqmm}{\ensuremath{\br(B_{q} \to \mu^+\mu^-)}}
\newcommand{\ssi}{\ensuremath{\sigma^{\rm SI}_p}}

\newcommand{\Och}{\ensuremath{\Omega_\chi h^2}}

\newcommand{\MW}{M_W}

\newcommand{\Mh}{M_h}

\newcommand{\MA}{M_A}

\newcommand{\mStop}{m_{\widetilde t}}

\newcommand{\MSbar}{{\overline {\rm MS}}}
\newcommand{\DRbar}{\ensuremath{\overline{\mathrm{DR}}}}

\newcommand{\mt}{m_t}
\newcommand{\mgl}{m_{\tilde g}}

\newcommand{\msqR}{m_{\tilde q_R}}

\newcommand{\neu}[1]{\tilde \chi^0_{#1}}
\newcommand{\mneu}[1]{m_{\tilde \chi^0_{#1}}}
\newcommand{\mste}{m_{\tilde t_1}}
\newcommand{\mstaue}{m_{\staue}}
\newcommand{\staue}{\tilde \tau_1}

\newcommand{\tb}{\tan\beta}

\newcommand{\gev}{\,\, \mathrm{GeV}}

\newcommand{\atlasfive}{ATLAS 5/fb jets + $\ETslash$}

\newcommand{\atlastwenty}{ATLAS 20/fb jets + $\ETslash$}

\newcommand{\rmu}{\ensuremath{{u}}}
\newcommand{\rmd}{\ensuremath{{d}}}
\newcommand{\rms}{\ensuremath{{s}}}
\newcommand{\rmc}{\ensuremath{{c}}}
\newcommand{\rmb}{\ensuremath{{b}}}
\newcommand{\rmt}{\ensuremath{{t}}}

\newcommand{\fTq}[1]{\ensuremath{f_{T_{#1}}}}
\newcommand{\fNTq}[1]{\ensuremath{f_{T_{#1}}^{(N)}}}

\newcommand{\SigmapiN}{\ensuremath{\Sigma_{\pi\!{\scriptscriptstyle N}}}}

\newcommand{\mup}{\ensuremath{m_{\rm u}}}
\newcommand{\md}{\ensuremath{m_{\rm d}}}
\newcommand{\ms}{\ensuremath{m_{\rm s}}}

\newcommand{\htb}[1]{{\color{blue}  #1}}

\newcommand{\htp}[1]{{\color{magenta}  #1}}

\newcommand{\ETslash}{/ \hspace{-.7em} E_T}

\graphicspath{{figs/}}

\title{
\vspace{-1.5cm}
\bf The CMSSM and NUHM1 after LHC Run 1 \\ \vspace{0.5em}}

\author{
{\bf O.~Buchmueller}\address[Imperial]
   {High\,Energy\,Physics\,Group,\,Blackett\,Laboratory,\,Imperial\,College,\,Prince\,Consort\,Road,\,London\,SW7\,2AZ,\,UK},
{\bf R.~Cavanaugh}\address[FNAL]
   {Fermi National Accelerator Laboratory, P.O. Box 500, 
    Batavia, Illinois 60510, USA}\hbox{$^{\rm ,}$}\address[UIC]
   {Physics Department, University of Illinois at Chicago, Chicago, 
    Illinois 60607-7059, USA},
{\bf A.~De Roeck}\address[CERN]
   {Physics Department, CERN, CH--1211 Gen\`eve 23, Switzerland}\hbox{$^{\rm ,}$}\address[Antwerpen]
   {Antwerp University, B--2610 Wilrijk, Belgium},
 {\bf M.J.~Dolan}\address[SLAC]
{Theory Group, SLAC National Accelerator Laboratory,
2575 Sand Hill Road, Menlo Park, \\ CA 94025-7090, USA},
{\bf J.R.~Ellis}\address[KCL]{Theoretical Particle Physics
  and Cosmology Group, Department of Physics, King's College London, London~WC2R~2LS, UK}\hbox{$^{\rm ,}$}\addressmark[CERN], 
{\bf H.~Fl\"acher}\address[Rochester]
   {H.H.~Wills Physics Laboratory, University of Bristol, Tyndall Avenue, Bristol BS8 1TL, UK},
{\bf S.~Heinemeyer}\address[Santander]
   {Instituto de F\'{\i}sica de Cantabria (CSIC-UC), 
    E--39005 Santander, Spain},
{\bf G.~Isidori}\address[Frascati]
{INFN, Laboratori Nazionali di Frascati, Via E. Fermi 40, 
I--00044 Frascati, Italy}\hbox{$^{\rm ,}$}\addressmark[CERN],
{\bf J.~Marrouche}\addressmark[Imperial],
{\bf D.~Mart\'inez~Santos}\address[NIKHEF]{NIKHEF\,and\,VU\,University\,Amsterdam,
Science\,Park\,105,\,NL-1098\,XG\,Amsterdam,\,The\,Netherlands},
{\bf K.A.~Olive}\address[Minnesota] 
{William I.\ Fine Theoretical Physics Institute, School of Physics and
 Astronomy, University of Minnesota, Minneapolis, Minnesota 55455, USA}, 
{\bf S.~Rogerson}\addressmark[Imperial],
{\bf F.J.~Ronga}\address[ETHZ]
   {Institute for Particle Physics, ETH Z\"urich, CH--8093 Z\"urich, 
   Switzerland},
{\bf K.J.~de~Vries}\addressmark[Imperial],
{\bf G.~Weiglein}\address[DESY]
   {DESY, Notkestrasse 85, D--22607 Hamburg, Germany}
}

\begin{document}

\begin{abstract}
We analyze the impact of data from the full Run~1 of
the LHC at 7 and 8~TeV on the CMSSM with $\mu > 0$ and $< 0$ and the 
NUHM1 with $\mu > 0$, incorporating the constraints imposed by
other experiments such as precision electroweak measurements, flavour
measurements, the cosmological density of cold dark matter and the direct search
for the scattering of dark matter particles in the LUX experiment. 
We use the following results from the LHC experiments: ATLAS searches
for events with $\ETslash$ accompanied by jets with the full 7 and 8~TeV data, the ATLAS and CMS
measurements of the mass of the Higgs boson, the CMS searches for heavy
neutral Higgs bosons and a combination of the LHCb and CMS measurements
of $\bsmm$ and $\bdmm$. Our results are based on samplings of the parameter spaces
of the CMSSM for both $\mu>0$ and $\mu<0$
and of the NUHM1 for $\mu > 0$ with 6.8$\times10^6$, 6.2$\times10^6$ and 1.6$\times10^7$ points, respectively,
obtained using the {\tt MultiNest} tool. The impact of the Higgs mass constraint
is assessed using {\tt FeynHiggs~2.10.0}, which 
provides an improved prediction for the masses of the MSSM Higgs
bosons in the region of heavy squark masses.
It yields in general larger values
of $\Mh$ than previous versions of {\tt FeynHiggs}, reducing the pressure on
the CMSSM and NUHM1. We find that the global $\chi^2$
functions for the supersymmetric models vary slowly over most of the
parameter spaces allowed by the Higgs mass and the $\ETslash$ searches,
with best-fit values that are comparable to the $\chi^2/{\rm dof}$ 
for the best Standard Model fit.
We provide 95\% CL lower limits on the masses of various sparticles and assess
the prospects for observing them during Run~2 of the LHC.

\begin{center}
{\tt KCL-PH-TH/2013-42, LCTS/2013-29, CERN-PH-TH/2013-297, \\
DESY 13-250, FTPI-MINN-13/43, UMN-TH-3316/13, SLAC-PUB-15861}
\end{center}

\end{abstract}

\maketitle

\section{Introduction}
\label{sec:intro}

In addition to establishing the mechanism for electroweak symmetry breaking, 
one of the primary objectives of experiments at the LHC has been to search
for possible physics beyond the Standard Model (SM), such as 
new particles that might alleviate
the naturalness problem and/or be associated with cosmological dark
matter. In contrast with the triumphant discovery at the LHC of a particle
that resembles 
the Higgs boson of the SM \cite{lhch}, and the observation of
\bsmm\ decay at a rate close to the SM prediction \cite{CMSBsmm,LHCbBsmm}, the first run of the LHC
has not revealed any convincing evidence for physics beyond the
SM. In particular, the LHC searches for jets + $\ETslash$
events~\cite{ATLAS20,CMS20} and for 
heavy Higgs bosons $H^\pm/H/A$~\cite{MSSMHiggsATLAS-CMS}
have drawn blanks so far.
In parallel, neither 
direct nor indirect searches for astrophysical dark matter have found
any convincing signals \cite{XENON100,LUX},
posing questions regarding the implications of those results 
for supersymmetric models.

\medskip
We have published previously several analyses of constrained versions of the minimal
supersymmetric extension of the Standard Model (MSSM) with universal
soft supersymmetry (SUSY)-breaking parameters $m_0$ for scalars and $m_{1/2}$
for fermions as well as a trilinear coupling $A_0$ at an input grand
unification scale and $\tb$, the ratio of the two vacuum expectation
values at the electroweak scale 
(the CMSSM~\cite{cmssm,AbdusSalam:2011fc}). We have also analyzed its
generalization to include common but non-universal 
soft supersymmetry-breaking Higgs masses $m_H$ (the
NUHM1~\cite{nuhm1,AbdusSalam:2011fc}). We have 
analyzed these models both before the start-up 
of the LHC and in the contexts of successive releases of LHC
data~\cite{mc1,mc2,mc3,mc35,mc4,mc5,mc6,mc7,mc75,mc8}. 

\medskip
Prior to
the LHC start-up, the discrepancy between the experimental measurement
of $\gmt$ \cite{newBNL} and theoretical calculations
(see~\cite{fredl-gm2,newerg-2} and references therein), 
favoured relatively light sparticle
masses, but these have not appeared in ATLAS and CMS $\ETslash$
searches, disfavouring 
small values of the CMSSM or NUHM1 SUSY-breaking mass parameters \cite{ATLAS20}. 
On the other hand, the discovery of a SM-like Higgs boson by
ATLAS and CMS~\cite{lhch} {with a mass consistent with the predictions of SUSY models has 
provided an important indirect constraint on SUSY model parameters
such as $m_0, m_{1/2}$, $A_0$ and $\tb$.
A significant r\^ole is also played by the observation by CMS and LHCb of \bsmm\ decay \cite{CMSBsmm,LHCbBsmm},
which imposes a complementary constraint on the CMSSM and NUHM1 parameter spaces.
Our most recent analyses \cite{mc8} of these models were based on the 7-TeV
\atlasfive\ data set \cite{ATLAS5,CMS5}. In this
paper we update our analyses to include the 8-TeV \atlastwenty\ data
set \cite{ATLAS20},  providing a complete  
study of the implications of the LHC Run~1 for the CMSSM and NUHM1
scenarios. On the basis of this study, we also discuss the prospects for discovering sparticles in the LHC Run~2.

\medskip
As described below, these constraints are analyzed in a frequentist
approach using an overhauled version of the 
{\tt MasterCode}~\cite{mcweb} framework to calculate  
the global $\chi^2$ function. For other recent post-LHC analyses of the 
CMSSM and NUHM, 
see~\cite{post-LHC,Stefaniak,125-other,eo6,elos,ehow+,moremuneg}. In this 
paper we sample the CMSSM and NUHM1 parameter spaces using the {\tt
  MultiNest} tool~\cite{0809.3437}, which is more efficient than the
Markov Chain Monte Carlo technique we used previously. We implement the
\atlastwenty\ constraint using scaling laws to extrapolate the
sensitivity to regions of the parameter space where documentation is not
available~\cite{mc8}. In our implementation of the  $\Mh$ constraint we
use a new version of {\tt FeynHiggs}, {\tt 2.10.0} \cite{newFH}, that
incorporates a resummation of leading and subleading logarithmic
corrections from the top/stop sector to provide improved results for
larger stop masses. Since,  in the SUSY models  we study, \bsmm\ and
\bdmm\ are expected to have the same ratio as in the SM, we combine
these constraints by assuming this SM ratio and incorporating the
experimental correlations between  \bsmm\ and \bdmm\ reported by the
LHCb and CMS Collaborations \cite{CMSBsmm,LHCbBsmm}. Recent cosmological
observations, including those by the Planck satellite~\cite{Planck},
have refined the estimate of the cold dark matter density, but this does
not have a relevant impact on our study. Concerning direct searches for
dark matter, the only constraint we apply is that from LUX~\cite{LUX} on
spin-independent dark matter scattering, which we incorporate taking due
account of the uncertainties in the hadronic scattering matrix element,
as discussed later. 

\medskip
We find that the global $\chi^2$ function varies relatively little
across most of the regions of the $(m_0, m_{1/2})$ planes that are
allowed by the $\ETslash$, Higgs and dark matter density
constraints on the CMSSM and NUHM1 
parameter spaces, with a global minimum at
large $m_0$ and $m_{1/2}$  
that is similar to the $\chi^2/{\rm dof}$ for the SM. Within the CMSSM,
there are four principal 
mechanisms for bringing the SUSY relic density $\Och$ into the range
favoured by Planck and other measurements~\cite{Planck}, namely coannihilation with the lighter
stau $\staue$ and other sleptons, coannihilation with the lighter stop ${\widetilde t_1}$,
rapid annihilation through the heavy Higgs bosons $H, A$ in the direct channel,
and annihilation in the focus-point region where the lightest neutralino $\neu1$
has an enhanced Higgsino component. In the following, we comment on the 
respective r\^oles of these mechanisms. Within the range of the CMSSM parameter space examined in
this paper, the $\Och$ constraint sets an upper bound on $m_{1/2}$ but not on $m_0$.
In the case of the NUHM1, more annihilation mechanisms may come into play, and we find
no upper bound on either $m_0$ or $m_{1/2}$.

\medskip
One of our key findings is that the LHC
measurement of $\Mh$ is not in tension with other constraints on the CMSSM and NUHM1 
parameter spaces except for $\gmt$. The $\Mh$ constraint does not impact them as strongly as had
previously been thought \cite{eo6,elos}, since 
the improved prediction incorporated in
{\tt FeynHiggs~2.10.0} \cite{newFH} yields in general
a higher value of $\Mh$ than previous versions of {\tt FeynHiggs} (as well as
{\tt SoftSUSY}) for the same values of the model parameters
\cite{ehow+}, as will be discussed in detail in 
Section \ref{mhconstraint}. 
The best-fit point in the CMSSM with $\mu > 0$ ($< 0$) has $\tb \sim 51 (36)$, 
and $\tb \sim 39$ is preferred in the NUHM1 with $\mu > 0$. All these points
have relatively large values of $m_0$ and $m_{1/2}$, but the likelihood functions of
these models are quite flat, and each of the models also has a local minimum of the
the likelihood function at low mass, with smaller $\tb$ and small $\Delta \chi^2 \leq 1$
relative to the global minimum. We present 95\% CL
lower limits on $\mgl$ {(the gluino mass), $\msqR$ 
(the average over the right-handed squark masses of the first two generations), $\mste$ (the light scalar top mass)
and $\mstaue$ (the light scalar tau mass) in each of these models.
In each case, we find that the lighter stop ${\tilde t_1}$ may be significantly lighter
than the other strongly-interacting sparticles.

\medskip
The structure of this paper is as follows. In Section~2 we discuss the
updated {\tt MasterCode} framework
and the more important changes in our implementations of the experimental constraints.
There are no significant changes in the ways we treat the constraints not discussed explicitly.
In Section~3 we describe the results of our fits within the CMSSM and NUHM1. Finally,
in Section~4 we summarize our conclusions and discuss the prospects for future
studies of these and other SUSY models, in particular during the LHC Run~2.


\section{Implementations of the Principal Experimental Constraints}

\subsection{The {\tt Mastercode} Framework}

\medskip
As described in our previous papers~\cite{mc1,mc2,mc3,mc35,mc4,mc5,mc6,mc7,mc75,mc8}, 
the {\tt MasterCode}~\cite{mcweb} is a framework that incorporates a code for the electroweak
observables based on~\cite{Svenetal}~\footnote{In this analysis we use the estimate
$\Delta \alpha_{\rm had} (M_Z) = 0.002756 \pm 0.0010$~\cite{Gfitter}.}
as well as the {\tt SoftSUSY~3.3.9}~\cite{Allanach:2001kg}, 
{\tt FeynHiggs~2.10.0}~\cite{newFH,FeynHiggs}, 
{\tt SuFla}~\cite{SuFla}, {\tt SuperIso~3.3}~\cite{SuperIso}, {\tt MicrOMEGAs~3.2}~\cite{MicroMegas} 
and {\tt SSARD}~\cite{SSARD} codes, which are interfaced using the SUSY Les Houches
Accord~\cite{SLHA}. The {\tt MasterCode} is used to construct
a global likelihood function that includes contributions from
electroweak precision observables, flavour measurements,
the cosmological dark matter density and direct searches for dark matter,
as well as the LHC Higgs mass measurement and $\ETslash$ searches.

\subsection{Implementation of {\tt MultiNest}}

\medskip
There has been a major overhaul of the {\tt MasterCode} since~\cite{mc8},
with the aim of simplifying its use and facilitating its application to different SUSY models. 
The most important change in its implementation has been to use the {\tt MultiNest} 
algorithm~\cite{0809.3437} to sample parameter spaces, instead of the Markov Chain Monte Carlo 
(MCMC) approach used previously. We find that {\tt MultiNest} is significantly more 
efficient for our purposes, and we have extensively checked that results obtained using the 
new version of the {\tt MasterCode} agree with those obtained from the previous version 
when the same input constraints are used.

\medskip
Although {\tt MultiNest}, like other sampling techniques such as MCMC, 
is geared towards Bayesian interpretation approaches, 
it can be used to sample well multi-dimensional parameter spaces, 
and thereby estimate efficiently and robustly frequentist confidence intervals. 
The main requirements for our purposes are that no nodes of high likelihood are missed, 
and that the regions with low $\chi^2$ are well sampled. For the scans used in this paper
we use the ranges $0 < m_0 < 7000 \gev$, $0 < m_{1/2} < 4000 \gev$, $2 < \tb < 68$ 
and $- 5000 \gev < A_0 < 5000 \gev$~%
\footnote{We use the same convention for
the sign of $A_0$ as in~\cite{mc8}, which is opposite to the convention used
in {\tt SoftSUSY}.}
in the CMSSM, for both signs of $\mu$, 
thereby extending significantly the $m_0$ range compared to~\cite{mc8}. 
In the case of the NUHM1, we use the same ranges for $m_{1/2}, \tb$ 
and $A_0$, sample $0<m_0<4000$ and study the range
$- 5\times10^7 \gev^2 < m_{H}^2 < 5\times10^7 \gev^2$, 
restricting our attention to $\mu > 0$. 
The total numbers of points sampled in the CMSSM with $\mu>0$ and $\mu<0$ and the NUHM1 are 
6.8$\times10^6$, 5.3$\times10^6$ and 1.6$\times10^7$, respectively. In all cases, 
the best-fit points were checked by running {\tt Minuit} on the parameter space, and
the differences in total $\chi^2$ between {\tt MultiNest} and {\tt Minuit} were $\ll 1$\%.}

\medskip
In this analysis we make several changes in our implementations of the constraints,
of which the most important are described in the following subsections.

\subsection{The \atlastwenty\ Constraint}

The ATLAS Collaboration has made public preliminary updates of their SUSY
searches using the entire available 8~TeV dataset, including the results of many different searches
targeting different $\ETslash$ final states and topologies. Here we follow the same prescription as in~\cite{mc8}, 
restricting ourselves to using the 0-lepton + 2 to 6 jets + $\ETslash$
search~\cite{ATLAS20}.
This is done in order to ensure that the limits presented by ATLAS in the CMSSM ${m_0, m_{1/2}}$
plane for $\tb = 30$ and $A_0 = 2 m_0$ can be extrapolated to other values of 
$\tb$ and $A_0$ in the ranges used in our scan. 
As in~\cite{mc8}, 
we have performed a dedicated validation to check that the 0-lepton $\ETslash$ limit reported
in~\cite{ATLAS20} is quite independent of  $\tb$ and $A_0$. 
As was to be to be anticipated given the similarity of the search methodologies 
between the ATLAS 0-lepton analyses at 7 and 8~TeV, we find very similar results to~\cite{mc8}. 
Therefore, we assume that the 95\% CL exclusion contour in the $(m_0, m_{1/2})$ plane
presented in~\cite{ATLAS20} may be used irrespective of $\tb$ and $A_0$, 
and apply a penalty term to points in our scan according to their distance from the stated 95\% CL limit, 
using the same scaling function as in~\cite{mc8}.


\subsection{The Higgs Mass Constraint}
\label{mhconstraint}
In view of the relatively large value of the Higgs mass~\cite{MH-ATLAS,MH-CMS}, 
$\Mh = 125.7 \pm 0.4 \gev$ (where the quoted uncertainty is purely
experimental)  
and the stronger lower limits on sparticle masses from direct LHC
searches \cite{ATLAS20} within the CMSSM and NUHM1, the calculation of the Higgs boson
masses using {\tt FeynHiggs} has been improved \cite{newFH} to achieve 
a higher accuracy for large stop mass scales. The calculations
implemented in {\tt FeynHiggs~2.8.7}, which we used previously~%
\footnote{
This version was an extension of the publicly available 
{\tt FeynHiggs~2.8.6}, which differed in the conversion of the trilinear
coupling $A_b$ from the \DRbar\ scheme to the on-shell (OS) scheme.
This issue was treated in an improved way in {\tt FeynHiggs~2.9.5}.
The implementation of {\tt FeynHiggs} used here slightly differs from the
public {\tt FeynHiggs~2.10.0}, with a small difference of
$\lsim 0.5 \gev$ in the $\Mh$ calculation. 
}%
~included the full one-loop contributions and the leading and subleading
two-loop corrections. The calculations 
included in the new version {\tt FeynHiggs~2.10.0} used here~\cite{newFH}
include a resummation to all orders of the leading and next-to-leading
logarithms of the type $\log(\mStop/\mt)$ (where $\mStop$ denotes
  the geometric average of the two scalar top masses), 
based on the relevant two-loop
Renormalization-Group Equations (RGEs)~\cite{SM2LRGE}, 
see~\cite{bse} and references therein for details.
The effects of this new correction start at the three-loop
order. It has been ensured 
that the resummed logarithms, which are obtained in the $\MSbar$ scheme,
are correctly matched onto the one- and two-loop corrections in the
on-shell scheme that were already included previously~\cite{newFH}. 
The main effect is an upward shift of $\Mh$ for stop masses in the
multi-TeV range, as well as the possibility of a refined estimate of the
theoretical uncertainty that is incorporated in our global fits. This shift in
$\Mh$ 
relaxes substantially the constraints from the Higgs mass on the CMSSM and NUHM1
and related models \cite{ehow+}.  

A numerical analysis in the CMSSM including leading three-loop
corrections to $\Mh$ using the code {\tt H3m}~\cite{mhiggsFD3l}) was presented
in~\cite{FKPS}. It was shown that the leading three-loop terms can have a
strong impact on the interpretation of the measured Higgs mass value in the
CMSSM. Here, with the new version of {\tt FeynHiggs}, we go beyond this
analysis by including (formally) subleading three-loop corrections as well as
a resummation to all orders of the logarithmic contributions to $\Mh$, see
above. 

\medskip
The new version of {\tt FeynHiggs} also includes an updated
estimate of the theoretical uncertainty, $\Delta \Mh|_{\rm FH}$, due to missing
higher-ordercontributions to $\Mh$~\cite{newFH}, which is typically in
the range 1.0 to 
1.5~GeV in the favoured regions of the parameter spaces we sample. The
theoretical uncertainty is to be incorporated in the global $\chi^2$ function
via a contribution of the form 
\begin{equation}
\label{Delta_mh}
\Delta \chi^2 (\Mh) \; = \; \frac{(M_{h, {\rm FH}} - M_{h, {\rm exp}})^2}
                 {(\Delta \Mh|_{\rm FH})^2 + (\Delta \Mh|_{\rm exp})^2} \, .
\end{equation}
Conservatively, in this paper we assume a fixed value $\Delta \Mh|_{\rm FH} = 1.5 \gev$
in our evaluation of (\ref{Delta_mh}), pending a more complete evaluation of
$\Delta \Mh|_{\rm FH}$ in a future version of {\tt FeynHiggs}.


\subsection{The \boldmath{\bsmm} and \boldmath{\bdmm} Constraints}

\medskip
To date, the most precise measurements of \bsmm\ and \bdmm\ have been provided by the 
CMS Collaboration~\cite{CMSBsmm}:
\begin{align}
\label{eq:BsmmBR}
\bsmm_{\rm CMS} &=(3.0_{-0.9}^{+1.0})\times10^{-9} \, , \nonumber \\
\bdmm_{\rm CMS} &= (3.5_{-1.8}^{+2.1})\times10^{-10} \, ,
\end{align}
and the LHCb Collaboration~\cite{LHCbBsmm}:
\begin{align}
\label{eq:BdmmBR}
\bsmm_{\rm LHCb} &= (2.9_{-1.0}^{+1.1})\times10^{-9} \, , \nonumber \\
\bdmm_{\rm LHCb} &= (3.7_{-2.1}^{+2.4})\times10^{-10} \, .
\end{align}
These numbers correspond to time averaged (TA) branching fractions,\footnote{The results from the ATLAS ~\cite{BsmmATLAS}, 
CDF~\cite{BsmmCDF} and D\O~\cite{BsmmD0} Collaborations are not considered in our study, as they
have significantly less precision than the results of CMS and LHCb.} and are in good agreement with
the SM TA expectations~\cite{Bobeth:2013uxa} (see also~\cite{BsmmSM}):
\begin{align}
\label{eq:BsmmSM}
\bsmm_{\rm SM} &=(3.65\pm0.23)\times10^{-9} \, , \nonumber \\
\bdmm_{\rm SM} &= (1.06\pm0.09)\times10^{-10} \, .
\end{align}
An official combination of the CMS and LHCb results can be found in the conference note~\cite{BsmmComb}:
\begin{align}
\label{eq:BsmmComb}
\bsmm_{\rm exp} &=(2.9\pm0.7)\times10^{-9} \, , \nonumber \\
\bdmm_{\rm exp} &= (3.6_{-1.4}^{+1.6})\times10^{-10} \, .
\end{align}
In all new physics (NP) models with minimal flavour violation
(MFV)~\cite{MFV}, including the CMSSM and the NUHM1, 
\bsmm\ and \bdmm\  can deviate from their corresponding SM predictions, but their ratio
remains fixed at the SM value~\cite{RmmMFV}:\footnote{The numerical value in (\ref{eq:BsmmMFV}) is obtained 
taking into account the latest SM inputs from Ref.~\cite{Bobeth:2013uxa}.}
\begin{equation}
\label{eq:BsmmMFV}
\left. \frac{\bsmm_{\rm NP}}{\bdmm_{\rm NP}} \right|_{\rm MFV} = 31.41\pm2.19 \, .
\end{equation}
We exploit this property to combine \bsmm\ and \bdmm\ measurements into a single constraint
in the CMSSM (NUHM1) parameter space. In particular, for each of the four measurements 
in (\ref{eq:BsmmBR}) and (\ref{eq:BdmmBR}) we determine the ratio
\begin{equation}
R_{\mu\mu} = \frac{\bqmm_{\rm exp} }{\bqmm_{\rm SM}} \quad (q = s, d)~, 
\end{equation}
that is independent of $q$ in the context of MFV models.

The four constraints are then combined into a single weighted mean (hereafter denoted $R_{\mu\mu}^{\rm exp}$), 
taking into account the correlations between the different measurements.
It should also be noted that {\tt SuFla} computes directly a theoretical prediction for $R_{\mu\mu}$,
allowing one to separate the theory uncertainties into three sources: SUSY theory uncertainties
(which are negligible), the uncertainty 
from (\ref{eq:BsmmMFV}), and those affecting the SM prediction of the
branching fractions. 

\medskip
The CMS Collaboration has provided an estimate of 
$R_{\mu\mu}^{\rm CMS} = 1.01_{-0.26}^{+0.31}$~\cite{Fabrizio} by
combining its \bsmm\ and \bdmm\ measurements (and using the SM values for the branching 
ratios in~\cite{BsmmSM}).
Here we construct a joint likelihood for the four measurements (\ref{eq:BsmmBR})--(\ref{eq:BdmmBR})
using correlation coefficients between \bsmm\ and \bdmm\ of $-50\%$  in
CMS and $+3\%$ in LHCb~\cite{Justine}. The log-likelihoods of quantities with asymmetric errors are 
approximated using a treatment equivalent to formula (4) in~\cite{Junk}. We then reparameterize the joint likelihood
as a function of the single parameter of interest, $R_{\mu\mu}^{\rm exp}$, imposing the constraint in Eq.~(\ref{eq:BsmmMFV}),
and assuming that the ratio of hadronization fractions of the $b$ quark (the ratio of probabilities of the $b$ 
quark to hadronize into a $B^0$ or a 
$B_s^0$), $f_d/f_s$, to be 
the same in both experiments.

\medskip
Our final estimate after profiling on the theory uncertainties and $f_d/f_s$ is:
\begin{equation}
\label{eq:RmumuEXP} 
R_{\mu\mu}^{\rm exp} = 0.92_{-0.20}^{+0.21}~.
\end{equation} 
We have checked that our approach reproduces with good accuracy both the results in 
(\ref{eq:BsmmComb}) and the $R_{\mu\mu}^{\rm CMS}$ value (using the SM values for the branching 
ratios in~\cite{BsmmSM}),
giving us confidence in our treatment. 
The contribution this function makes to the global $\chi^2$ function
is shown as the solid blue line in Fig.~\ref{fig:bsdmm}, where it is compared
with the contribution calculated previously in~\cite{mc8} (dashed red line).

\begin{figure*}[htb!]
\begin{center}
\resizebox{8cm}{!}{\includegraphics{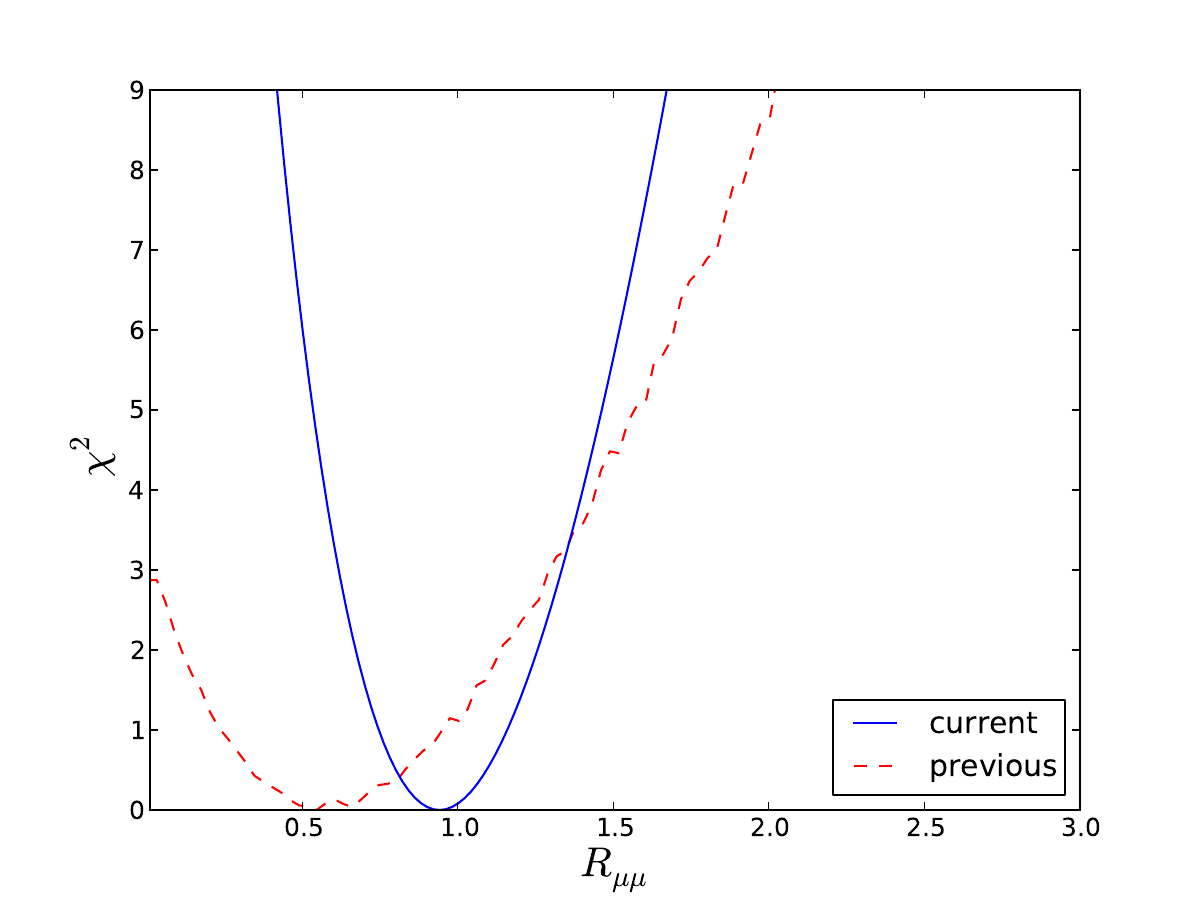}}
\end{center}
\caption{\it The contribution to the global $\chi^2$ function of the LHCb and CMS measurements
of \bsmm\ and \bdmm\ reported in (\ref{eq:BsmmBR}, \ref{eq:BdmmBR}), as calculated using the
prescription described in the text (blue solid line) compared with the contribution calculated previously 
in~\cite{mc8} (dashed red line).
}
\label{fig:bsdmm}
\end{figure*}

\subsection{The Dark Matter Constraints}

\medskip
There are two important dark matter constraints on the CMSSM and NUHM1
parameter spaces. One is the cosmological dark matter density $\Och = 0.1198 \pm 0.0026$
estimated from Planck data~\cite{Planck}, and the other is the upper limit
on the spin-independent elastic cold dark matter scattering cross section $\ssi$ from the
LUX experiment~\cite{LUX}, which is stronger by a factor $\sim 2$ than
that from the XENON100 experiment~\cite{XENON100}
in the range of neutralino masses relevant to this study.
Upper limits on the spin-dependent cross section do not impinge on the parameter
spaces of the models we study.

\medskip
Previously, we used {\tt Micromegas~2.4.5} to calculate \Och, which we checked gave
results similar to the independent {\tt SSARD} code in the regions of interest.
Here we use {\tt Micromegas~3.2} \cite{MicroMegas}. The recent results from the Planck satellite~\cite{Planck}
refine the previous observational estimate of \Och, but this does not alter significantly the
implications for other observables.

\medskip
We compute the elastic scattering cross section, $\ssi$ using \cite{SSARD}.
There are, however,  important uncertainties in the calculation of $\ssi$ and these are
now incorporated in the present analysis also computed using \cite{SSARD}.
There are two major sources for these uncertainties which we review here briefly.
The first is the uncertainty is related to the shift in the nucleon mass due to finite quark masses,
$\sigma_0 = 36 \pm 7$ MeV. The second is due to the uncertainty in 
the $\pi$-nucleon sigma term, $\SigmapiN$, which we take here as $50 \pm 7$ MeV.

The spin-independent matrix element for $\neu1$-nucleon scattering is 
proportional to a parameter $f_N$ that can be written as 
\begin{equation} \label{eqn:fN}
  \frac{f_N}{m_N}
    = \sum_{q=\rmu,\rmd,\rms} \fNTq{q} \frac{\alpha_{3q}}{m_{q}}
      + \frac{2}{27} f_{TG}^{(N)}
        \sum_{q=\rmc,\rmb,\rmt} \frac{\alpha_{3q}}{m_q} \, ,
\end{equation}
where the parameters $\fNTq{q}$ are defined by
\begin{equation} \label{eqn:Bq}
  m_N \fNTq{q}
  \equiv \langle N | m_{q} \bar{q} q | N \rangle \, ,
\end{equation}
with~\cite{Shifman:1978zn,Vainshtein:1980ea}
\begin{equation} \label{eqn:fTG}
  f_{TG}^{(N)} = 1 - \sum_{q=\rmu,\rmd,\rms} \fNTq{q} \, .
\end{equation}
An expression for $\alpha_{3q}$ in terms of supersymmetric model
parameters is given in~\cite{EFO}: it does not contribute significantly to the uncertainty
in the calculation of the cross section, which is dominated by uncertainties in hadronic parameters.

\medskip
These  matrix elements are all directly proportional to $\SigmapiN$.
It is well known that the elastic cross section is very sensitive to the strange 
scalar density in the nucleon, 
\begin{equation} \label{eqn:y}
    y = 1 - \sigma_0/\SigmapiN \; .
\end{equation}
Indeed, $f_{T_s}$ is proportional to $\SigmapiN y$, and hence the
uncertainties in both $\SigmapiN$ and $\sigma_0$ enter. 

Our calculation of the uncertainty in the elastic cross section propagates the independent
uncertainties in $\SigmapiN$, $\sigma_0$ as well as uncertainties in the quark mass ratios $\md/\mup$ and $\ms/\md$, though
the latter two are much smaller than the former two. For a more complete discussion of these
uncertainties, see \cite{eos}.
Thus, while the uncertainty in \ssi\ is often attributed to the uncertainty in $\SigmapiN$,
there is an almost equally large contribution to the uncertainty in \ssi\
coming from $\sigma_0$, particularly in the determination of the important strangeness contribution,
$ \fTq{\rms}$. 

\medskip

We display in Fig.~\ref{fig:chi2_contributions_Xenon} the contribution to the global $\chi^2$ function
that we calculate on the basis of the LUX 90\%~CL~upper limit on
the spin-independent cross section $\ssi$~\cite{LUX}, 
without (red points) and with (blue points) taking into account the uncertainty in the calculation of $\ssi$.
The horizontal blue bar is a representative example of the effect of the theoretical uncertainty
in the hadronic matrix element on the calculation of \ssi\ for one of the
CMSSM points.

\begin{figure*}[htb!]
\begin{center}
\resizebox{8cm}{!}{\includegraphics{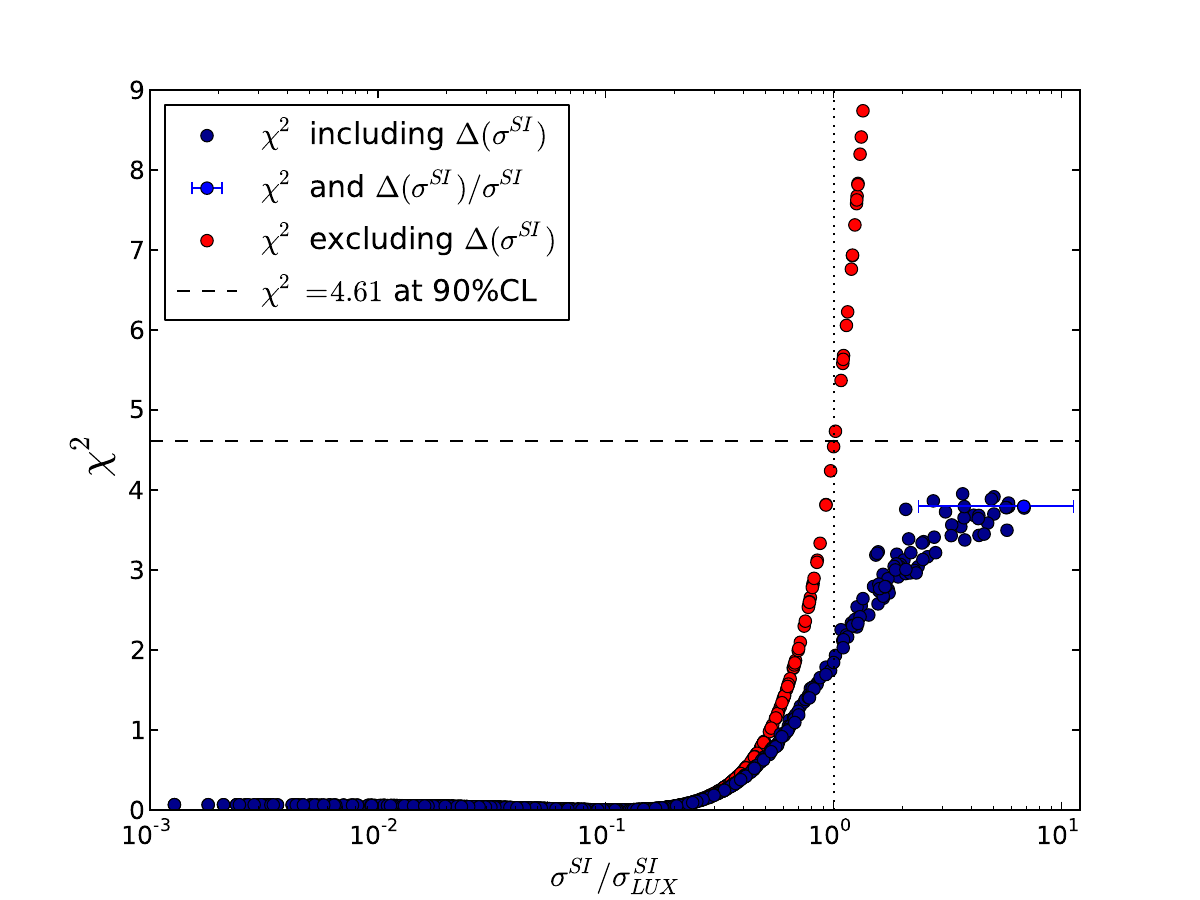}}
\end{center}
\caption{\it The contribution to the global $\chi^2$ function
that we calculate on the basis of the LUX 90\%~CL~upper limit on
the spin-independent cross section $\ssi$~\cite{LUX}, 
without (red points) and with (blue points) taking into account the theoretical uncertainty in the calculation of $\ssi$.
The horizontal blue bar exhibits the effect of this uncertainty
in the hadronic matrix element on the calculation of \ssi\ for a specific CMSSM point.}
\label{fig:chi2_contributions_Xenon}
\end{figure*}


\section{Results}

\subsection{CMSSM Fits}

\medskip
We now present the results of our new CMSSM fit using
the above new inputs, considering first the case of $\mu > 0$
and then the case of $\mu < 0$. In each case, we
present first some illustrative parameter planes, and then some
one-dimensional likelihood functions.

\subsubsection{Parameter Planes for \boldmath{$\mu > 0$}}

\medskip
In each of the parameter planes in Fig.~\ref{fig:CMSSM}, the best-fit point is indicated by a
green star, the $\Delta \chi^2 = 2.30$ contour that corresponds
approximately to the 68\% CL is shown as a red line, and the $\Delta \chi^2 = 5.99$
contour that corresponds approximately to the 95\% CL is shown as blue line. The results 
of the current fit are indicated by solid lines and solid
stars. The results 
of fits to the same set of data constraints as used in our previous paper~\cite{mc8}, but
using current theoretical codes including {\tt FeynHiggs~2.10.0} and treating the dark
matter scattering uncertainty as in Section~2.6, are shown as dashed
lines and open stars. This comparison of the two results allows to
determine the effects of {\em new data}, independent of any code
update. The effects of the new codes are discussed later.

\begin{figure*}[htb!]
\resizebox{8.5cm}{!}{\includegraphics{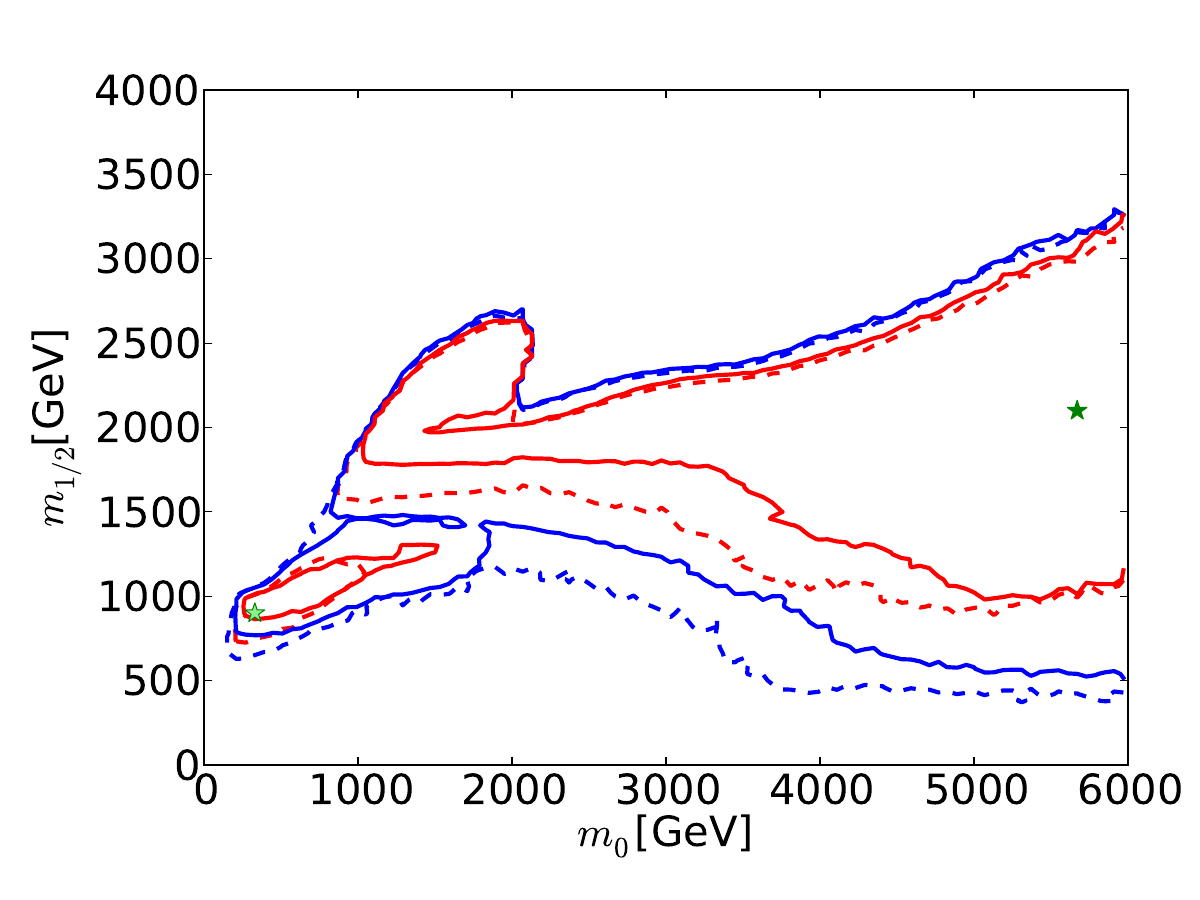}}
\resizebox{8.5cm}{!}{\includegraphics{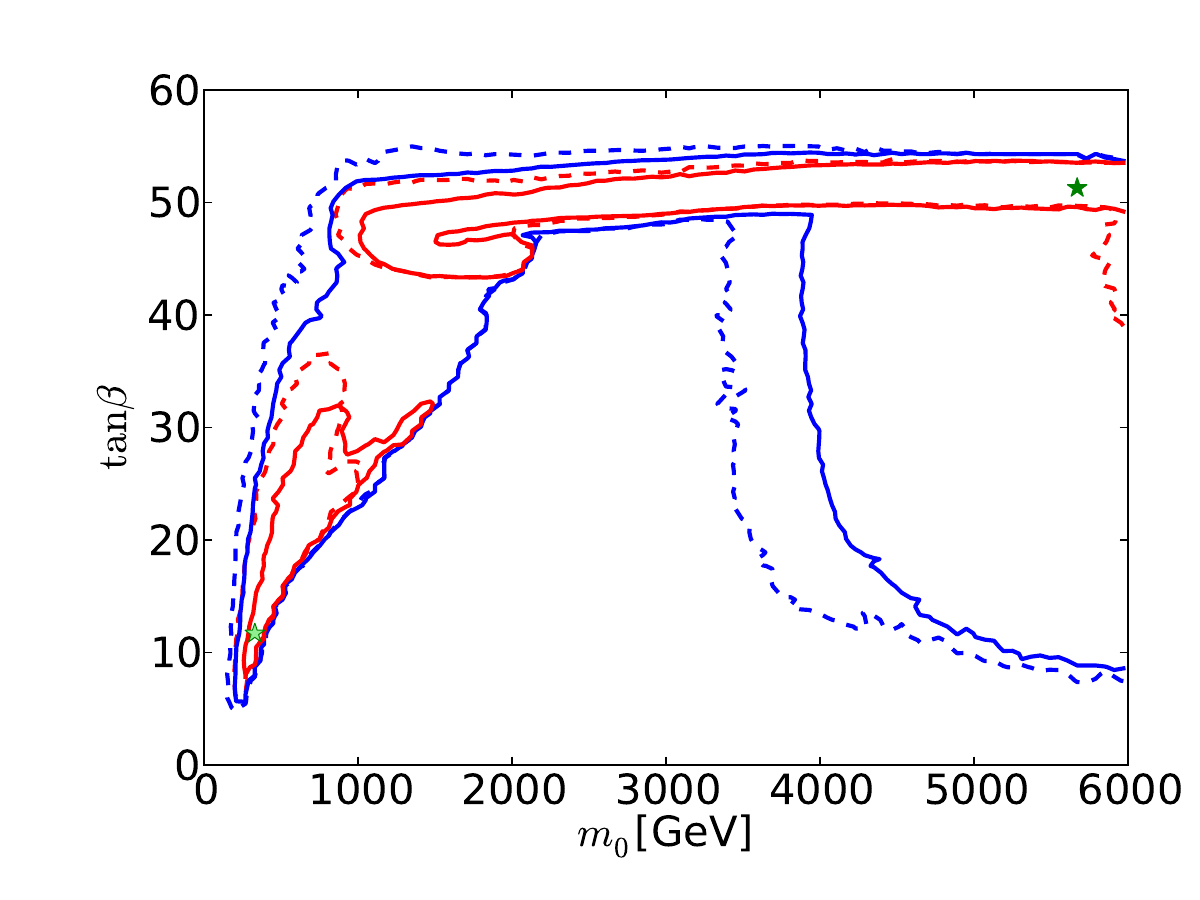}}
\resizebox{8.5cm}{!}{\includegraphics{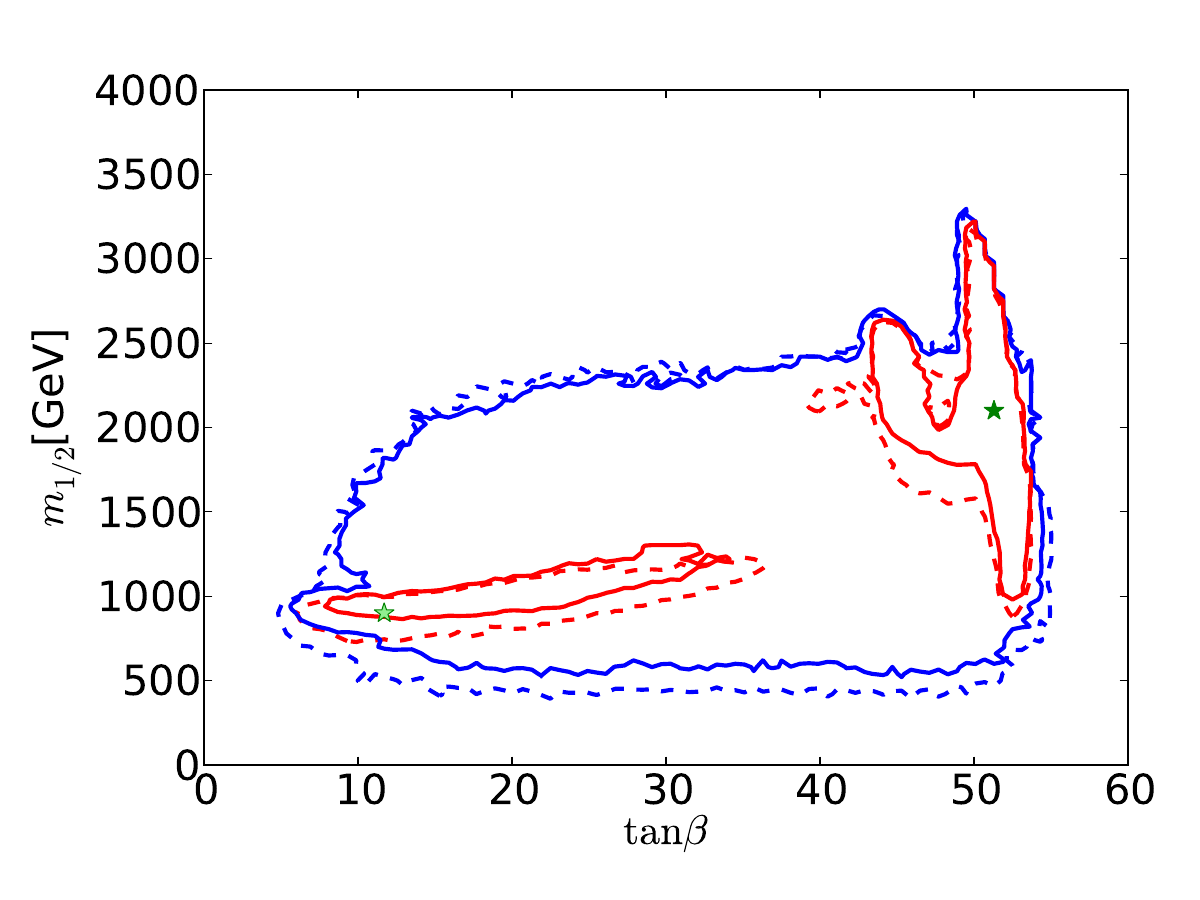}}
\resizebox{8.5cm}{!}{\includegraphics{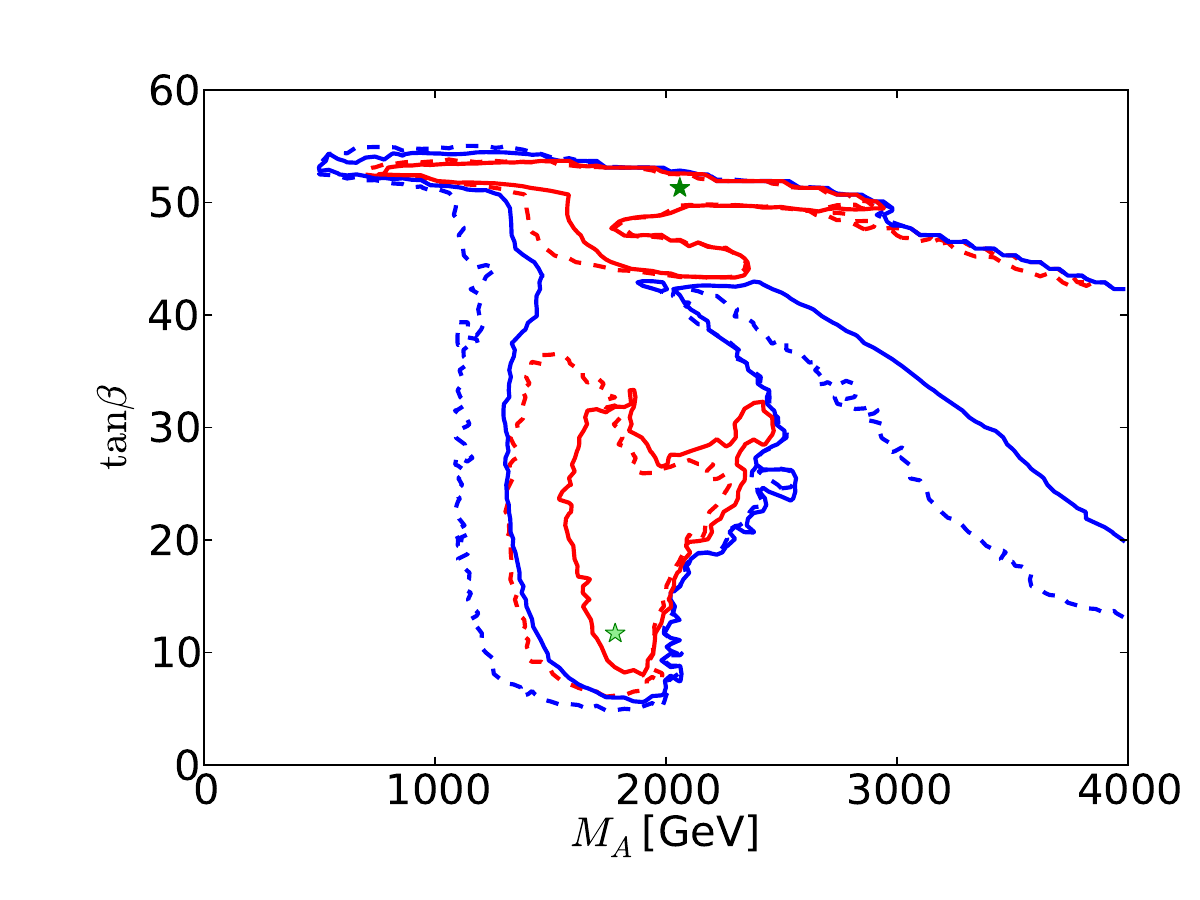}}
\vspace{-1cm}
\caption{\it A compilation of parameter planes in the CMSSM for $\mu > 0$, including the $(m_0, m_{1/2})$ plane
(upper left), the $(m_0, \tb)$ plane (upper right), the $(\tb, m_{1/2})$ plane (lower left), and the
$(\MA, \tb)$ plane (lower right), after implementing the \atlastwenty, \bsdmm, $\Mh$,
$\Och$, LUX constraints
and other constraints as described in the text.
The results of the current CMSSM fit are indicated by solid lines
and filled   stars, and a fit to previous data~\cite{mc8} using the same implementations of
the $\Mh$, \ssi\ and other constraints is indicated by dashed lines and open stars.
The red lines denote $\Delta \chi^2 = 2.30$ contours (corresponding approximately to
the  68\%~CL), and the red lines denote $\Delta \chi^2 = 5.99$ (95\%~CL)
contours.} 
\label{fig:CMSSM}
\end{figure*}

\medskip
We see in the upper left panel of Fig.~\ref{fig:CMSSM} 
that the current CMSSM fit has two disjoint $\Delta \chi^2 = 2.30$
contours in the $(m_0, m_{1/2})$ plane, one enclosing an `island' at relatively low masses, 
centred around $(m_0, m_{1/2}) \sim (500, 1000) \gev$,
and a larger `continent' extending from $(m_0, m_{1/2}) \sim (500, 1500) \gev$ to larger
mass values, beyond the range $m_0 < 6000 \gev$ studied here. As we discuss below, the
low-mass `island' lies in the stau-coannihilation region, where the \gmt\ contribution to
the global $\chi^2$ function is reduced, whereas in the high-mass `continent'
the relic density is brought within the cosmological range by rapid $\neu1$
annihilations due to direct-channel heavy Higgs boson resonances.
Our current best-fit point lies in the outer 68\% CL region and has
$(m_0, m_{1/2}) \sim (5650, 2100) \gev$. There is a single $\Delta \chi^2 = 5.99$ contour enclosing both
the inner and outer 68\% CL regions. We note that the global $\chi^2$
function is quite flat in the outer region, and very similar to the $\chi^2$
value for the SM~\footnote{We estimate this as in~\cite{mc8}, i.e., by using
{\tt MasterCode} to calculate the $\chi^2$ for the CMSSM point with 
$(m_0, m_{1/2}) = 15$~TeV,  $\tb = 10$ and $A_0 = 100 \gev$,
discarding the contributions from $\Mh$ and $\Och$ and evaluating the
\ssi\ contribution as 0.14, which yields $\chi^2 = 36.5$.}. 

\medskip
The lower limit on $m_{1/2}$ at small $m_0 \sim 500 \gev$
is provided mainly by the ATLAS 20/fb search  for events with $\ETslash$ and
2 to 6 jets, whereas at large $m_0 \gsim 3000 \gev$ there are several relevant ATLAS limits
using different event topologies with jets, leptons, $b$ quarks and $\ETslash$. These are
quite sensitive to $\tb$ and $A_0$, and have little impact on the preferred regions of the
CMSSM parameter space, so we do not
include them in our analysis. The lower limit on
$m_0$ and the low-mass `island' corresponds to the stau LSP boundary and the nearby coannihilation 
strip. The region at large $m_0$ and $m_{1/2}$ containing the best-fit point is in the rapid-annihilation 
funnel region, with the upper bound on $m_{1/2}$
being provided by the cosmological constraint on $\Och$. The region at small $m_{1/2}$ and large $m_0$
is in the focus-point region.

\begin{table*}[!tbh!]
\renewcommand{\arraystretch}{1.5}
\begin{center}
\begin{tabular}{|c|c||c|c|c|c|c|c|} \hline
Model & Data set & Minimum & Prob- & $m_0$ & $m_{1/2}$ & $A_0$ & $\tb$ \\
  &    & $\chi^2$/d.o.f.& ability & (GeV) & (GeV) & (GeV) & \\ 
\hline \hline
CMSSM & ATLAS 7 TeV & 
32.6/23 & 8.8\% &340 &910 & 2670 & 12\\
$\mu > 0$ &ATLAS$_{\rm 20/fb}$ (low) &
35.8/23 & 4.3\% &670 &1040 & 3440 & 21\\
&ATLAS$_{\rm 20/fb}$ (high)    & 
35.1/23 & 5.1\% &5650 &2100 & -780 & 51\\
\hline
CMSSM &ATLAS$_{\rm 20/fb}$ (low) &
38.9/23 & 2.0\% &330 &970 & 3070 & 10\\
$\mu < 0$ &ATLAS$_{\rm 20/fb}$ (high)    & 
36.6/23 & 3.6\% &6650 &2550 & -3150 & 39\\
\hline
NUHM1 & ATLAS 7 TeV &
30.5/22 & 10.7\% &370 &1120 & 5130 & 8\\
$\mu > 0$ & ATLAS$_{\rm 20/fb}$ (low) &
33.3/22 & 5.8\% &470 &1270 & 5700 & 11\\
& ATLAS$_{\rm 20/fb}$ (high) & 
32.7/22 & 6.6\% &1380 &3420 & -3140 & 39\\
\hline \hline
``SM" & ATLAS$_{\rm 20/fb}$ (high) & 36.5/24 & 5.0\% & - & - & - & - \\

\hline
\end{tabular}
\caption{\it The best-fit points found in global CMSSM fits for both signs of $\mu$ and an NUHM1 fit with $\mu > 0$,
  using the \atlastwenty\ constraint~\cite{ATLAS20}, and the combination of the
  CMS~\cite{CMSBsmm} and LHCb~\cite{LHCbBsmm} constraints on  \bsdmm~\cite{BsmmComb},
  as well as an update of the {\tt FeynHiggs} calculation of $\Mh$ and a more 
  conservative treatment of the hadronic matrix element uncertainties in \ssi,
  as discussed in the text. The results for the CMSSM with $\mu > 0$ and
  the NUHM1 
  are compared with those found previously in global fits based
  on the ATLAS 7-TeV $\ETslash$ data and the previous experimental constraint on \bsdmm,
  and with a current SM fit made using the procedure discussed in the text.
  We list the parameters of the best-fit
  points in both the low- and high-mass `islands' in Figs.~\protect\ref{fig:CMSSM}, \ref{fig:CMSSMneg} and \ref{fig:NUHM1}.
  We note that the
  overall likelihood function is quite flat in both the CMSSM and the NUHM1, so that the precise
  locations of the best-fit points are not very significant, and we do not quote uncertainties.
  For completeness, we note that in the best NUHM1
  fits   $m_H^2  = - 2.54 \times 10^7 \gev^2$ at the low-mass point and
 $m_H^2 \equiv = 1.33 \times 10^7 \gev^2$ at the high-mass point.
 }
\label{tab:bestfits}
\end{center}
\end{table*}

\medskip
Looking now at the $(m_0, \tb)$ plane in the upper right panel of
Fig.~\ref{fig:CMSSM}, we see that the low-mass coannihilation  `island' corresponds to
values of $\tb \lsim 30$, whereas the lower-$\chi^2$ part of the high-mass continent
corresponds to a band with larger values of $\tb \sim 50$ in the rapid-annihilation funnel
region, connected to a `continental shelf' in the focus-point region extending to lower
$\tb$ when $m_0 \gsim 3500 \gev$. The best-fit point has $\tb \sim 50$
and lies in the funnel region.

\medskip
The lower left panel of Fig.~\ref{fig:CMSSM} shows the $(\tb, m_{1/2})$
plane.
As already mentioned, the only \atlastwenty\ limit we
use is that on $\ETslash$ + 2 to 6 jets: other limits using topologies with leptons and/or $b$ jets
could have an impact when $m_{1/2} \gsim 500 \gev$, depending in particular on the value
of $\tb$ and/or $A_0$. We have not attempted to model these limits, but note that they 
would not affect the 68\% CL region displayed.

\medskip
Finally, the lower right panel of Fig.~\ref{fig:CMSSM} displays the $(\MA, \tb)$
plane of the CMSSM. We see that in the low-mass coannihilation `island' typical values
of $\MA \sim 1500$ to 2500~GeV. The best-fit point has a similar value of $\MA$,
but with a much larger value of $\tb$. The band at large $\tb$ corresponds
to the rapid-annihilation funnel region. It is clear that the larger values of
$m_0$ seen in the other panels correspond to large values of $\MA \sim 2500 \gev$
and more.

\medskip
We compare  in Fig.~\ref{fig:mc8old} the results of the current analysis (solid lines and filled stars)
with the results that were shown in~\cite{mc8} using the previous data
set and the previous implementations of 
the constraints (dashed lines and open stars). As already mentioned,
the strengthened ATLAS $\ETslash$ constraints with 20/fb of data at
8~TeV have had little impact except 
to strengthen the lower limit on $m_{1/2}$ at low $m_0 \sim 500 \gev$. 
At larger $m_0$, the range of $m_{1/2}$ values is broader than that shown
in~\cite{mc8} in part because we use here new versions of {\tt SoftSUSY} and {\tt MicrOMEGAs}, 
which makes it possible to find the correct dark matter density in a larger range of parameters.
This is the case, in particular, for the best-fit point we now find, which 
lies in the outer 68\% CL region and has
$(m_0, m_{1/2}) \sim (5650, 2100) \gev$.
Our new treatment of the uncertainty in \ssi\
discussed in Section~2.6, combined with  larger Higgs mass found in {\tt FeynHiggs~2.10.0},
has the effect of disfavouring the focus-point region~ \cite{FP} less than in~\cite{mc8},
leading to an expansion in the region allowed at the 95\% CL at large $m_0$ and small $m_{1/2}$.
The extension of the CMSSM 95\% CL region to larger $m_0$ in the left panel of Fig.~\ref{fig:mc8old}
is due to the extended sampling range we use here: the {\tt MultiNest} technique
used here does not have a big impact beyond improving the density of sampling.

\begin{figure*}[htb!]
\resizebox{8.5cm}{!}{\includegraphics{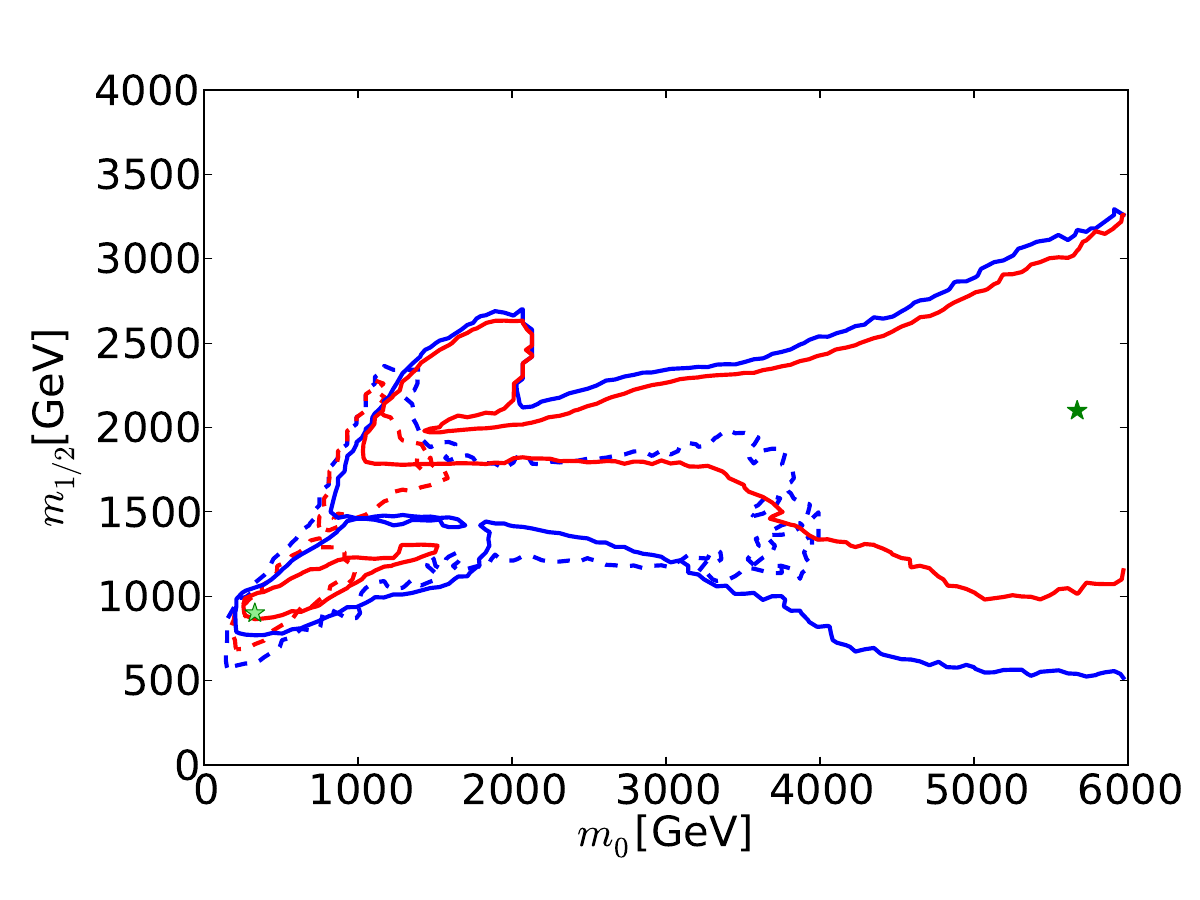}}
\resizebox{8.5cm}{!}{\includegraphics{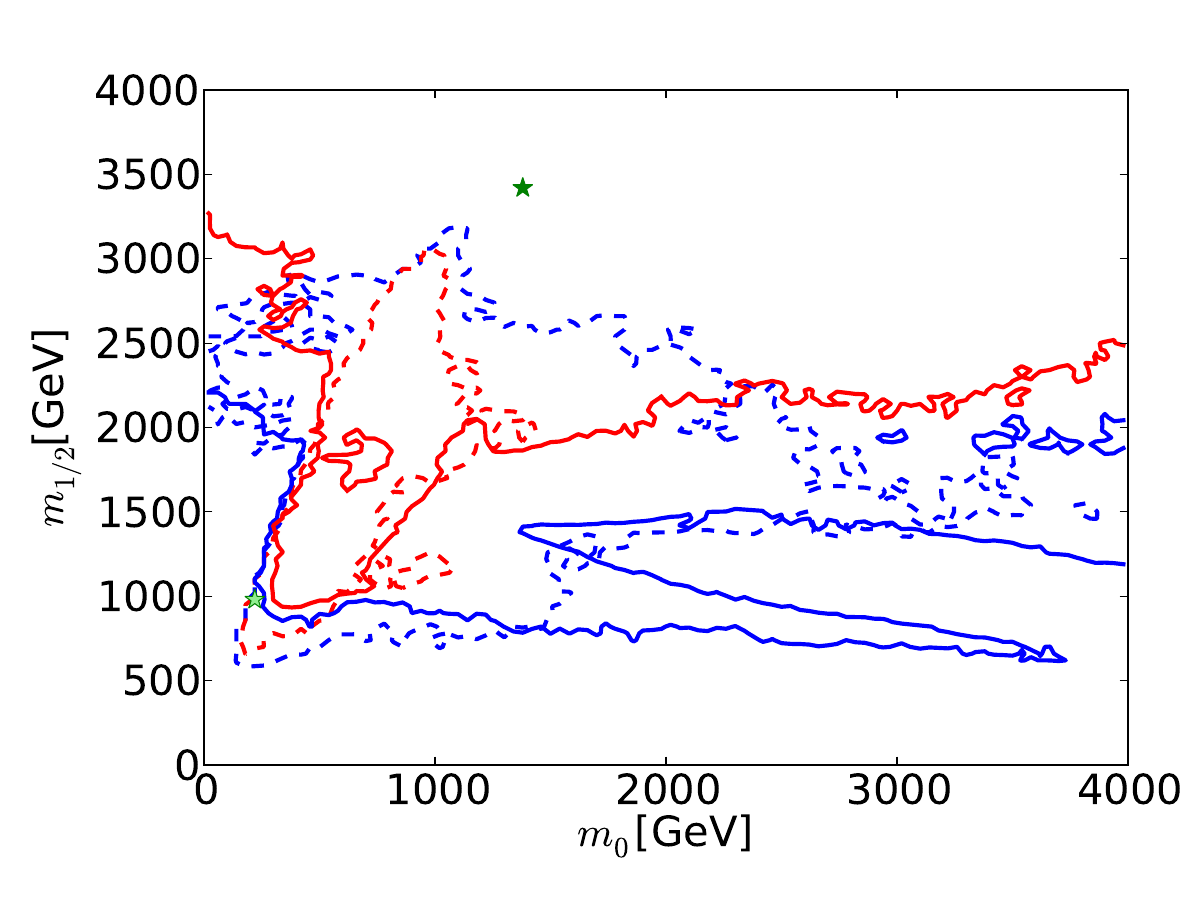}}
\vspace{-1cm}
\caption{\it The $(m_0, m_{1/2})$ planes in the CMSSM (left) and the NUHM1 (right) for $\mu > 0$, comparing 
the results of the current CMSSM fit (solid lines and filled  stars), with the results
shown in~\cite{mc8} (dashed lines and open stars).
The red lines denote $\Delta \chi^2 = 2.30$ contours (corresponding approximately to
the  68\%~CL), and the red lines denote $\Delta \chi^2 = 5.99$ (95\%~CL)
contours.}  
\label{fig:mc8old}
\end{figure*}

The global likelihood function calculated in~\cite{mc8} had two local minima
with almost equal values of $\chi^2$. The new ATLAS constraint and the new
implementations of the $\Mh$ and \ssi\ constraints combine to slightly
disfavour the local minimum in the low-mass `island' by only $\sim 0.7$ in $\chi^2$ compared
to the global minimum at $(m_0, m_{1/2}) \sim (5650, 2100) \gev$, where
the main contribution comes from the ATLAS $\ETslash$ constraint. 
As in the
case of our previous analysis~\cite{mc8}, the \bsdmm\ constraint does not 
play a large r\^ole in the current fit. Its main importance is at large $\tb$
and small $m_0, m_{1/2}$ and $\MA$, but the low-mass `island' has small
$\tb$ and the current best-fit point has large values of $m_0, m_{1/2}$
and $\MA$. 


\subsubsection{Characteristics of the Best-Fit Points for \boldmath{$\mu > 0$}}

\medskip
Table~\ref{tab:bestfits} summarizes the values of $\chi^2$ and the locations of the
best-fit points found in the current analysis in the low-mass (`island') and high-mass
(`continent') regions of the CMSSM parameter space with $\mu > 0$%
~\footnote{We discuss later the best-fit points for the CMSSM with 
$\mu < 0$ and for the NUHM1 with $\mu > 0$.}%
. We see that although the minimum value of $\chi^2$ in the `continent'
is smaller than in the `island',  the difference is less than unity, and
hence not significant. As already mentioned, the `island' best-fit point is in the stau-coannihilation
region, whereas rapid annihilation via direct-channel $H/A$ poles is dominant
at the best-fit point in the high-mass `continent'.

\medskip
Comparing with the best fit  found previously in
the CMSSM using the ATLAS 7-TeV $\ETslash$ constraint and the previous
\bsdmm\  measurement, we see that the best-fit $\chi^2$  has increased
by about 2.1, and the `island' $\chi^2$ by about 3. Thus the
pressure  exerted by the \atlastwenty\ and \bsdmm\ constraints does not
change  significantly the overall picture for the CMSSM. Specifically,
the values of $m_0$ and $m_{1/2}$  at the new best-fit `island' point
are not very different from those at the previous CMSSM best-fit  point,
though the values of $A_0$ and $\tb$ have changed substantially. Also
shown for comparison is the value of $\chi^2$ for the ``SM", as calculated
using the {\tt MasterCode} by setting $m_0 = m_{1/2} = 15$~TeV. We see
that the CMSSM is unable to reduce $\chi^2$ much below the ``SM"
value, with a similar fit probability.

\begin{table*}[htb!]
\renewcommand{\arraystretch}{1.2}
\begin{center}
\small{
\begin{tabular}{|c||c|c||c|c||c|c||c|} \hline
Observable & $\Delta \chi^2$ & $\Delta \chi^2$ & $\Delta \chi^2$ & $\Delta \chi^2$ & $\Delta \chi^2$ & $\Delta \chi^2$ & $\Delta \chi^2$ \\
& CMSSM & CMSSM & CMSSM & CMSSM & NUHM1 & NUHM1 & Standard \\
& $\mu > 0$ (high) & $\mu > 0$ (low) &$\mu < 0$ (high) & $\mu < 0$ (low) & $\mu > 0$ (high) &  $\mu > 0$ (low) & Model \\
\hline \hline
Global                                  & 35.1 & 35.8 & 36.6  & 38.9  & 32.7 & 33.3 & 36.5 \\
\hline \hline
BR$_{\rm b \to s \gamma}^{\rm exp/SM}$  & 0.52 & 1.58 &  0.37 &  0.00 & 0.54 & 0.02 & 0.57 \\
\hline                                                                                    
BR$_{\rm B \to \tau\nu}^{\rm exp/SM}$   & 1.77 & 1.63 &  1.63 &  1.61 & 1.65 & 1.66 & 1.60 \\
\hline                                                                                    
$\epsilon_K$                            & 1.94 & 1.88 &  1.94 &  1.87 & 1.94 & 1.94 & 1.96 \\
\hline                                                                                    
$ a_{\mu}^{\rm exp} - a_{\mu}^{\rm SM}$ & 10.71& 9.34 &  11.42&  12.65& 10.50& 9.63 & 11.19\\
\hline                                                                                    
$\MW$                                   & 1.35 & 0.22 &  2.15 &  0.04 & 0.00 & 0.11 & 1.38 \\
\hline                                                                                    
$\Mh$                                   & 0.00 & 0.04 &  0.03 &  0.53 & 0.00 & 0.22 & (1.5)\\
\hline                                                                                    
$ R_\ell$                               & 1.10 & 1.04 &  1.10 &  1.00 & 1.07 & 1.00 & 1.09 \\ 
\hline                                                                                    
$ A_{\rm fb}({b})$                      & 6.56 & 6.79 &  6.05 &  7.61 & 5.45 & 6.93 & 6.58 \\ 
\hline                                                                                    
$ A_\ell({\rm SLD})$                    & 3.59 & 3.40 &  3.99 &  2.81 & 4.59 & 3.30 & 3.55 \\ 
\hline                                                                                    
$\sigma_{had}^0$                        & 2.52 & 2.55 &  2.56 &  2.51 & 2.59 & 2.56 & 2.54 \\ 
\hline                                                                                    
LUX                                     & 0.03 & 0.07 &  0.66 &  0.07 & 0.00 & 0.07 & - \\ 
\hline                                                                                    
ATLAS 20/fb                             & 0.04 & 2.52 &  0.02 &  3.35 & 0.02 & 1.15 & - \\
\hline                                                                                    
$B_{s,d} \to \mu^+ \mu^-$               & 0.51 & 0.46 &  0.13 &  0.11 & 0.22 & 0.35 & 0.15 \\ 
\hline
\end{tabular}
\caption{\it Summary of the contributions of the most relevant
  observables to the global $\chi^2$ function at the best-fit high- and low-mass points in the
  CMSSM (with both signs of $\mu$) and NUHM1 (with $\mu > 0$), including the recently-updated observables
  \atlastwenty, \bsdmm\ and the LUX upper limit on dark matter scattering. As noted in parentheses,
  within the SM, $\Delta \chi^2 \sim 1.5$ is found in~\cite{Gfitter} due to the (small) tension between the measured
value of $\Mh$ and the precision electroweak data. 
}.
  \label{tab:chi2}}
\end{center}

\end{table*}

\medskip
Table~\ref{tab:chi2} gives more details of the contributions to the
global $\chi^2$ function from different observables in the CMSSM at the
high- and low-mass best-fit points, compared with our implementation of
the SM. At both the high- and low-mass points, the $\Mh$ measurement 
makes a small contribution to the global $\chi^2$ function.
We see that the low-mass point has less tension with $(g-2)_\mu$, 
and is favoured by both $M_W$ and ${\rm BR}(B_{s,d}\to\mu^+\mu^-)$, 
in particular, whereas the high-mass point is preferred by ${\rm BR}(b\rightarrow s\gamma)$ 
and ATLAS 20/fb jets + $\ETslash$, in particular.
The ``SM" fit is noticeably worse for $\gmt$ and $\MW$.


\subsubsection{One-Dimensional Likelihood Functions for \boldmath{$\mu > 0$}}

\medskip
We now present the one-dimensional $\chi^2$ likelihood functions for various
particle masses and other observables when $\mu > 0$, which are shown as
continuous lines in Fig.~\ref{fig:CMSSM1D} (the dotted lines are
  discussed below).
The upper left panel displays the $\chi^2$ function for $\mgl$. We see
that it falls essentially monotonically for $\mgl \gsim 1000 \gev$, a
feature that masks the structures seen in the upper left  panel of
Fig.~\ref{fig:CMSSM}. The one-dimensional projection merges the low-mass
`island' and the high-mass `continent' that are separated in the 
$(m_0, m_{1/2})$ plane of the CMSSM. It is to be expected that the
$\chi^2$ function continues close to zero also at larger values of
$\mgl$. 

\begin{figure*}[htb!]
\resizebox{8cm}{!}{\includegraphics{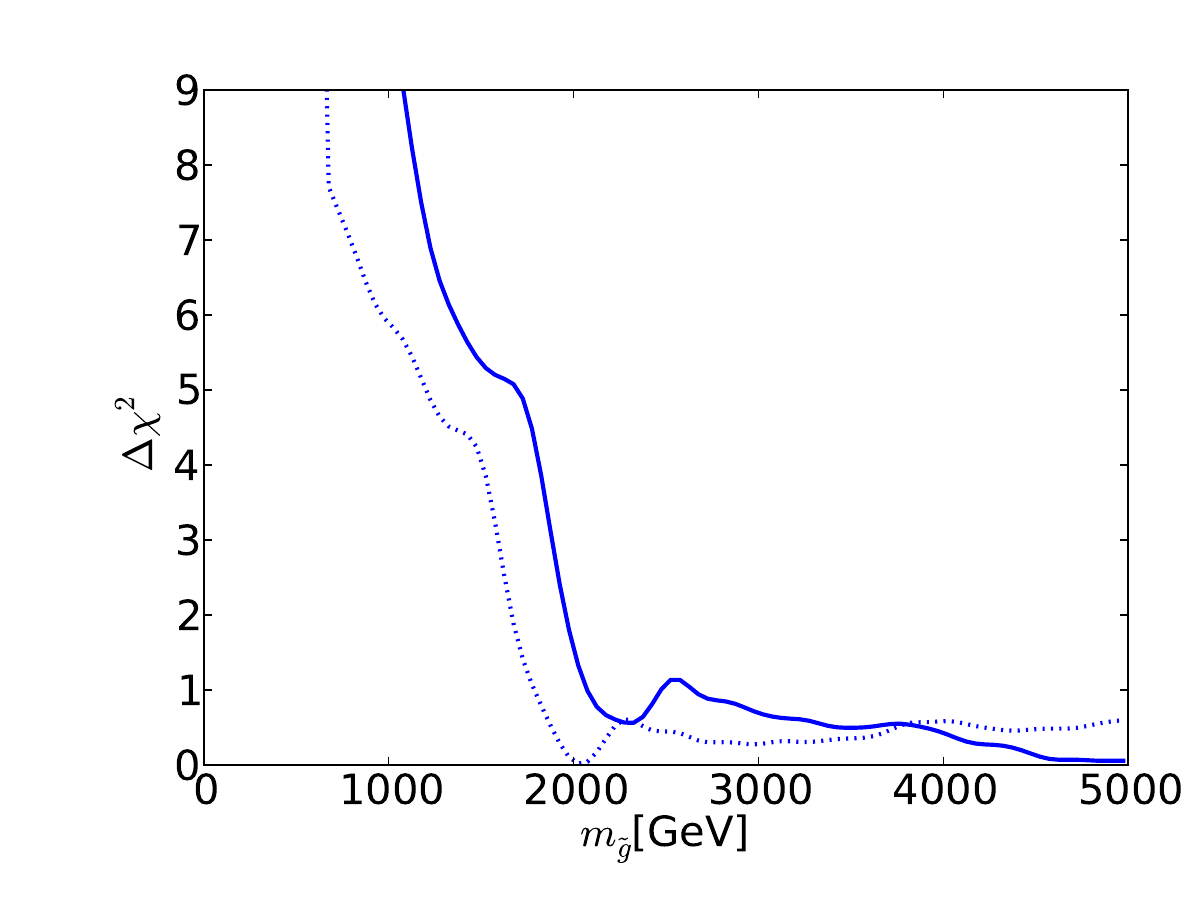}}
\resizebox{8cm}{!}{\includegraphics{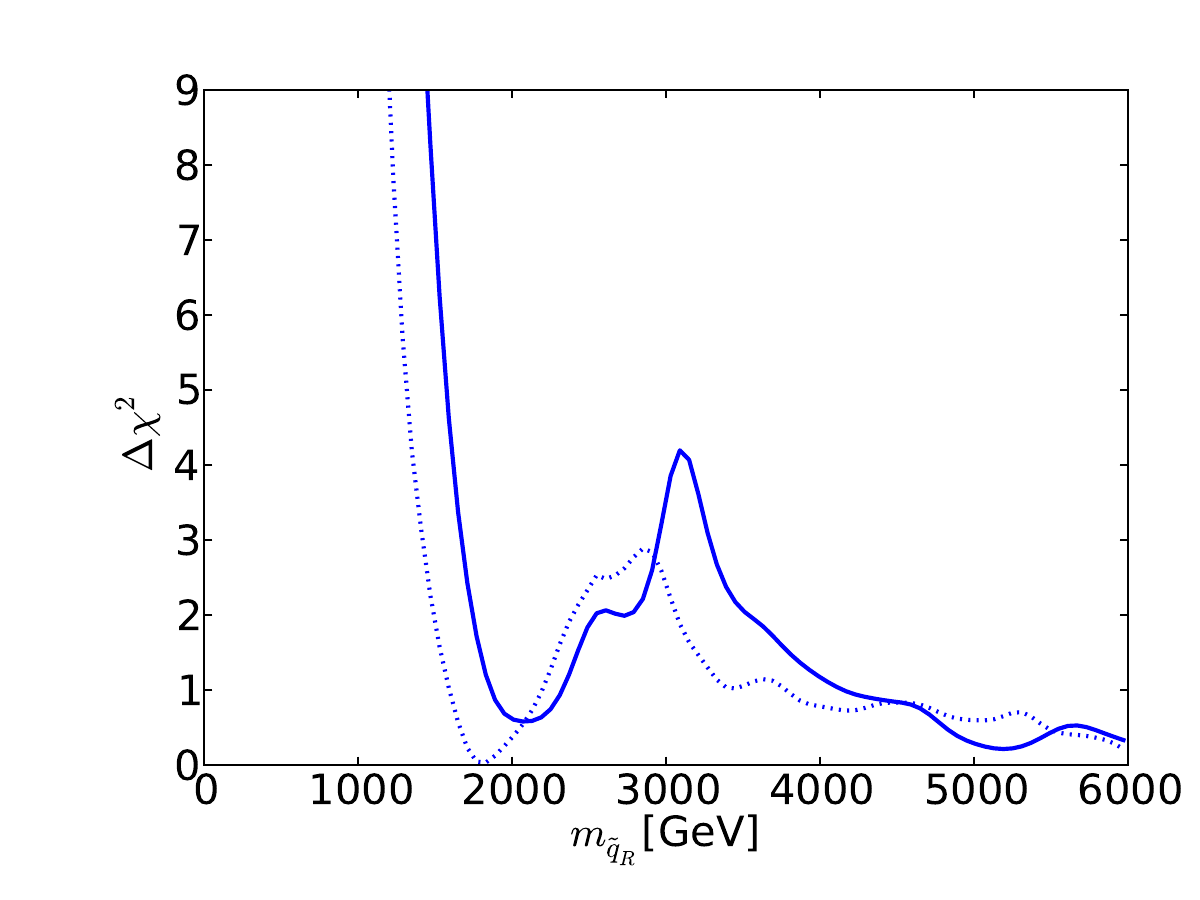}}
\resizebox{8cm}{!}{\includegraphics{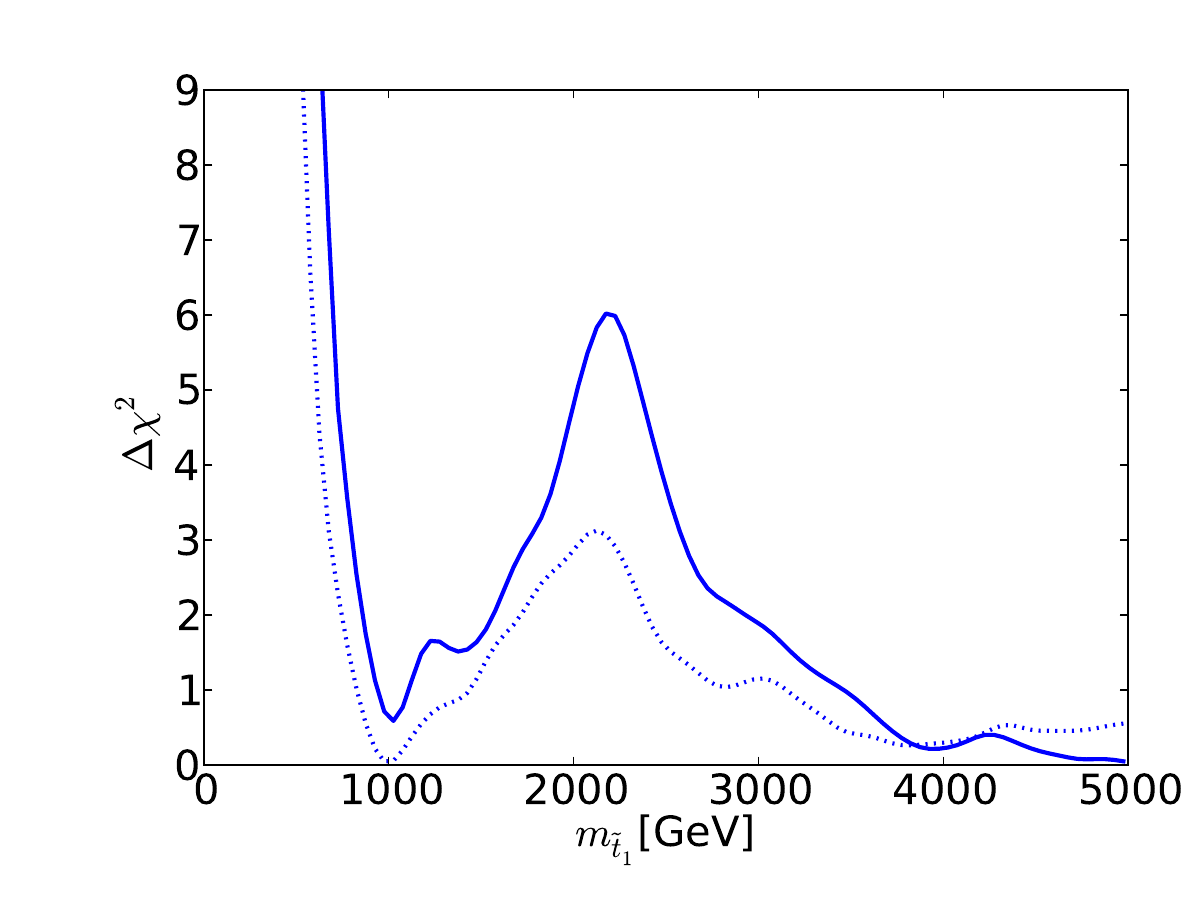}}
\resizebox{8cm}{!}{\includegraphics{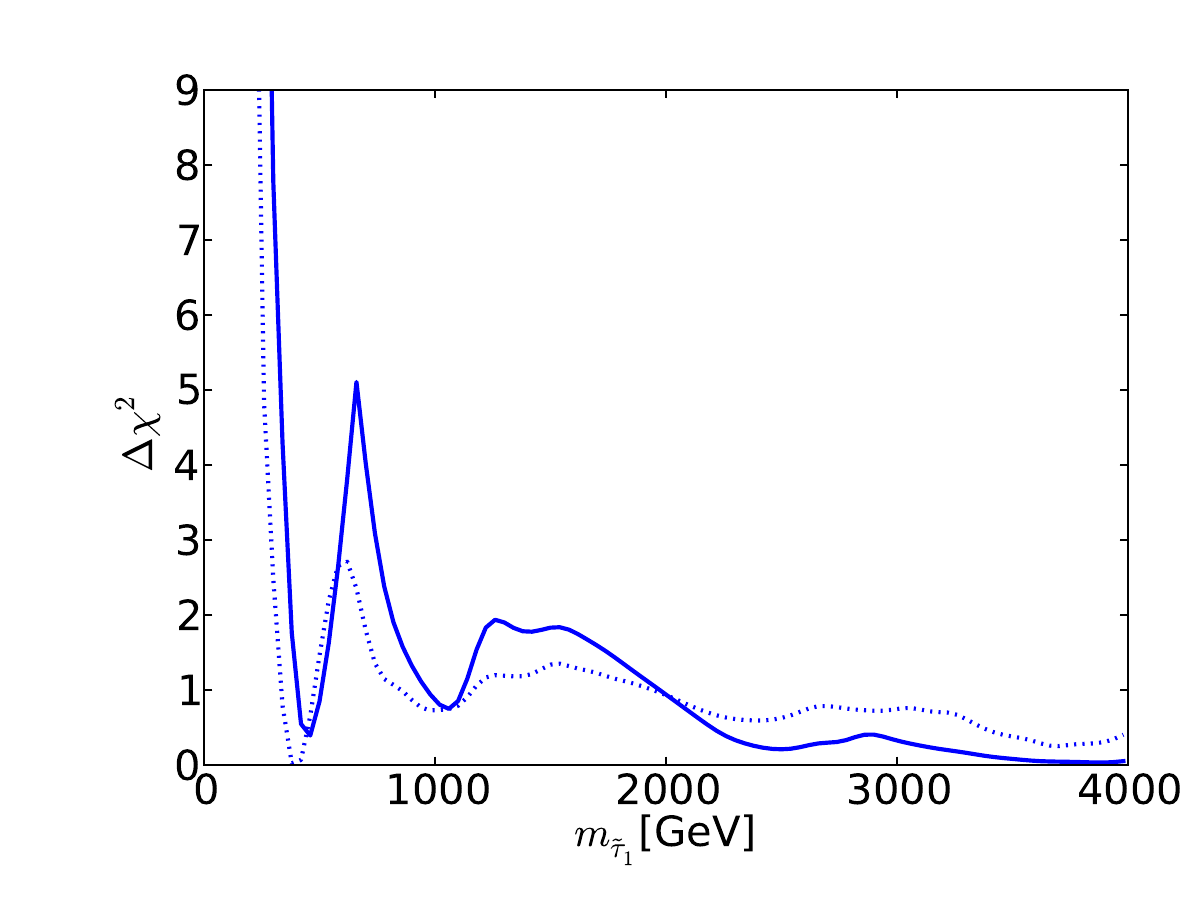}}
\vspace{-1cm}
\caption{\it The one-dimensional $\chi^2$ likelihood functions in the CMSSM for $\mu > 0$ for
$\mgl$ (upper left), $\msqR$ (upper right), $\mste$ (lower left) and $\mstaue$
(lower right). In each panel, the solid line is derived from a global analysis of the
present data, and the dotted line is obtained from a reanalysis of the data used
in~\cite{mc8}, using the implementations of the $\Mh$ and \ssi\ constraints
discussed in Section~2.} 
\label{fig:CMSSM1D}
\end{figure*}

\medskip
The $\chi^2$ function for $\msqR$ seen in the upper right panel of Fig.~\ref{fig:CMSSM1D}
exhibits more structure, with a local minimum at $\msqR \sim 2200 \gev$, a local maximum
at $\msqR \sim 3000 \gev$, and then an essentially monotonic fall at larger $\msqR$.
The appearance of the local minimum can be understood by remembering that
$\msqR^2 \sim m_0^2 + 5 m_{1/2}^2$, so that the value of $\msqR$ is fixed along
elliptical contours in the $(m_0, m_{1/2})$ plane. The local minimum in the $\chi^2$
function for $\msqR$ corresponds to an ellipse passing through the red `island' in 
the upper left panel of Fig.~\ref{fig:CMSSM}, and the local maximum corresponds to
an ellipse passing between the `island' and the `continent'. However, we should emphasize that 
neither the local minimum nor the local maximum is very significant, since they have
$\Delta \chi^2 \sim 1, 4$ relative to the minimum value of $\chi^2$.

\medskip
Similar features are seen in the $\chi^2$ function for the mass of the lighter stop squark,
$\mste$, as seen in the lower left panel of Fig~\ref{fig:CMSSM1D}. 
However, in this case the local minimum appears at a lower mass $\mste \sim 1000 \gev$,
and the local maximum is also at a lower mass $\mste \sim 2000 \gev$, reflecting the fact
that the isomass contours for $\mste$ and $\msqR$ are different. As in many other models,
we find that the ${\tilde t_1}$ is likely to be considerably lighter
than the other strongly-interacting sparticles.
This is due to a large mixing in the scalar top sector, driven by
  the relatively large value of $\Mh$.

\medskip
Similar local structures can be seen in the $\chi^2$ function for the lighter stau,
$\mstaue$, as seen in the lower right panel of Fig.~\ref{fig:CMSSM1D}. In this case, the
local minimum is at $\mstaue \sim 450 \gev$, nearly degenerate with the lightest neutralino,
and placing the ${\tilde \tau_1}$ and other sleptons
beyond the reach of an $e^+ e^-$ collider with $E_{\rm CM} \lsim 900 \gev$. We also find that
$\Delta \chi^2 > 9$ for $\mstaue < 300 \gev$. However, we emphasize that these
observations are very model-dependent.

\medskip
We now comment briefly on the differences between the one-dimensional likelihood
functions found in our analysis of the current data, and those found 
using the same implementations of the $\Mh$ and \ssi\ constraints for the data 
set used in~\cite{mc8}, shown in Fig.~\ref{fig:CMSSM1D} as dotted
lines. The current
likelihood functions for $\mgl, \msqR, \mste$ and $\mstaue$ are generally higher at small masses,
where the ATLAS $\ETslash$ search has the most impact, but are similar at high masses.

\medskip
Fig.~\ref{fig:CMSSMHiggs} displays the $\chi^2$ functions for the mass of the lightest SUSY
Higgs boson, $\Mh$, shown in the left panel, and the mass of the pseudoscalar Higgs boson, $\MA$,
shown in the right panel. We see that the likelihood for $\Mh$ is well maximized close to the
measured Higgs mass.
The likelihood for $\MA$ is very flat for $\MA \gsim 1000 \gev$,
with $\Delta \chi^2$ rising rapidly to reach $> 9$ for $\MA < 500 \gev$, and is very similar to
the likelihood found using the same data set as in~\cite{mc8}.

\begin{figure*}[htb!]
\resizebox{8cm}{!}{\includegraphics{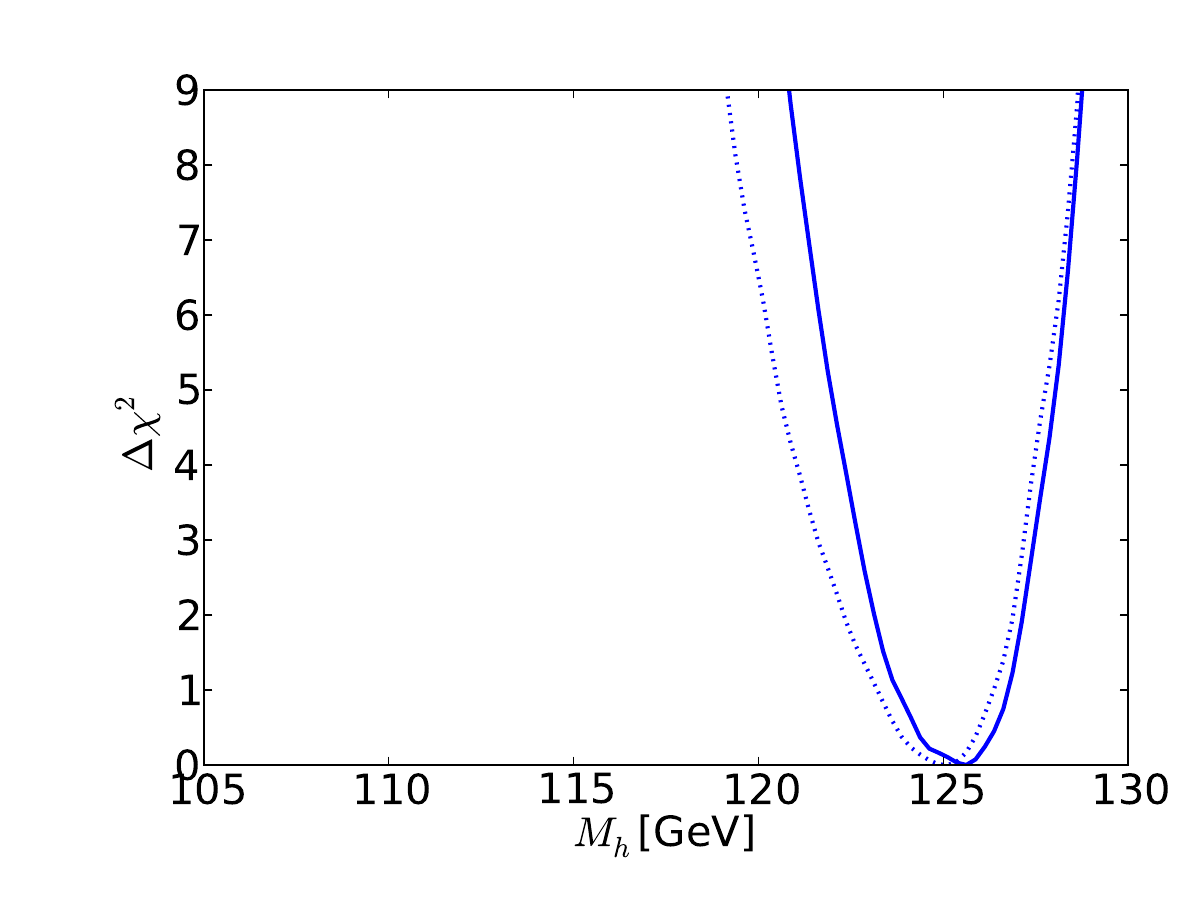}}
\resizebox{8cm}{!}{\includegraphics{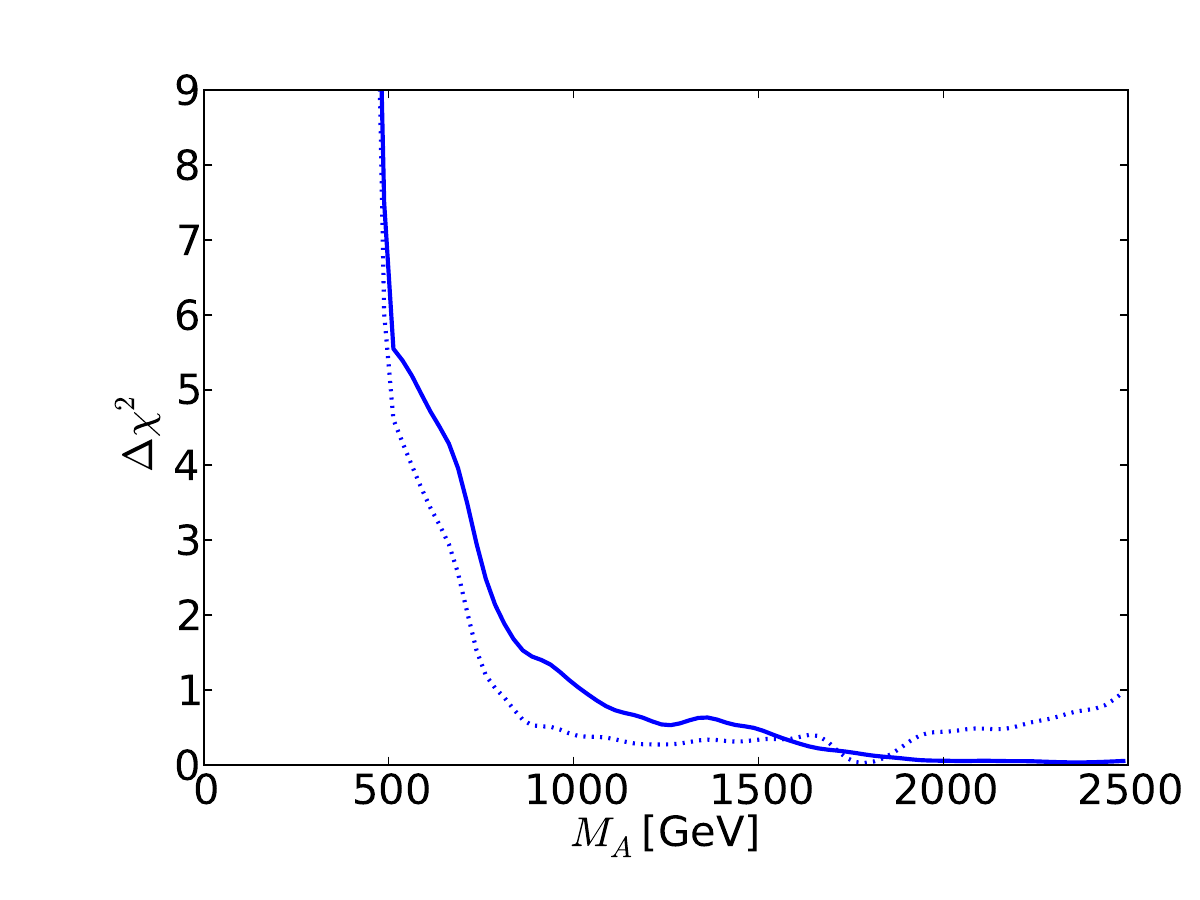}}
\vspace{-1cm}
\caption{\it The one-dimensional $\chi^2$ likelihood functions in the CMSSM for $\mu > 0$ for
$\Mh$ (left) and $\MA$
(right). In each panel, the solid line is derived from a global analysis of the
present data, and the dotted line is derived from 
a reanalysis of the data used
in~\cite{mc8}, using the implementations of the $\Mh$ and \ssi\
constraints discussed in Section~2.}
\label{fig:CMSSMHiggs}
\end{figure*}

\medskip
On the basis of these one-dimensional likelihood functions we can establish 95\% CL
lower limits on $\mgl, \msqR, \mste$ and $\mstaue$ for the CMSSM with $\mu > 0$, 
which are listed in the second column of Table~\ref{tab:lowerlimits}.
Reflecting the relatively large values of $m_0$ favoured in this analysis, we see that
the lower limit on $\msqR$ is considerably stronger than that on $\mgl$. On the other hand,
the ${\tilde t_1}$ could be substantially lighter than the other strongly-interacting sparticles.

\begin{table*}[!tbh!]
\renewcommand{\arraystretch}{1.5}
\begin{center}
\begin{tabular}{|c||c|c|c|} \hline
 & CMSSM & CMSSM & NUHM  \\
Sparticle &  $\mu > 0$  & $\mu < 0$ & $\mu > 0$ \\ 
\hline \hline
${\tilde g}$ & 1810  & (2100) (3200) 3540 & 1920 \\
${\tilde q_R}$ & 1620 & (1900) 6300 & 1710 \\
${\tilde t_1}$ & 750 & (950) 4100 & (650) 1120 \\
$\staue$ & 340 & (400) 4930 & 380 \\
$\MA$ & 690 & (1900) 3930 & 450 \\
\hline
\end{tabular}
\caption{\it  The 95\% CL lower limits (in GeV) on various sparticle masses
in the CMSSM with both signs of $ \mu$ and the NUHM1 with $\mu > 0$.
We emphasize that these limits are specific to the models studied.
In the case of the CMSSM with $\mu < 0$ and the NUHM1, the parentheses indicate the approximate 
locations of small mass ranges where the $\chi^2$ function dips briefly below the 95\% CL.
}
\label{tab:lowerlimits}
\end{center}
\end{table*}

\medskip
The left panel of Fig.~\ref{fig:CMSSMothers} displays the likelihood function for \bsdmm,
which is seen to be minimized close to the SM value. The rise at larger
\bsdmm\ is largely due to the direct experimental constraint on this quantity, but the steep rise at
lower \bsdmm\ is due to the other constraints on the CMSSM, which are hard to reconcile
with $R_{\mu \mu } < 1$. 
The rise at large \bsdmm\
found from the data set used in~\cite{mc8} is less steep, reflecting the evolution in the
measurement of \bsdmm. The right panel of Fig.~\ref{fig:CMSSMothers} displays the
$(\mneu1 , \ssi)$ plane, again with solid (dashed) lines representing the current analysis
(the constraints of~\cite{mc8}), respectively, with the filled (open) green star denoting the corresponding best-fit point
whereas the red (blue) lines representing 68 (95)\%
CL contours, respectively.
We see that a range $10^{-47} {\rm cm}^2 \lsim \ssi\ \lsim 10^{-43}$~cm$^2$ is allowed at the 95\% CL,
and the best-fit point yields a value in the middle part of this range
$\sim 10^{-45}$~cm$^2$. The mass  of $\mneu1$ at the best fit point is $935 \gev$.

\begin{figure*}[htb!]
\resizebox{8cm}{!}{\includegraphics{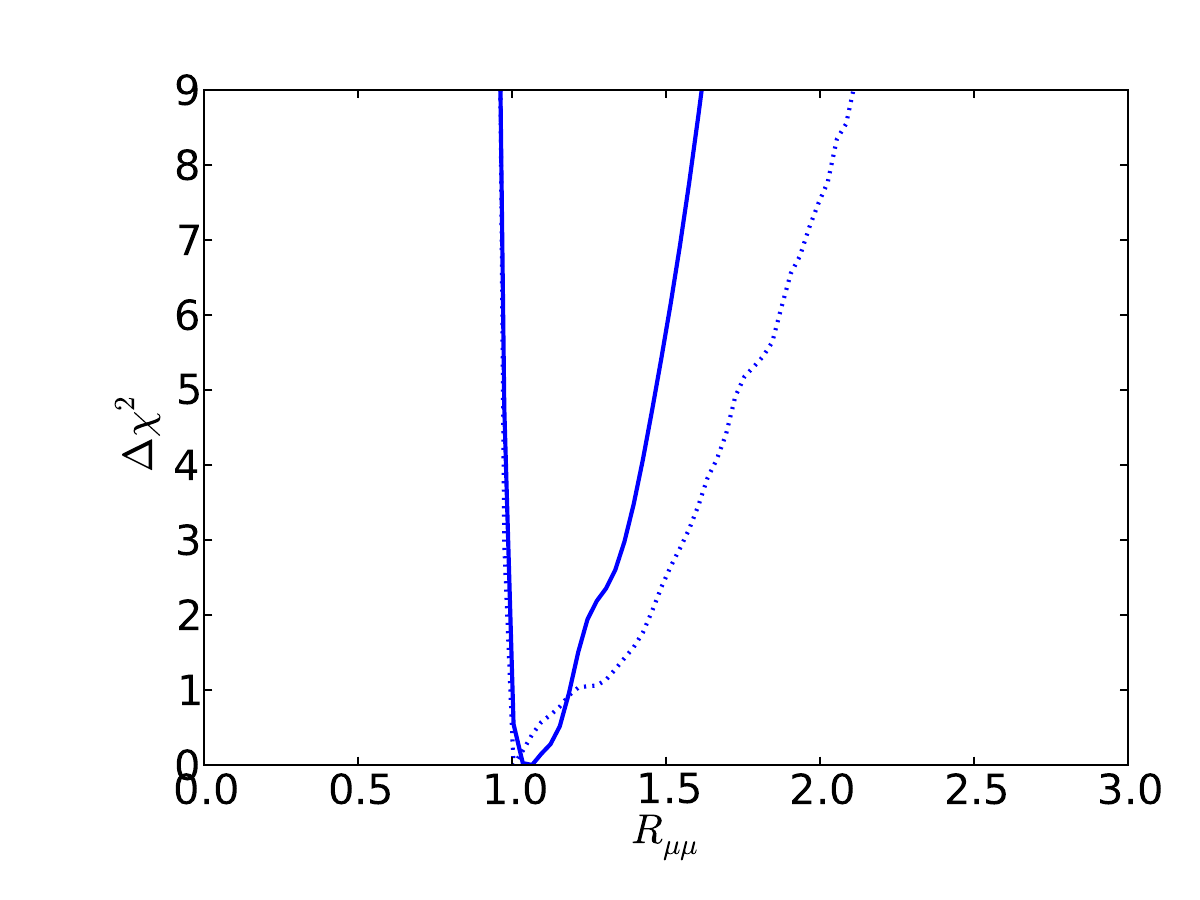}}
\resizebox{8cm}{!}{\includegraphics{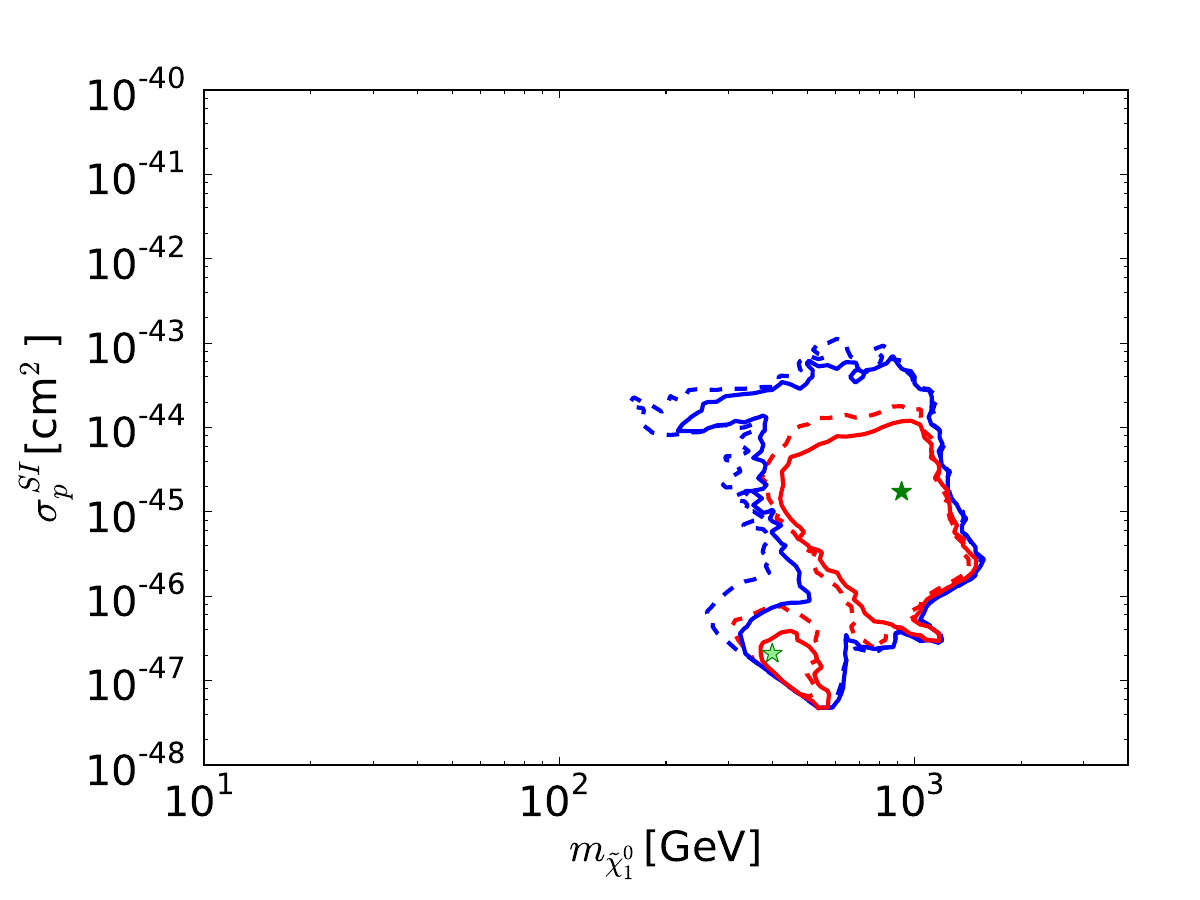}}
\vspace{-1cm}
\caption{\it The one-dimensional $\chi^2$ likelihood function in the CMSSM for $\mu > 0$ for
\bsdmm\ (left) and the $(\mneu1 , \ssi)$ plane (right).
In both panels, the solid lines are derived from a global analysis of the
present data, and the dotted lines are derived from 
a reanalysis of the data used
in~\cite{mc8}, using the implementations of the $\Mh$ and \ssi\
constraints discussed in Section~2.
In the right panel, the red lines denote
the $\Delta \chi^2 = 2.30$ contours, the blue lines denote the $\Delta \chi^2 = 5.99$
contours in each case, and the filled (open) green star denotes the corresponding best-fit point.} 
\label{fig:CMSSMothers}
\end{figure*}


\subsubsection{Comparisons between Analyses}

\medskip
We restrict our attention here to the only other analysis that
incorporates the latest \atlastwenty\ constraint. Preliminary results
from a new global frequentist analysis of the CMSSM with $\mu > 0$
within the {\tt F{\small ITTINO}} framework have recently been
presented~\cite{Stefaniak}. The best-fit point found in~\cite{Stefaniak}
is very similar to the best-fit point we find in the low-mass region of
the CMSSM with $\mu > 0$. However, the regions of the parameter space
favoured at the 68 and 95\% CL in the {\tt F{\small ITTINO}} analysis do
not extend to values of $(m_0, m_{1/2})$ as large as those we find in
the present analysis. In addition to \atlastwenty, this analysis
also uses {\tt H{\small IGGS}S{\small SIGNALS}} to derive constraints
from the Higgs mass and signal strength measurements. The latter do not
change substantially the results, since the Higgs rate predictions in
the favoured regions of the CMSSM parameter space, which are in
the in the decoupling regime 
\footnote{The fact that the light CMSSM Higgs boson should be
SM-like was already a {\em pre}-LHC prediction of the
model~\cite{CMSSM-SM-like-Higgs}.}%
, are quite similar to those in the SM and do not vary
significantly%
\footnote{However, adding many channels of Higgs production and decay
  properties whose measurements agree with the predictions for a
  SM Higgs boson does yield a better $\chi^2/{\rm dof}$.}%
.


\subsection{CMSSM with \boldmath{$\mu < 0$}}

\medskip
The case $\mu < 0$ has been studied less than $\mu > 0$ (but see, e.g., \cite{Arbey:2012bp,moremuneg}),
for various reasons:
It {\it worsens} the discrepancy  between the
experimental value of $\gmt$ and the SM calculation,
it is in general {\it more restricted} by \bsg\, and it yields a {\it smaller}
value of $\Mh$ for fixed values of the other CMSSM parameters.
However, since the \atlastwenty\ and other constraints require
relatively large values of $m_0$ and $m_{1/2}$ where the
SUSY contribution to $\gmt$ and \bsg\ are small,
it is appropriate to reconsider the $\mu < 0$ case.

\subsubsection{Parameter Planes with \boldmath{$\mu < 0$}}

\medskip
We see in the upper left panel of Fig.~\ref{fig:CMSSMneg} that 
there are three regions of the $(m_0, m_{1/2})$
plane that are allowed at the 95\% level, two small `reefs' at relatively low masses $(m_0, m_{1/2}) \sim (300, 1000)$
and $(600, 2000) \gev$ and a more extensive
`continent' at larger masses $m_0 \gsim 4000 \gev$.
The lower-mass `reef' is in the stau-connihilation region, as in the $\mu > 0$ case,
but the higher-mass `reef' is in the stop-coannihilation region. Compared to the
high-mass `continent' in the rapid-annihilation funnel and focus-point regions, the `reef' has smaller
contributions to the global $\chi^2$ function for some electroweak and
flavour observables, but is disfavoured by \atlastwenty. The
best-fit point in the CMSSM for $\mu < 0$ is shown as a yellow star: it is located
in the high-mass `continent', in the focus-point region.

\begin{figure*}[htb!]
\resizebox{8.5cm}{!}{\includegraphics{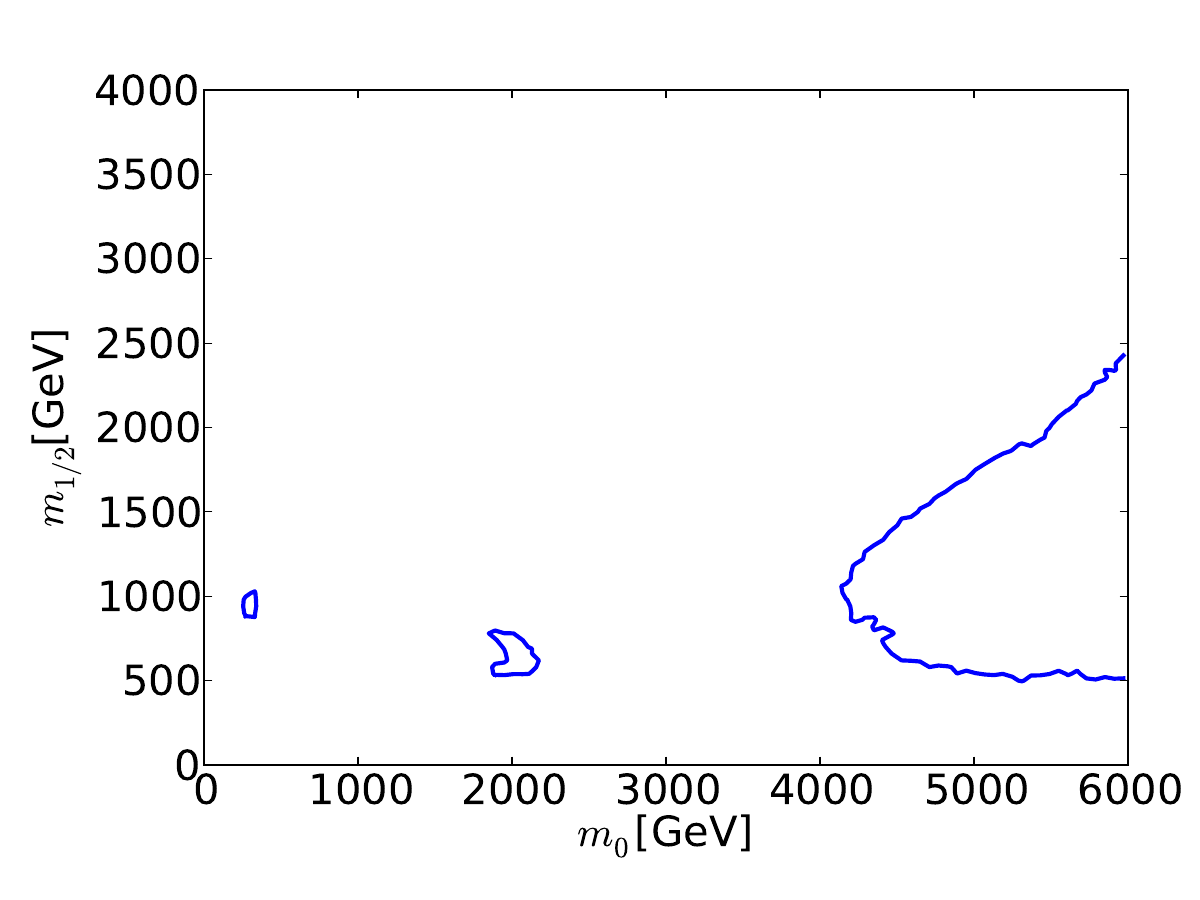}}
\resizebox{8.5cm}{!}{\includegraphics{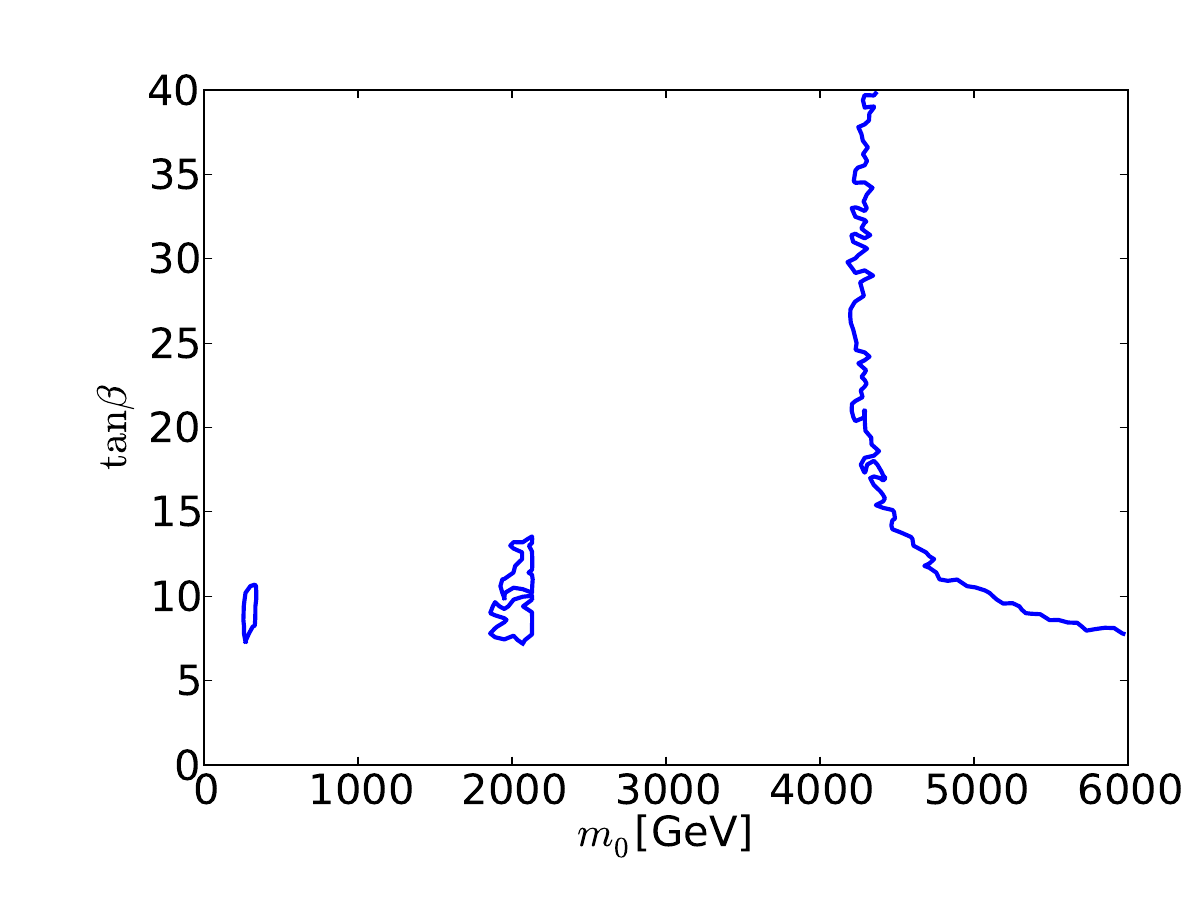}}
\resizebox{8.5cm}{!}{\includegraphics{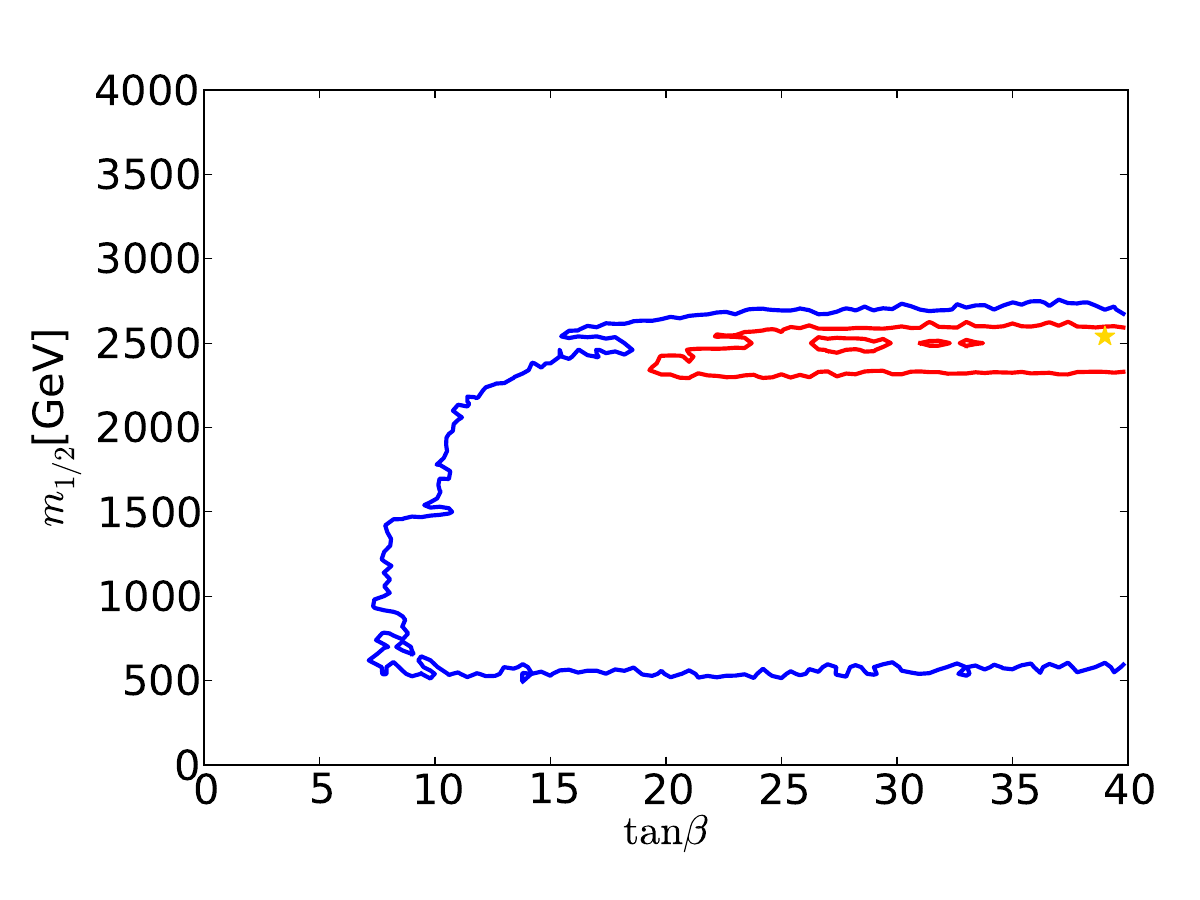}}
\resizebox{8.5cm}{!}{\includegraphics{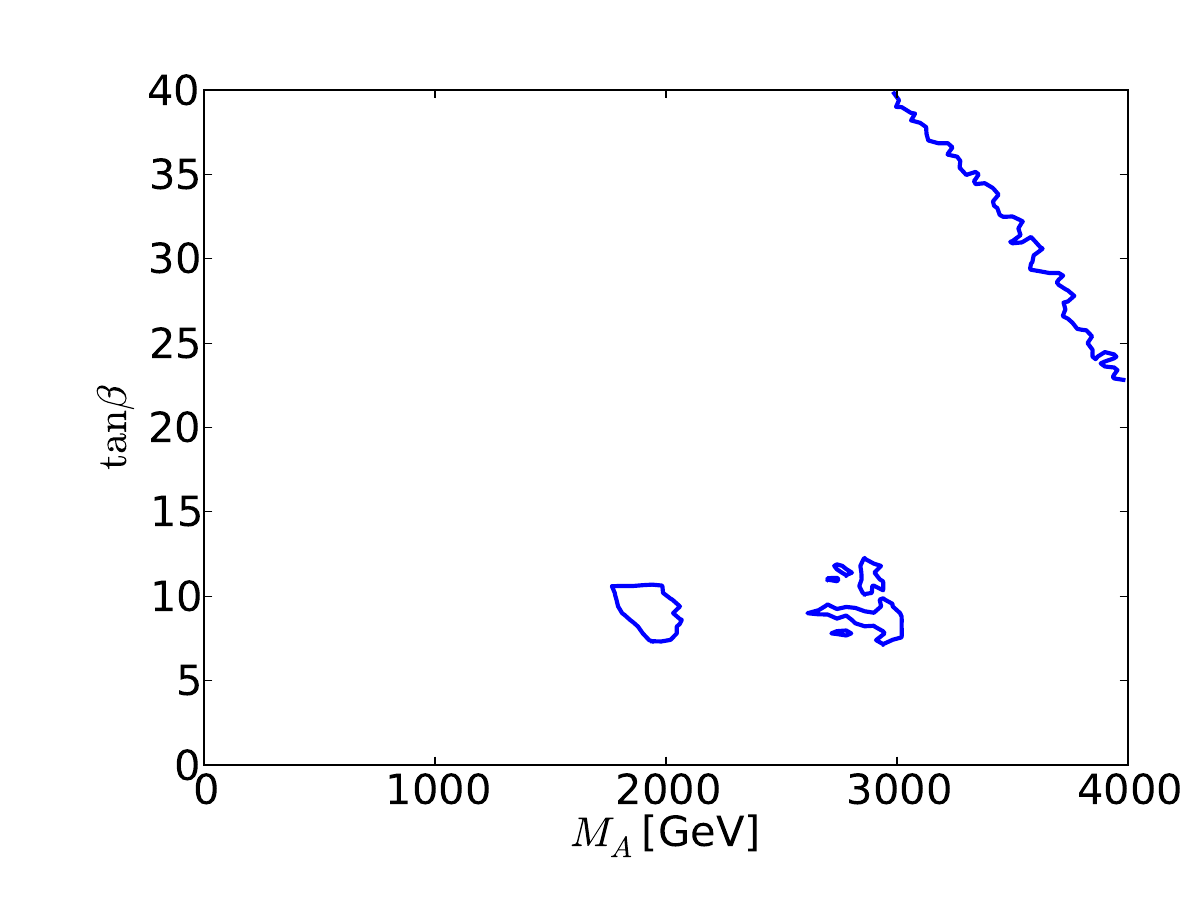}}
\vspace{-1cm}
\caption{\it As in Fig.~\protect\ref{fig:CMSSM}, but for $\mu < 0$ in the CMSSM. For the reason discussed in the text,
only the ranges $\tb \le 40$ are displayed. The yellow star in the lower left panel marks the best-fit point in the CMSSM
with $\mu < 0$, which is out of the ranges of the other panels.} 
\label{fig:CMSSMneg}
\end{figure*}

\medskip
The $(m_0, \tb)$ plane for $\mu < 0$ is shown in the upper right panel
of Fig.~\ref{fig:CMSSMneg}~\footnote{Here and in subsequent panels, we restrict
attention to $\tb \le 40$. The electroweak vacuum conditions can be
satisfied for larger values of $\tb$, but the ranges of $m_0$ and $A_0$
studied here give incomplete sampling in this case.}. Here we see that the
low-mass `reefs' are restricted to $5 \lsim \tb \lsim 15$, whereas the `continent'
extends over all $\tb \gsim 8$.
In the lower left panel of Fig.~\ref{fig:CMSSMneg}, we see in the 
$(\tb, m_{1/2})$ plane that the `reefs' and `continent'
merge in this projection of the CMSSM parameter space with $\mu < 0$. We also see that
$500~\gev \lsim m_{1/2} \lsim 2500 \gev$ is allowed at the 95\% CL for the range $m_0 < 6000 \gev$
studied here~\footnote{As in the
$\mu > 0$ case, we have not studied in detail the sensitivity to $\tb$ of the lower bound on $m_{1/2}$
due to \atlastwenty\ searches with leptons and/or $b$ jets, which are not important near the `reef' or the best-fit
point for $\mu < 0$.}. The small region within the red 68\% contour does not appear in the other
panels, because it corresponds to values of $m_0 > 6000 \gev$ and 
$\MA > 4000 \gev$, which are not displayed in the other panels of Fig.~\ref{fig:CMSSMneg}.
Finally, in the lower right panel of Fig.~\ref{fig:CMSSMneg} 
we see in the $(\MA, \tb)$ plane that only in the `reefs' are values
of $\MA \lsim 3000 \gev$ are allowed at the 95\% CL when $\tb \le 40$~\footnote{However, our 
incomplete sampling at larger $\tb$ shows that $\MA \sim 1000 \gev$ is allowed
for $\tb \sim 50$.}. The `reef's are again clearly separated at relatively small
values of $\tb$, with a restricted range of $\MA \in (2000, 3000) \gev$.

\subsubsection{Characteristics of the Best-Fit Points for \boldmath{$\mu < 0$}}

\medskip
We display in Table~\ref{tab:bestfits} the characteristics of the best-fit points in the CMSSM with
$\mu < 0$ in the low-mass `reef' region and the high-mass `continent'. Unlike the case of the CMSSM
with $\mu > 0$, ${\tilde t_1}$ coannihilation is important at the best-fit point in the `reef' region, and
${\tilde \chi^\pm_{1}}, \neu2, \neu3 $ coannihilation at the best-fit `continental' point. In both cases, the global
$\chi^2$ function is somewhat higher than in the corresponding regions for $\mu > 0$, by
$\sim 3.1$ in the low-mass region and by $\sim 1.5$ in the high-mass region. The main origins
of the differences can be seen in Table~\ref{tab:chi2}. The high-mass model with $\mu < 0$ receives
larger contributions from $\gmt$, $\MW$ and $\ssi$, whereas there are larger contributions
from $\gmt$ and $\Mh$ in the low-mass case, compensated only partially by smaller $\chi^2$ contributions from \bsg\ and \bsdmm.
As a result, the best-fit CMSSM points with $\mu < 0$ have higher
$\chi^2$ and lower fit probabilities than the SM.


\subsubsection{One-Dimensional Likelihood Functions for \boldmath{$\mu < 0$}}

\medskip
We display in Fig.~\ref{fig:CMSSM1Dneg} the one-dimensional $\chi^2$ functions
for various sparticle masses in the CMSSM with $\mu < 0$. We see in the upper left panel
that the $\chi^2$ function for $\mgl$ falls essentially monotonically as $\mgl \to 5000 \gev$ towards $\Delta \chi^2 \sim 2.5$  
relative to the global minimum. The best fit for $\mu < 0$ has $\Delta \chi^2 \sim 1.8$ at
$\mgl \sim 5300 \gev$, and hence is not seen in this plot.

\begin{figure*}[htb!]
\resizebox{8cm}{!}{\includegraphics{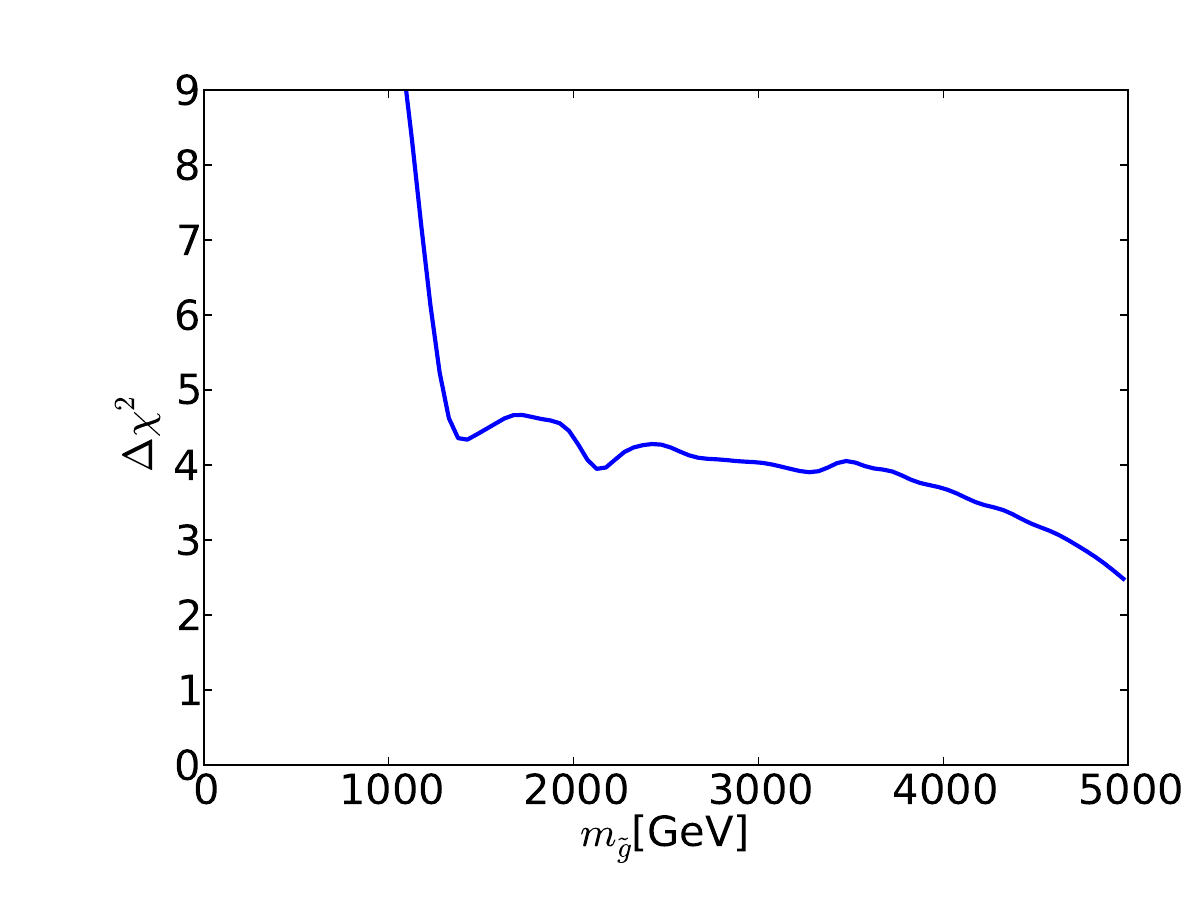}}
\resizebox{8cm}{!}{\includegraphics{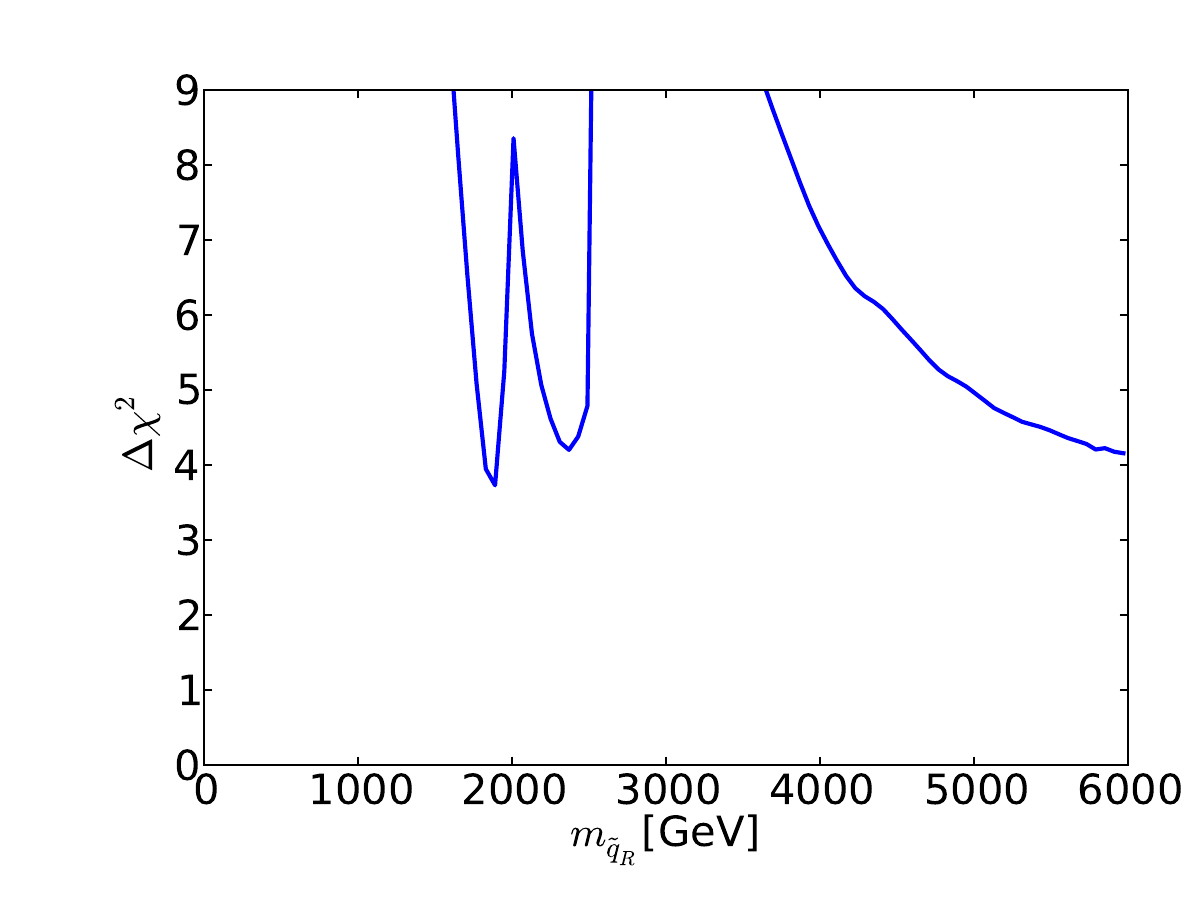}}
\resizebox{8cm}{!}{\includegraphics{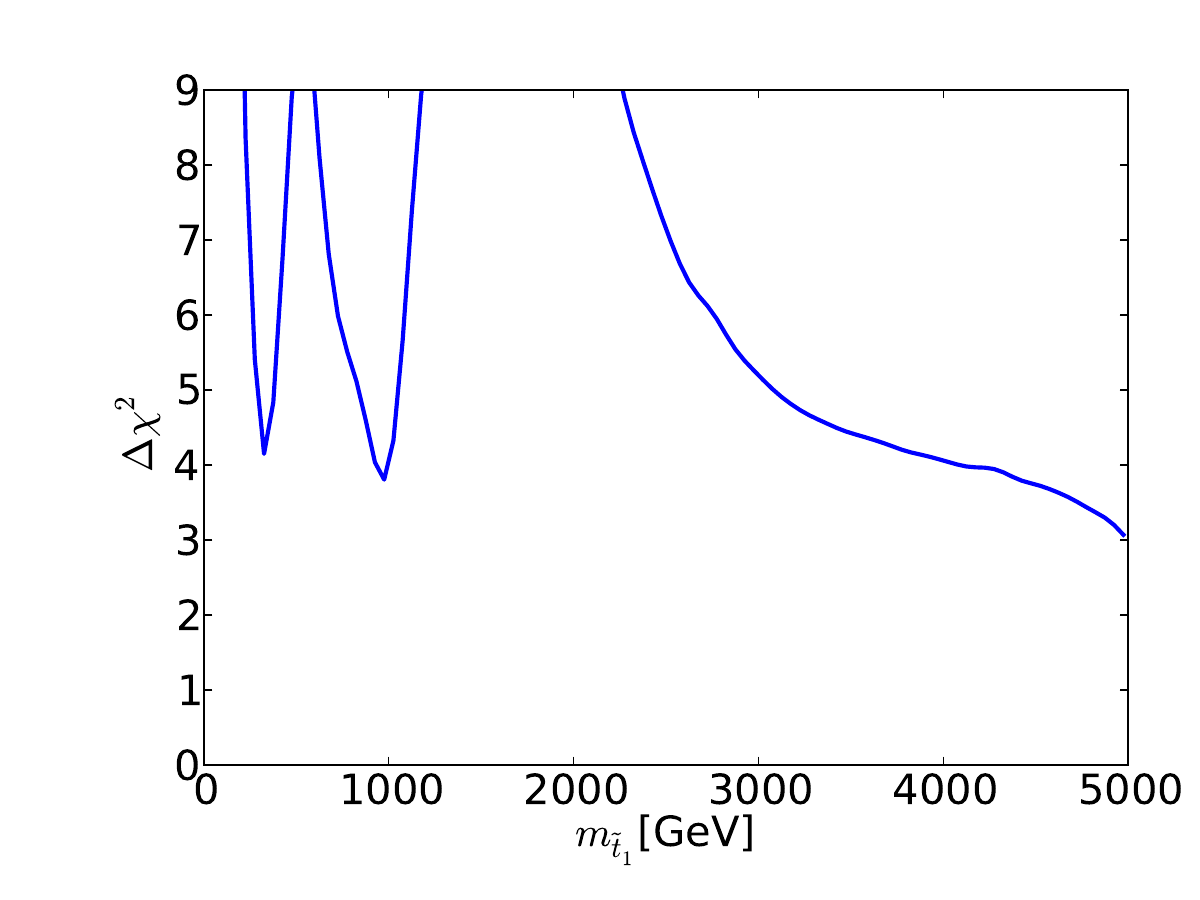}}
\resizebox{8cm}{!}{\includegraphics{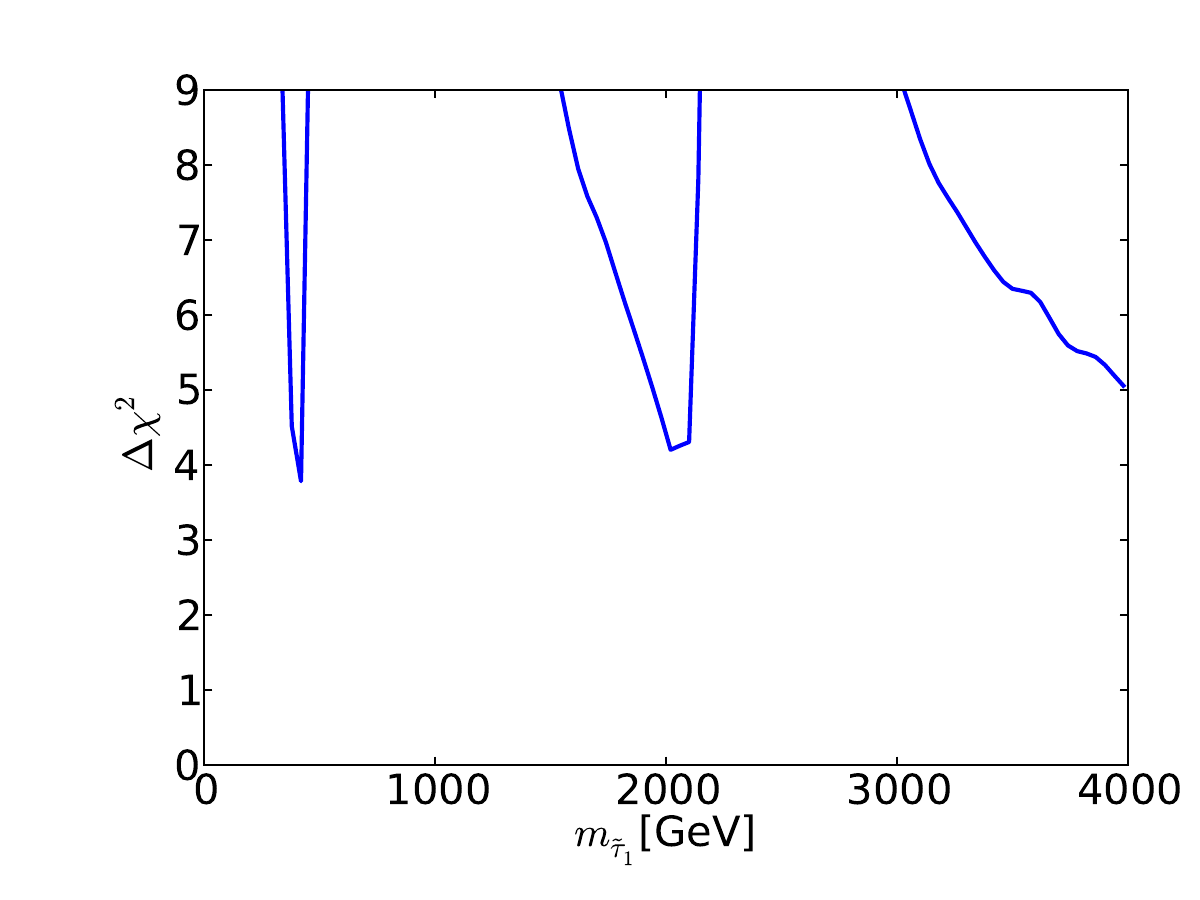}}
\vspace{-1cm}
\caption{\it As in Fig.~\protect\ref{fig:CMSSM1D}, but for $\mu < 0$ in
  the CMSSM. $\Delta \chi^2$ refers to the the difference in $\chi^2$
  with respect to the global minimum.} 
\label{fig:CMSSM1Dneg}
\end{figure*}

\medskip
On the other hand, the one-dimensional $\chi^2$ function for $\msqR$, shown
in the upper right panel of Fig.~\ref{fig:CMSSM1Dneg} has a very different form. 
After falling initially to $\Delta \chi^2 \sim 4$, there is a local maximum at $\msqR \sim 2000 \gev$ with
$\Delta \chi^2 \sim 8$. This is followed by a
region where $\Delta \chi^2$ falls again to $\sim 4$,
followed by a sharp rise to $\Delta \chi^2 > 9$. Finally, the $\chi^2$ function falls again below
$\Delta \chi^2 = 9$ when $\msqR > 3800 \gev$ and continues falling with increasing $\msqR$. 
The low-mass structures are in the `reef' regions,
and the high-mass fall is in the `continental' region.
Similar features are seen in the $\chi^2$ function for $\mste$, but at lower masses, 
in the lower left panel of Fig.~\ref{fig:CMSSM1Dneg}.
The $\chi^2$ function for $\mstaue$ shown in the lower right panel of Fig.~\ref{fig:CMSSM1Dneg}
exhibits sharp local minima at $\mstaue \sim 400$ and $2000 \gev$, followed again by a decrease
across the `continent' at large masses.

\medskip
We display in Fig.~\ref{fig:CMSSMHiggsneg} the one-dimensional $\chi^2$
functions for $\Mh$ (left panel) and $\MA$ (right panel) as calculated using {\tt FeynHiggs~2.10.0}. 
We see that $\Mh$ has a well-defined minimum at $\Mh \sim 126 \gev$. The fact that low
values of $\Mh \lsim 122 \gev$ do not acquire a heavier $\chi^2$ penalty is due to the
theoretical uncertainty in the calculation of $\Mh$ that we take to be $1.5 \gev$.
The $\chi^2$
function for $\MA$ has a local minimum at $\MA \sim 2000 \gev$ followed by a rise
to a local maximum at $\MA \sim 2300 \gev$ and then a decrease towards
$\Delta \chi^2 \sim 4$ when $\MA \sim 4000 \gev$.

\begin{figure*}[htb!]
\resizebox{8cm}{!}{\includegraphics{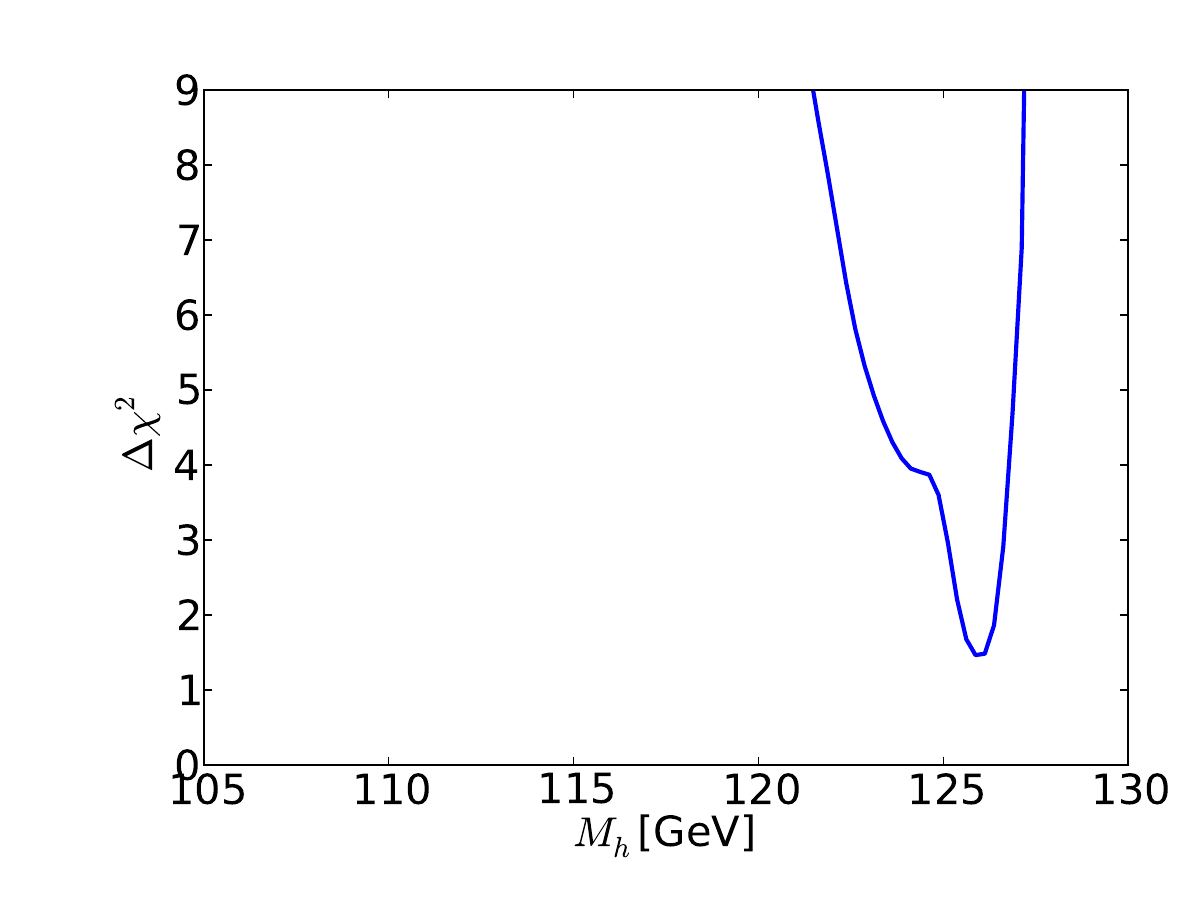}}
\resizebox{8cm}{!}{\includegraphics{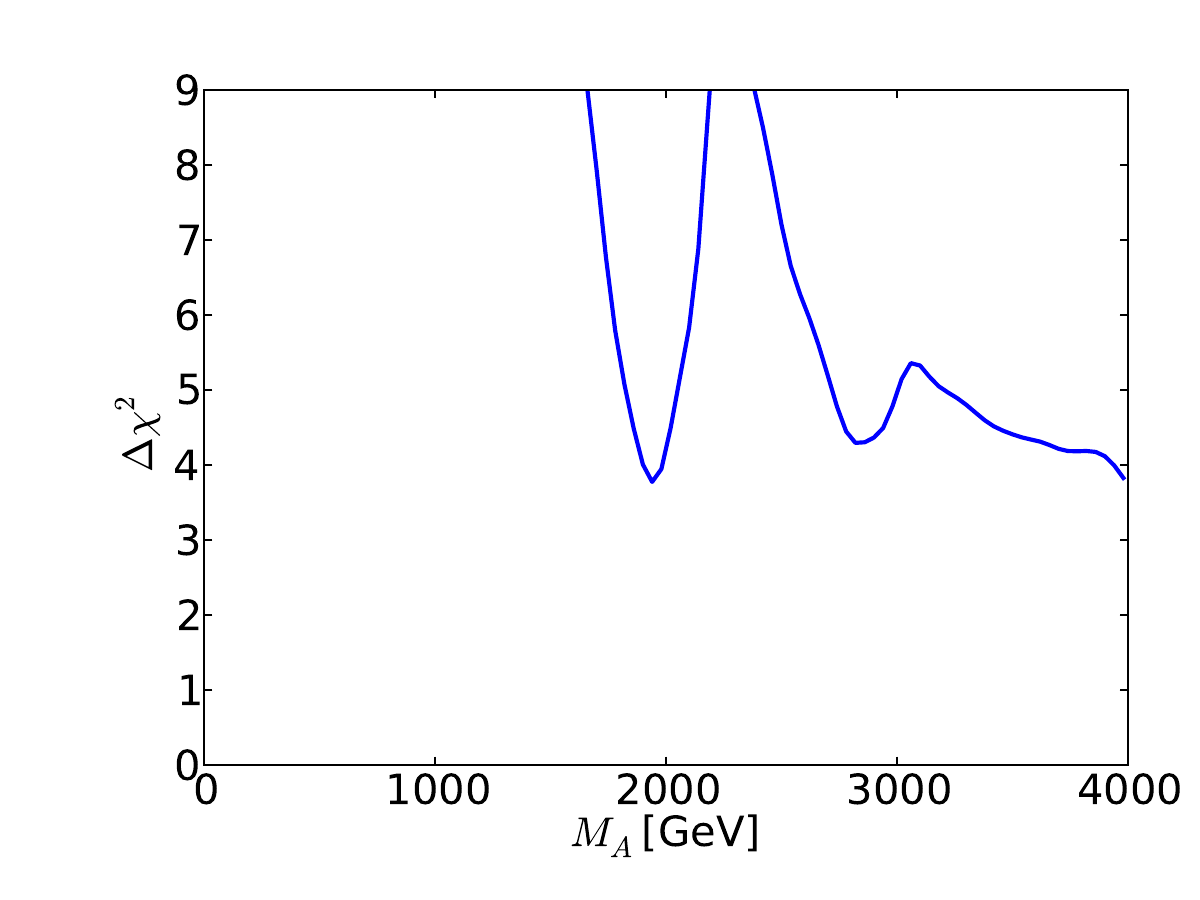}}
\vspace{-1cm}
\caption{\it As in Fig.~\ref{fig:CMSSMHiggs}, but for $\mu < 0$ in the CMSSM.} 
\label{fig:CMSSMHiggsneg}
\end{figure*}

\medskip
These one-dimensional likelihood functions can be used to set 95\% lower limits
on various sparticle masses by requiring $\Delta \chi^2 < 4$ relative to the global
minimum for the CMSSM, which occurs for $\mu > 0$ as discussed earlier.
These lower limits are tabulated in the third column of Table~\ref{tab:lowerlimits}.
We indicate in parentheses the approximate locations of limited ranges of
masses where the $\chi^2$ function dips briefly below the 95\% CL.

\medskip
Fig.~\ref{fig:CMSSMothersneg} shows the one-dimensional $\chi^2$ functions for
\bsdmm\ (left panel) and \ssi\ (right panel) for $\mu < 0$. We see that \bsdmm\ is
expected to be very similar to the SM value, reflecting the previous
observation that the lowest $\chi^2$ values for $\mu > 0$ are attained in the `continent' at large
sparticle masses and large $\MA$, and the secondary minima in the `reefs' at low masses
has small values of $\tb$. We also see that the preferred values of \ssi\ for $\mu < 0$
are $\sim 10^{-44}$ to $10^{-45}$~cm$^2$ at large $\mneu1$, whereas \ssi\ is $\lsim 10^{-48}$~cm$^2$
in the `reef' region.

\begin{figure*}[htb!]
\resizebox{8cm}{!}{\includegraphics{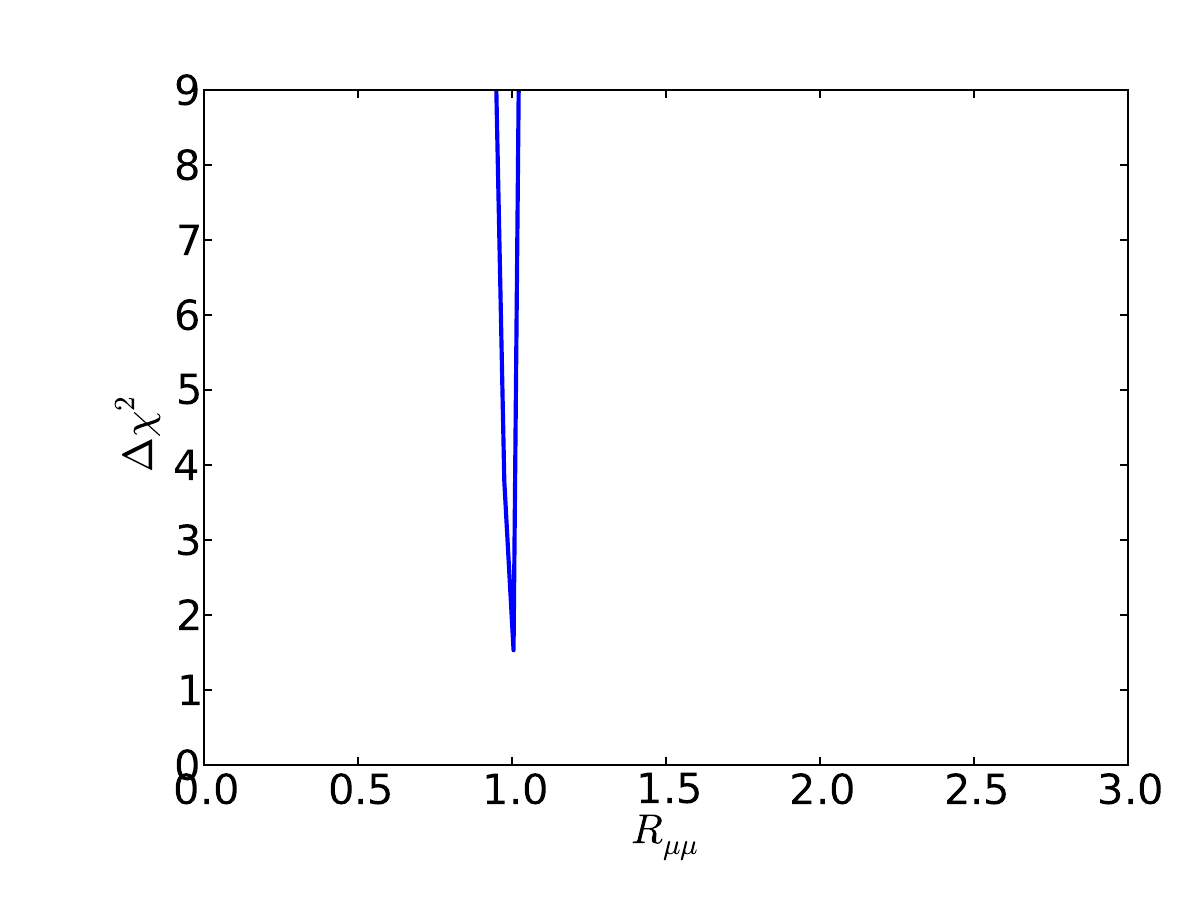}}
\resizebox{8cm}{!}{\includegraphics{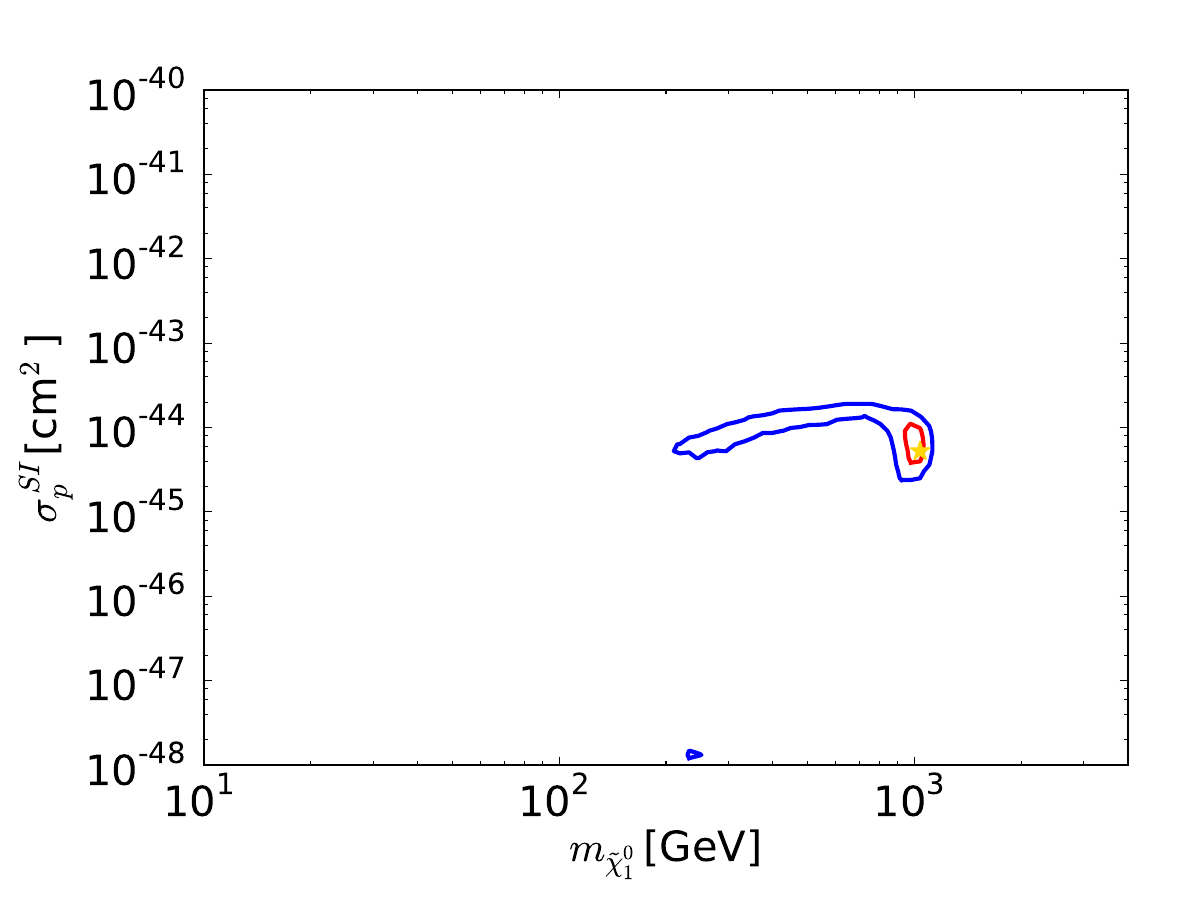}}
\vspace{-1cm}
\caption{\it
As in Fig.~\ref{fig:CMSSMothers}, but for $\mu < 0$ in the
  CMSSM.
} 
\label{fig:CMSSMothersneg}
\end{figure*}

\subsection{The NUHM1 with \boldmath{$\mu > 0$}}

\medskip
We now turn our attention to the NUHM1, concentrating on the case $\mu > 0$,
since our study of the CMSSM indicates that this sign is still preferred by the data,
albeit less strongly than in \cite{mc8}. 

\subsubsection{NUHM1 Parameter Planes}

\medskip
Fig.~\ref{fig:NUHM1} displays our selection of NUHM1 parameter
planes, with the same conventions for solid/dashed lines as in Fig.~\ref{fig:CMSSM}.
We see in the upper left panel that the likelihood function is relatively flat
for $m_{1/2} \gsim 2000 \gev$, and that there is a low-mass `peninsula' extending down to
$(m_0, m_{1/2}) \sim (500, 1200) \gev$, which is analogous to the `island' in the CMSSM. 
The 68\% CL region extends to values of $m_{1/2} > 4000 \gev$,
which was not the case in the CMSSM. This is because the NUHM1 is able to satisfy the $\Och$
constraint for larger values of $m_{1/2}$ than are possible in the CMSSM, thanks to the extra
degree of freedom associated with the soft SUSY-breaking contribution to the
Higgs masses. This permits values of $\mu$ or $\MA$ that allow $\Och$ to fall within the
astrophysical range even if $m_{1/2}$ is large. We also note that the
NUHM1 can satisfy the electroweak vacuum conditions in regions of the parameter space with
$m_0^2 < 0$, though we have not studied this possibility in any detail. 

\begin{figure*}[htb!]
\resizebox{8.5cm}{!}{\includegraphics{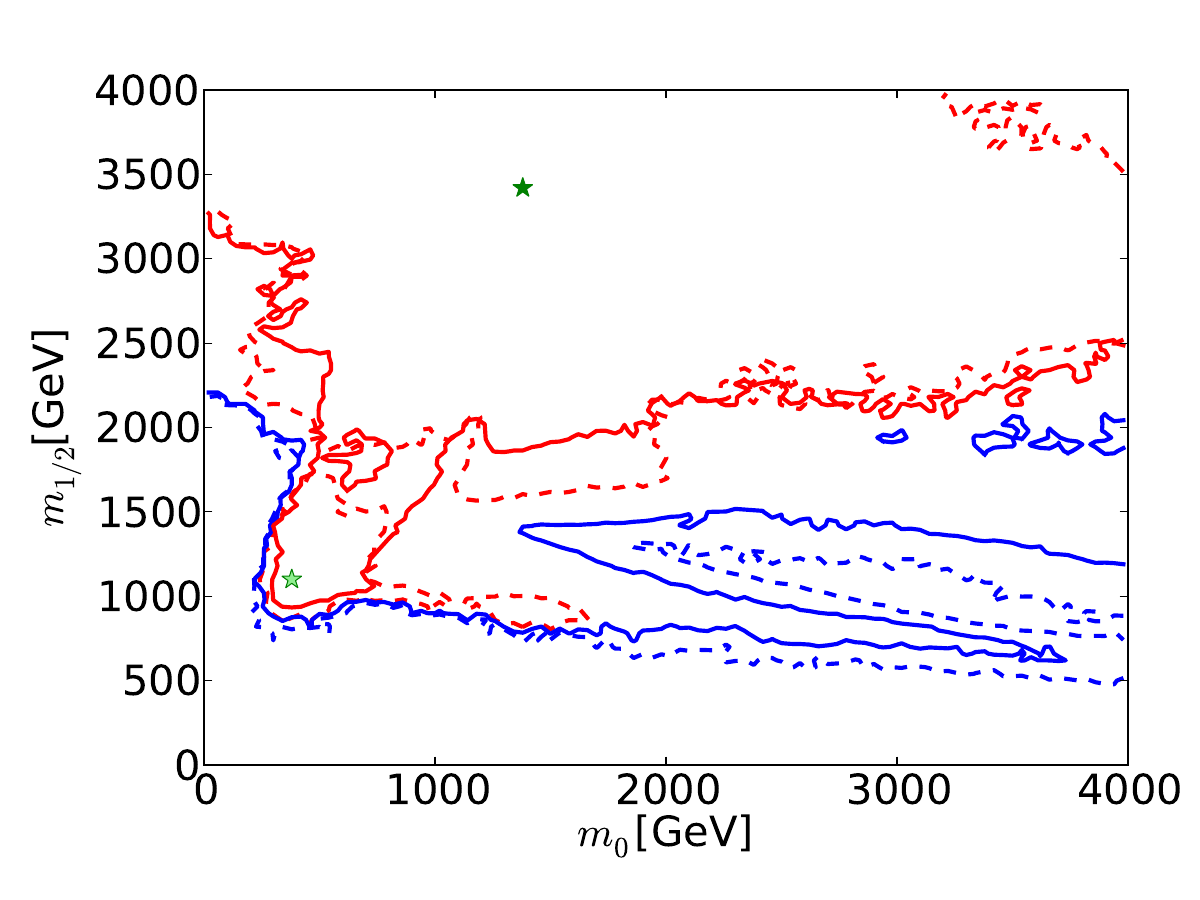}}
\resizebox{8.5cm}{!}{\includegraphics{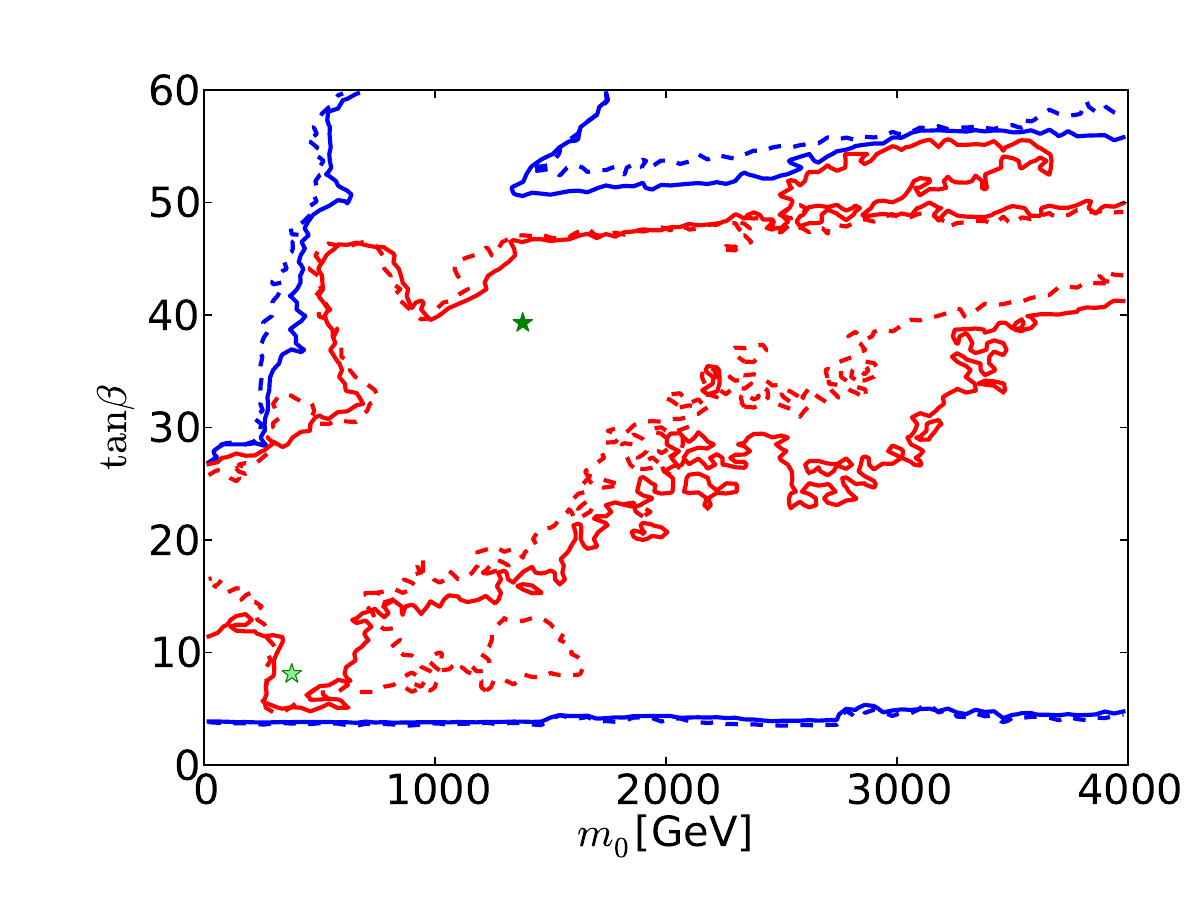}}
\resizebox{8.5cm}{!}{\includegraphics{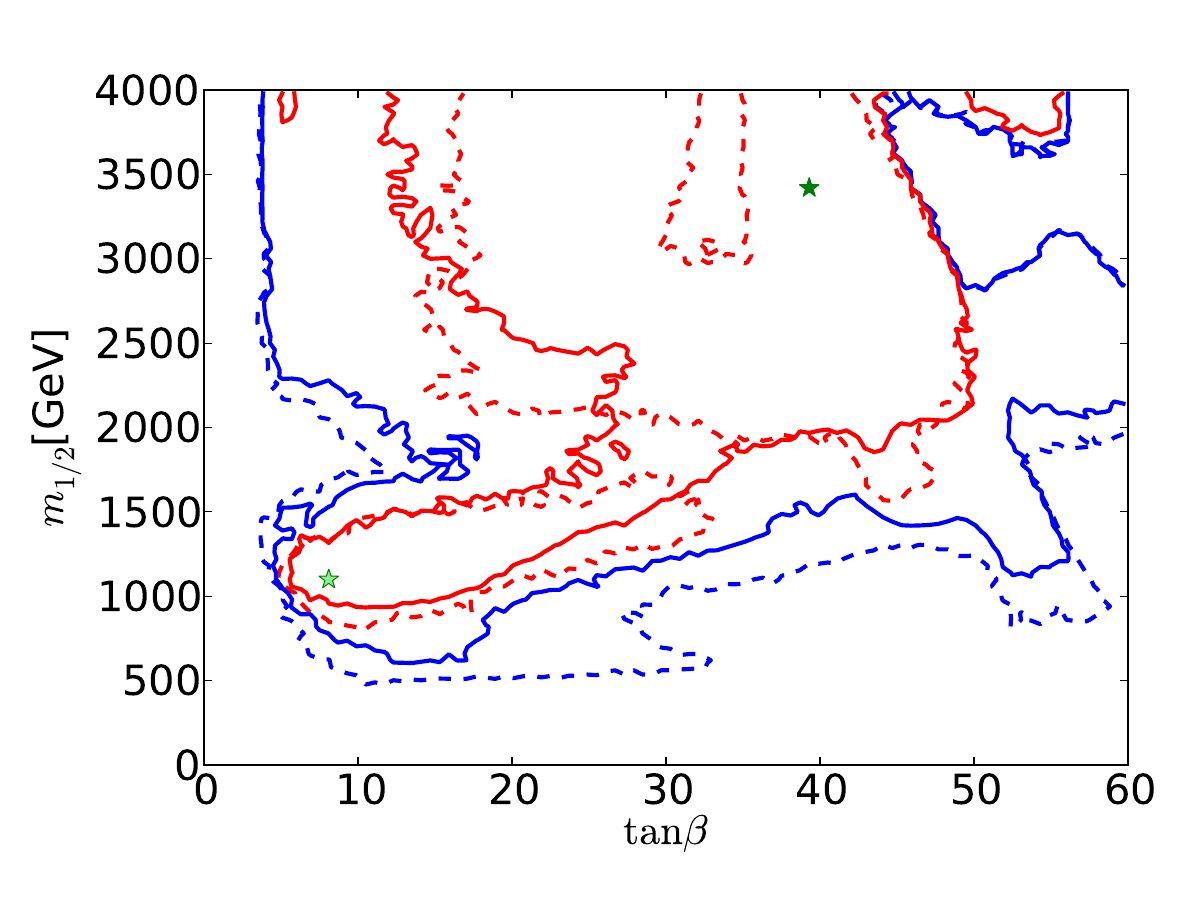}}
\resizebox{8.5cm}{!}{\includegraphics{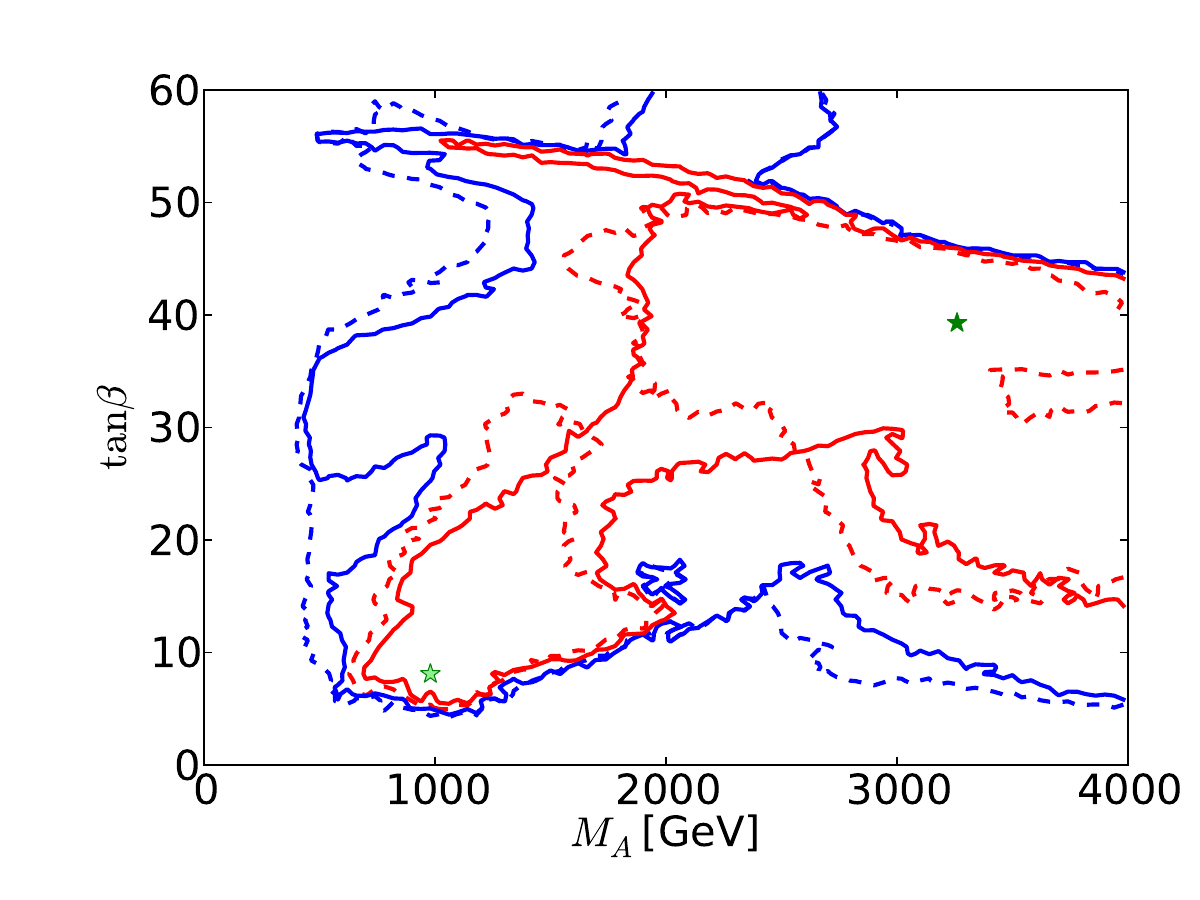}}
\vspace{-1cm}
\caption{\it As in Fig.~\protect\ref{fig:CMSSM}, but for the NUHM1 with $\mu > 0$. 
} 
\label{fig:NUHM1}
\end{figure*}

The differences  in Fig.~\ref{fig:NUHM1} between the results of the current analysis (solid lines and filled stars)
with our current implementations of the data constraints used in~\cite{mc8} are relatively minor. On the other hand,
looking back at the right panel of Fig.~\ref{fig:mc8old} where our current NUHM1 results are compared
with those shown previously in~\cite{mc8}, cf, the dashed lines
and open star, we see that both the 68\%
and 95\% CL regions now extend to much larger $m_{1/2}$. This is largely the result of
sampling an extended range in $m_H^2$, as well as using
{\tt FeynHiggs~2.10.0} to calculate $\Mh$. 
As in the CMSSM case shown in the left panel
of Fig.~\ref{fig:mc8old}, the extension of the 95\% CL region to lower $m_{1/2}$ at large $m_0$
is due to the new implementation of the dark matter scattering constraint discussed in Section~2.6.

\medskip
The upper right panel of Fig.~\ref{fig:NUHM1} displays the $(m_0, \tb)$ plane in the NUHM1.
We see a general trend for the preferred range of $\tb$ to increase with the value of $m_0 \gsim 1000 \gev$.
Values of $\tb$ as low as $\sim 5$ are allowed in the `peninsula' region.
In the $(\tb, m_{1/2})$ plane shown in the lower left panel of Fig.~\ref{fig:NUHM1},
we see that values of $\tb \sim 5$ to 30 are preferred when $m_{1/2} \lsim 2000 \gev$,
whereas larger values of $m_{1/2}$ are associated with $\tb \gsim 15$. Finally, we see in the
lower right panel of Fig.~\ref{fig:NUHM1} that values of $\MA \gsim 500 \gev$ are generally preferred,
with most of the favoured region appearing in a lobe with $\MA \gsim 2000 \gev$.


\subsection{Characteristics of the Best-Fit Points in the NUHM1}

The best-fit point in the `continental' region has nearly-degenerate $\neu1$, $\neu2$ and $\tilde \chi^\pm_1$,
 since $\mu \ll m_{1/2}$ and the LSP is nearly a pure higgsino, and the $\staue$ is $\sim 20 \gev$ heavier in this case. 
Thus  $\tilde \chi^\pm_1$, $\tilde{\chi}^{0}_{1,2}$ coannihilation is important in
fixing $\Och$, but $\staue$ coannihilation is not negligible. As could be expected from the
shape of the 68\% CL region in the lower right panel
of Fig.~\ref{fig:NUHM1}, whilst $\tilde \chi^\pm_1$, $\neu2$ coannihilation is important in most of the
`continental' region, different dynamical processes are important in different regions of the NUHM1
parameter space. For example, $\staue$ coannihilation and rapid annihilation via direct-channel poles
are both important in the lobe where $\MA \sim 1000 \gev$ and $\tan \beta \sim 10$ that includes
the best-fit point to the previous data set (open star). On the other hand,
only rapid annihilation via direct-channel poles is important in the lobe where $\MA \sim 1000 \gev$ and $\tan \beta \sim 30$,
and only $\tilde \chi^\pm_1$, $\neu2$ coannihilation is important in the narrow strip where
$\MA \sim 1000 \gev$ and $\tan \beta \sim 55$. Finally, both $\tilde \chi^\pm_1$ coannihilation and
rapid annihilation via direct-channel poles are important
in the lobe where
$\MA \gsim 2000 \gev$ and $\tan \beta \lsim 60$. 

\medskip
We see in Table~\ref{tab:chi2}, comparing the contributions to the global $\chi^2$ 
functions for the high-mass points in the NUHM1 and the CMSSM with $\mu > 0$, 
that the NUHM1 point has a noticeably smaller $\chi^2$ contribution from $\MW$. 
Comparing the low-mass points in the NUHM1 and the CMSSM with $\mu > 0$, 
we see that the NUHM1 point has smaller $\chi^2$ contributions from \bsg\ and \atlastwenty, in particular. 
The $\Mh$ constraint does not make an important contribution to $\chi^2$ at either of the NUHM points.

\subsection{One-Dimensional Likelihood Functions in the NUHM1}

\medskip
Fig.~\ref{fig:NUHM11D} displays the one-dimensional $\chi^2$ functions
for various sparticle masses. The likelihood function for $\mgl$ (upper
left panel) decreases essentially monotonically until $\mgl \sim 2600 \gev$, which
is followed by a local maximum at $\mgl \sim 3500 \gev$. The global minimum is at 6800~GeV, and hence not visible on this plot.
The $\chi^2$ function for $\msqR$ shown in the upper right panel of Fig.~\ref{fig:NUHM11D}
has similar behaviour. On the other hand,
the $\chi^2$ function for $\mste$, shown in the lower left panel of Fig.~\ref{fig:NUHM11D},
manifests an important local minimum at $\mste \sim 700 \gev$ followed by a local
maximum at $\mste \sim 1000 \gev$ before exhibiting a second local minimum and local maximum
at $\mste \sim 2000$ and $2700 \gev$, respectively. Finally, the $\chi^2$ function for $\mstaue$,
seen in the lower right panel of Fig.~\ref{fig:NUHM11D}, exhibits a low-mass local minimum at $\mstaue \sim 500 \gev$
associated with the above-mentioned `peninsula' followed by a local maximum at $\mstaue \sim 700 \gev$, 
and then falls to a shallow
minimum at $\mstaue \sim 1000 \gev$, eventually rising slowly at larger masses.

\begin{figure*}[htb!]
\resizebox{8cm}{!}{\includegraphics{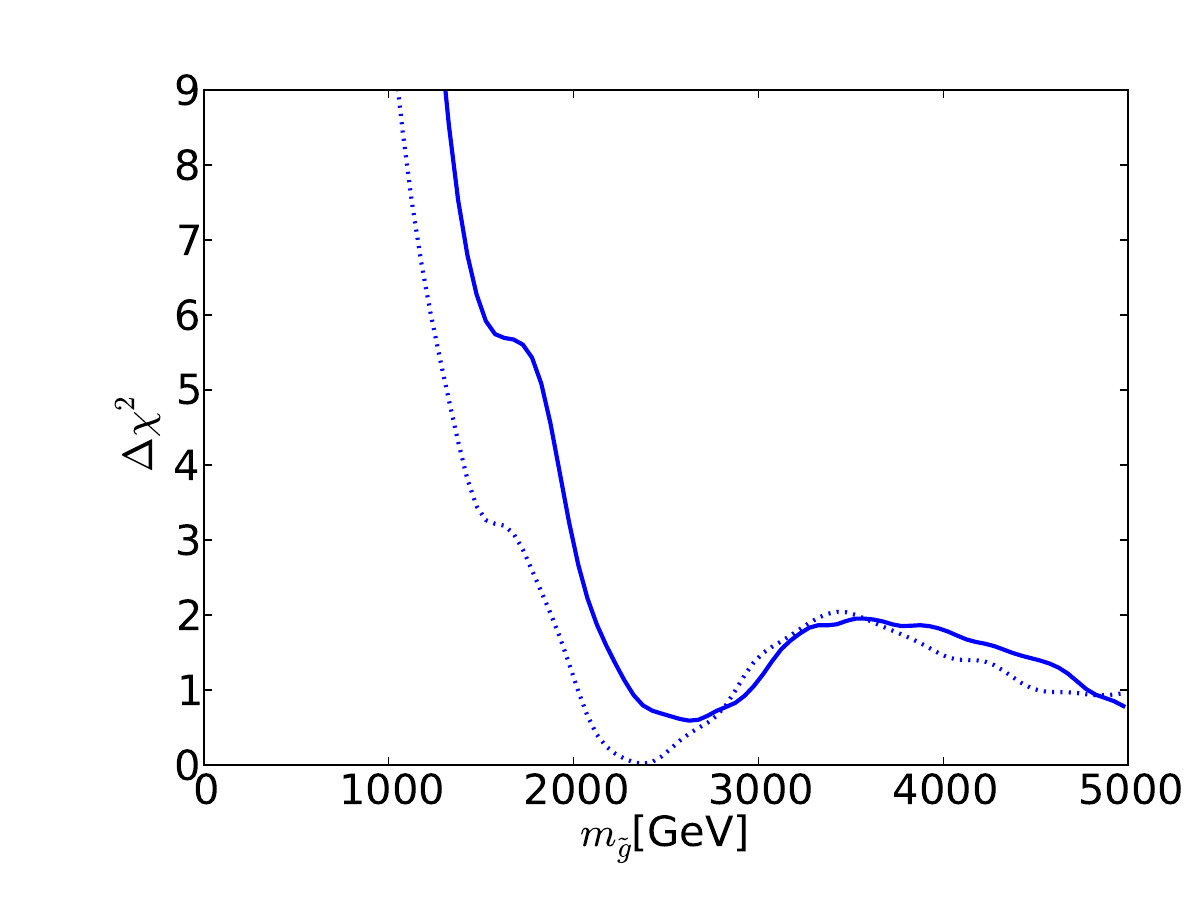}}
\resizebox{8cm}{!}{\includegraphics{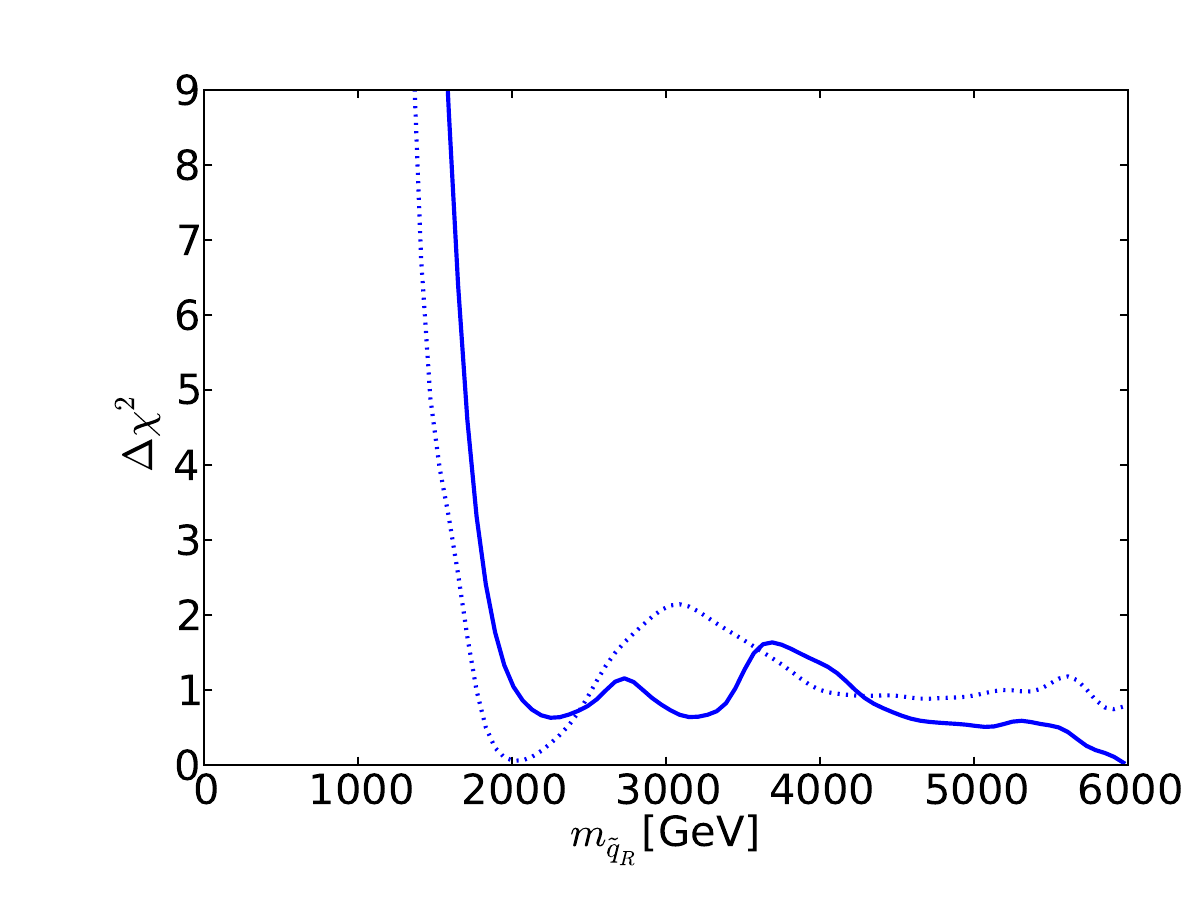}}
\resizebox{8cm}{!}{\includegraphics{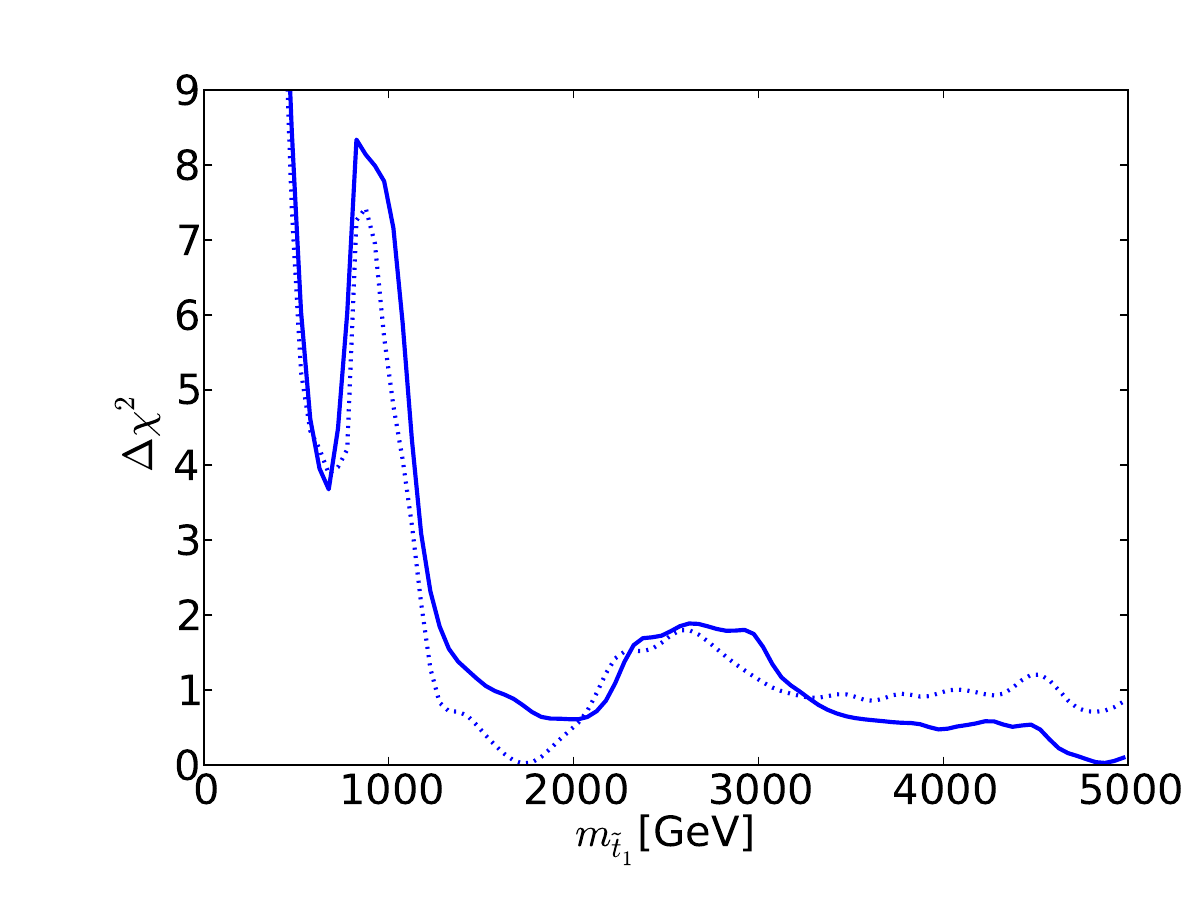}}
\resizebox{8cm}{!}{\includegraphics{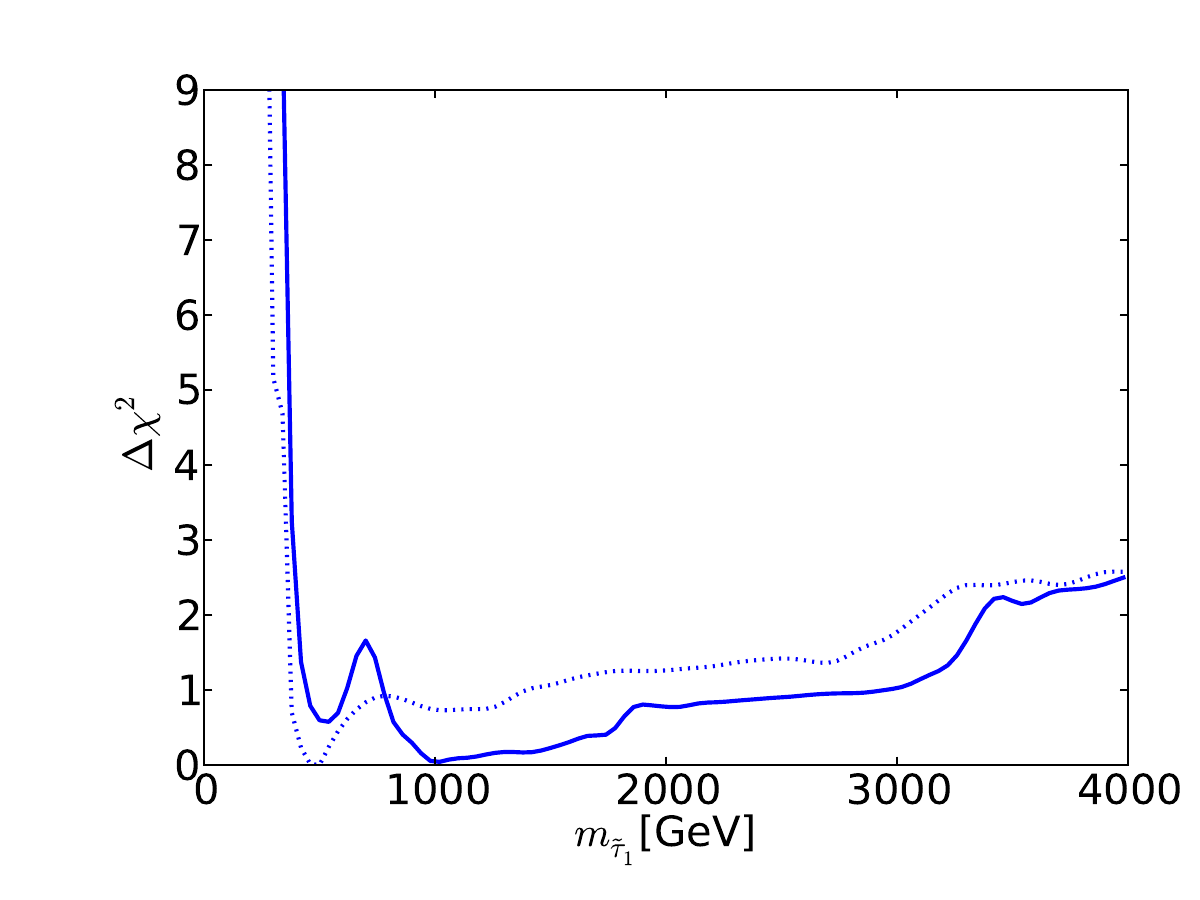}}
\vspace{-1cm}
\caption{\it As in Fig.~\protect\ref{fig:CMSSM1D}, but for the NUHM1 with 
$\mu > 0$. }
\label{fig:NUHM11D}
\end{figure*}

\medskip
Turning now to the one-dimensional $\chi^2$ functions for the SUSY
Higgs bosons shown in Fig.~\ref{fig:NUHM1Higgs}, we see in the left panel that
the likelihood function for the mass of the lightest supersymmetric Higgs boson $\Mh$
is maximized very close to the experimental value, though with tail extending to
lower and higher masses reflecting the theoretical uncertainty in the calculation. As for $\MA$,
we see in the right panel of Fig.~\ref{fig:NUHM1Higgs} that the likelihood function
is rather flat for $\MA \gsim 1000 \gev$.
The 95\% CL lower bounds on $\mgl, \msqR, m_{\tilde t_1}, \mstaue$ and $\MA$
inferred from the one-dimensional $\chi^2$ functions in Figs.~\ref{fig:NUHM11D}
and \ref{fig:NUHM1Higgs} are tabulated in Table~\ref{tab:lowerlimits}. As in the
CMSSM cases studied, the ${\tilde t_1}$ may be significantly lighter than the
other strongly-interacting sparticles.

\begin{figure*}[htb!]
\resizebox{8cm}{!}{\includegraphics{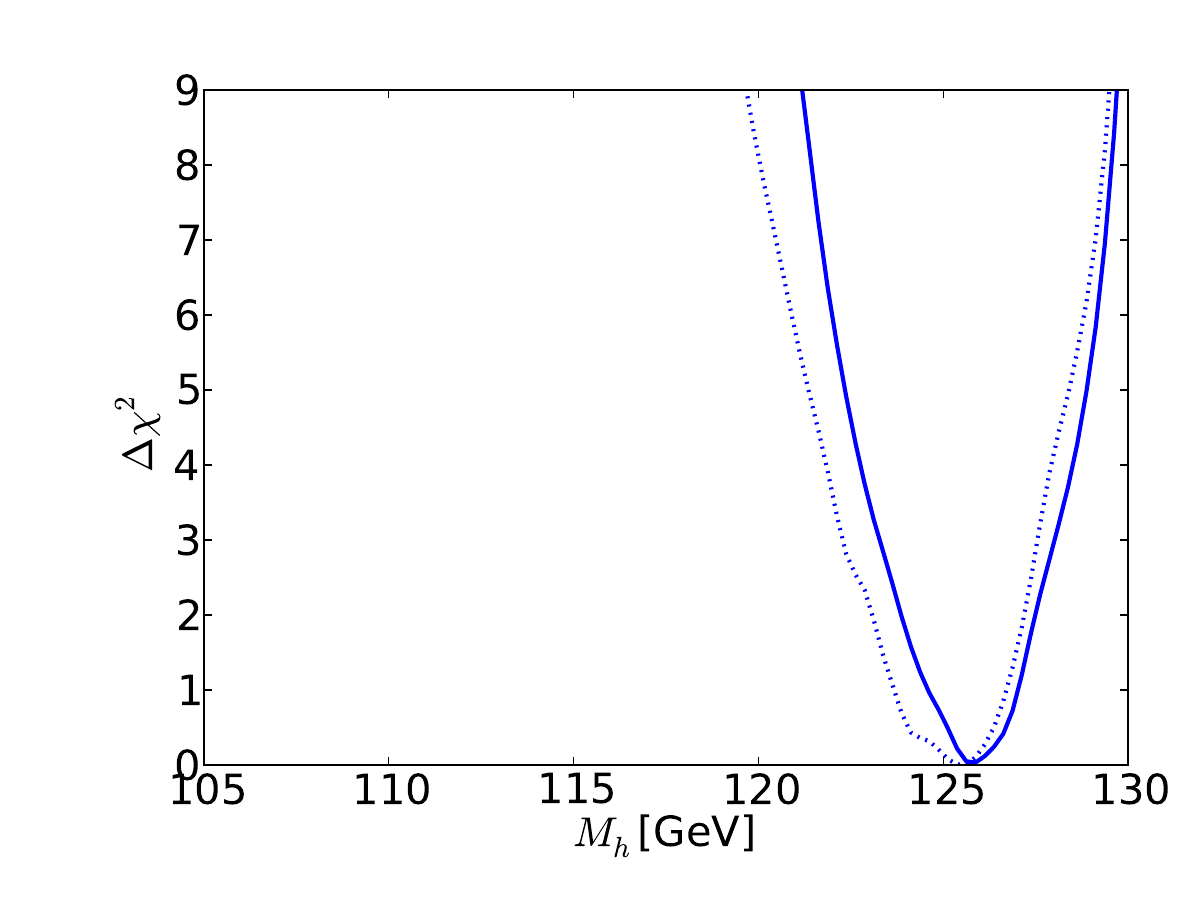}}
\resizebox{8cm}{!}{\includegraphics{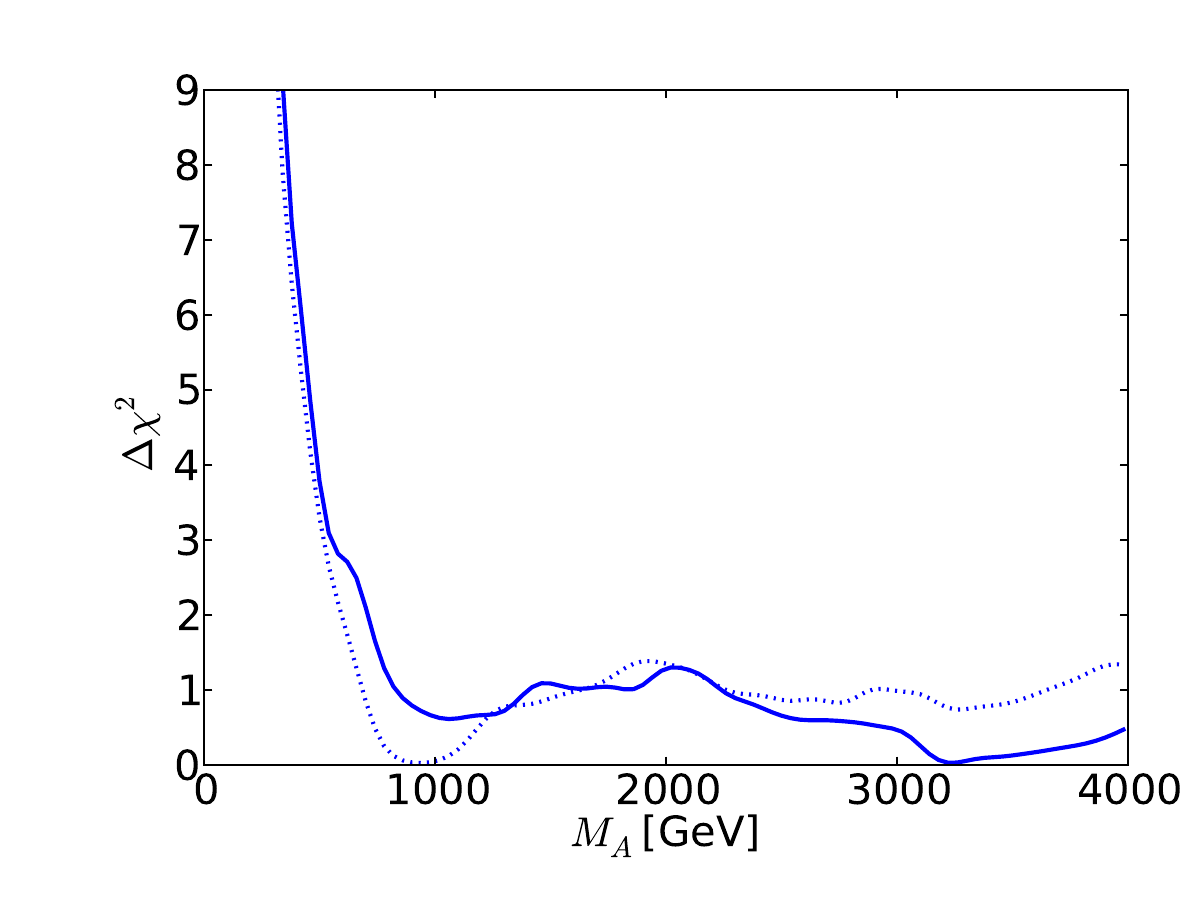}}
\vspace{-1cm}
\caption{\it As in Fig.~\ref{fig:CMSSMHiggs}, but for the NUHM1 with $\mu > 0$.} 
\label{fig:NUHM1Higgs}
\end{figure*}

\medskip
Finally, we see in Fig.~\ref{fig:NUHM1others} that the one-dimensional $\chi^2$
function for \bsmm\ is minimized close to the SM value. The NUHM1
offers very little scope for values of \bsmm\ below this, but values larger than
in the SM are not so strongly disfavoured.
The right plot of Fig.~\ref{fig:NUHM1others} shows the NUHM1 results in
  the $(\mneu1, \ssi)$ plane. Similar ranges of $\mneu1$ and
  $\ssi$ are favoured as for the CMSSM with $\mu > 0$.

\begin{figure*}[htb!]
\resizebox{8cm}{!}{\includegraphics{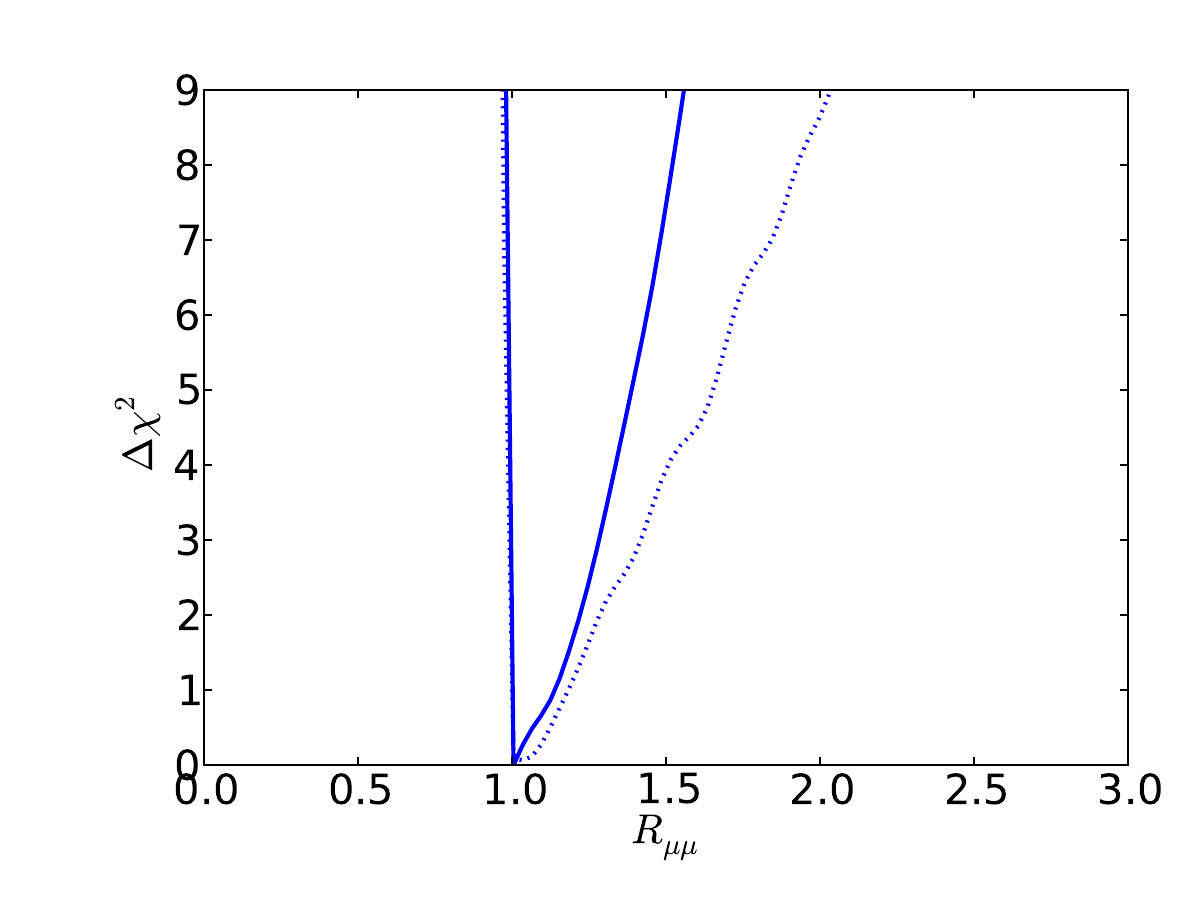}}
\resizebox{8cm}{!}{\includegraphics{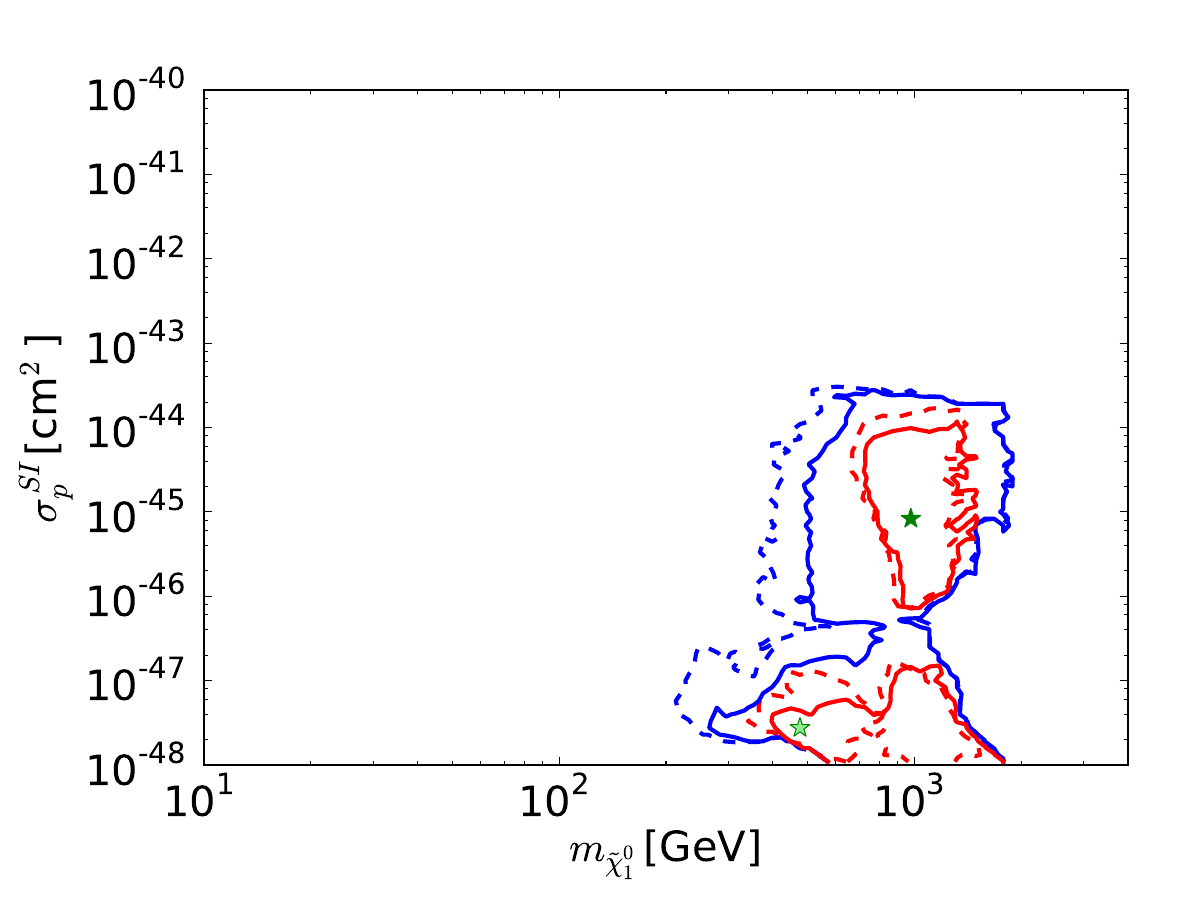}}
\vspace{-1cm}
\caption{\it As in Fig.~\ref{fig:CMSSMothers}, but for the NUHM1 with $\mu > 0$.
}
\label{fig:NUHM1others}
\end{figure*}

\section{Summary and Prospects}

\medskip
We have presented in this paper analyses of the CMSSM with both signs of $\mu$ and the NUHM1
with $\mu >0$ that take into account all the relevant constraints from the first run of the LHC with
$\sim 5$/fb of luminosity at 7~TeV and $\sim 20$/fb of luminosity at 8~TeV, as well as flavour and
precision electroweak observables and the first results from the LUX search for
spin-independent dark matter scattering~\cite{LUX}. We have sampled the model parameter spaces using
the {\tt MultiNest} technique, made SUSY model calculations of $\Mh$ using version {\tt 2.10.0}
of the {\tt FeynHiggs} code, and taken account of uncertainties in these calculations and in the
estimation of hadronic matrix elements for dark matter scattering.

\medskip
It is a general feature of our analysis that we find larger values of $m_0$ and $m_{1/2}$ to be
allowed than were found in our previous analyses, largely because of  our updated interpretation
of the experimental $\Mh$ constraint using {\tt FeynHiggs~2.10.0} and the newer version of
{\tt MicrOMEGAs} that we use.
The parameters of the best fits we find in the CMSSM and NUMH1 are displayed in Table~\ref{tab:bestfits}:
we note that they also have larger values of $m_0$ and $m_{1/2}$ than were favoured previously.
Also shown for comparison are the model parameters for local minima of the global $\chi^2$
functions at low masses, which are disfavoured by the \atlastwenty\ constraint, in particular.
We note that all the favoured CMSSM and NUHM1 model
points can accommodate the measured value of
$\Mh$. None of the SUSY models studied has a global $\chi^2$ value that is
much lower than the SM. This is because none of the SUSY models discussed
reduces significantly the contributions to the global $\chi^2$ functions from the observables
that make the largest contributions to the global $\chi^2$ functions in the SM fit, namely $\gmt$,
$A_{\rm fb}({b})$, $ A_\ell({\rm SLD})$, $\sigma_{\rm had}^0$, $\epsilon_K$ and $\btn$, as seen
in Table~\ref{tab:chi2}.

\medskip
The 95\% CL lower limits on sparticle masses found in our 
CMSSM and NUHM1 analysis are displayed in Table~\ref{tab:lowerlimits}.
We see that gluino masses above $\sim 1300 \gev$ are preferred in
the models analyzed. The right-handed squark
mass is restricted to even higher values, because of the preferred 
values of $m_0$, whereas the
lighter stop squark may be significantly lighter.
The lighter stau slepton may also be relatively light
in the CMSSM and NUHM1 with $\mu > 0$. On the other hand, the heavier Higgs bosons $A, H$ and
$H^\pm$ are all expected to have masses above $500 \gev$ in these
models.

\medskip
Estimates of the discovery reach of the LHC at 14~TeV have been provided in~\cite{ATLAS-HL-LHC}.
With 300/fb of luminosity, the 5-$\sigma$ discovery reach for squarks and gluinos should extend to $(\msqR, \mgl)
\sim (3500, 2000)$~GeV in the CMSSM with $\mu > 0$, would include the low-mass `reef' in the CMSSM
with $\mu < 0$, and would reach the first local minimum of the $\chi^2$ function in the NUHM1 with
$\mu > 0$, at $(\msqR, \mgl) \sim (2500, 3000)$~GeV. The discovery range with 3000/fb of luminosity
would extend a few hundred GeV further, and would be very similar to the 95\% CL exclusion reach
with 300/fb. The reach for 95\% CL exclusion with 3000/fb would extend several hundred GeV further still,
e.g., to $(\msqR, \mgl) \sim (4000, 2700)$~GeV in the CMSSM with $\mu > 0$.

\medskip
We conclude that large parts of the preferred parameter regions
of the CMSSM and NUHM1 are
accessible in future runs of the LHC, although the strongly-interacting sparticle
masses might be so high as to escape the searches at the LHC. 
That said, we re-emphasize that all the likelihood estimates made in this paper and
the estimates of the LHC physics reach are specific to the models studied, and are quite model-dependent.
The approach we have followed here for constructing the global likelihood function can easily be extended
to other models, a subject to which we will turn in future work.

\subsubsection*{Acknowledgements}

The work of O.B., J.E., J.M., K.A.O.\ and K.J.de V.\ is
supported in part by the London Centre for Terauniverse Studies (LCTS),
using funding from the European Research Council 
via the Advanced Investigator Grant 267352. 
The work of S.H.\ is supported 
in part by CICYT (grant FPA 2010--22163-C02-01) and by the
Spanish MICINN's Consolider-Ingenio 2010 Program under grant MultiDark
CSD2009-00064. The work of K.A.O.\ is supported in part by DOE grant
DE-FG02-94ER-40823 at the University of Minnesota.



\begin{thebibliography}{99}

\bibitem{lhch}
G.~Aad {\it et al.}  [ATLAS Collaboration],
  Phys.\ Lett.\ B {\bf 716}, 1 (2012)
  [arXiv:1207.7214 [hep-ex]];
   S.~Chatrchyan {\it et al.}  [CMS Collaboration],
  Phys.\ Lett.\ B {\bf 716}, 30 (2012)
  [arXiv:1207.7235 [hep-ex]].

\bibitem{CMSBsmm}
 S.~Chatrchyan {\it et al.}  [CMS Collaboration],
 Phys.\ Rev.\ Lett.\  {\bf 111} (2013) 101804
 [arXiv:1307.5025 [hep-ex]].

\bibitem{LHCbBsmm}
 R.Aaij {\it et al.}  [LHCb Collaboration],
 Phys.\ Rev.\ Lett.\  {\bf 111} (2013) 101805
 [arXiv:1307.5024 [hep-ex]].

\bibitem{ATLAS20}
ATLAS Collaboration, \\
{\tt http://cds.cern.ch/record/1547563/} {\tt files/ATLAS-CONF-2013-047.pdf}.

\bibitem{CMS20}
CMS Collaboration, \\
{\tt https://twiki.cern.ch/twiki/bin/view/} {\tt CMSPublic/PhysicsResultsSUS.}

\bibitem{MSSMHiggsATLAS-CMS} S.~Bressler,
  talk given at ``Higgs Couplings 2013'', Freiburg, October 2013;
  see {\tt http://indico.cern.ch/getFile.py/access?\\
  contribId=31\&sessionId=4\&resId=0\&\\
  materialId=slides\&confId=253774.}

 \bibitem{XENON100}
  E.~Aprile {\it et al.}  [XENON100 Collaboration],
  Phys.\ Rev.\ Lett.\  {\bf 107} (2011) 131302
  [arXiv:1104.2549 [astro-ph.CO]].

\bibitem{LUX}
D.~S.~Akerib {\it et al.}  [LUX Collaboration],
  arXiv:1310.8214 [astro-ph.CO].

 \bibitem{cmssm}
M.~Drees and M.~M.~Nojiri,
Phys.\ Rev.\ D {\bf 47} (1993) 376 [arXiv:hep-ph/9207234];
  G.~L.~Kane, C.~F.~Kolda, L.~Roszkowski and J.~D.~Wells,
  Phys.\ Rev.\  D {\bf 49} (1994) 6173
  [arXiv:hep-ph/9312272];
H.~Baer and M.~Brhlik,
Phys.\ Rev.\ D {\bf 53} (1996) 597 [arXiv:hep-ph/9508321];
  Phys.\ Rev.\  D {\bf 57} (1998) 567
  [arXiv:hep-ph/9706509];
  J.~R.~Ellis, T.~Falk, K.~A.~Olive and M.~Schmitt,
Phys.\ Lett.\ B {\bf 388} (1996) 97
[arXiv:hep-ph/9607292];
V.~D.~Barger and C.~Kao,
Phys.\ Rev.\ D {\bf 57} (1998) 3131
[arXiv:hep-ph/9704403];
L.~Roszkowski, R.~Ruiz de Austri and T.~Nihei,
JHEP {\bf 0108} (2001) 024
[arXiv:hep-ph/0106334];
A.~Djouadi, M.~Drees and J.~L.~Kneur,
JHEP {\bf 0108} (2001) 055
[arXiv:hep-ph/0107316];
U.~Chattopadhyay, A.~Corsetti and P.~Nath,
Phys.\ Rev.\ D {\bf 66} (2002) 035003
[arXiv:hep-ph/0201001];
J.~R.~Ellis, K.~A.~Olive and Y.~Santoso,
New Jour.\ Phys.\  {\bf 4} (2002) 32
[arXiv:hep-ph/0202110].

\bibitem{AbdusSalam:2011fc}
  S.~S.~AbdusSalam,  {\it et al.},
  Eur.\ Phys.\ J.\ C {\bf 71}, 1835 (2011)
  [arXiv:1109.3859 [hep-ph]].


   \bibitem{nuhm1}
H.~Baer, A.~Mustafayev, S.~Profumo, A.~Belyaev and X.~Tata,
  Phys.\ Rev.\  D {\bf 71}, 095008 (2005)
  [arXiv:hep-ph/0412059];
            H.~Baer, A.~Mustafayev, S.~Profumo, A.~Belyaev and X.~Tata,
               JHEP {\bf 0507} (2005) 065,
               hep-ph/0504001;
         J.~R.~Ellis, K.~A.~Olive and P.~Sandick,
  Phys.\ Rev.\  D {\bf 78}, 075012 (2008)
  [arXiv:0805.2343 [hep-ph]].
  


\bibitem{mc1}
O.~Buchmueller {\it et al.},
  Phys.\ Lett.\  B {\bf 657} (2007) 87
  [arXiv:0707.3447 [hep-ph]].

\bibitem{mc2}
  O.~Buchmueller {\it et al.},
  JHEP {\bf 0809} (2008) 117
  [arXiv:0808.4128 [hep-ph]].

\bibitem{mc3}
  O.~Buchmueller {\it et al.},
  Eur.\ Phys.\ J.\  C {\bf 64} (2009) 391
  [arXiv:0907.5568 [hep-ph]].

\bibitem{mc35}
  O.~Buchmueller {\it et al.},
  Phys.\ Rev.\  D {\bf 81} (2010) 035009
  [arXiv:0912.1036 [hep-ph]].

\bibitem{mc4}
  O.~Buchmueller {\it et al.},
  Eur.\ Phys.\ J.\  C {\bf 71} (2011) 1583
  [arXiv:1011.6118 [hep-ph]].

\bibitem{mc5}
  O.~Buchmueller {\it et al.},
  Eur.\ Phys.\ J.\  C {\bf 71} (2011) 1634
  [arXiv:1102.4585 [hep-ph]].

\bibitem{mc6}
O.~Buchmueller {\it et al.},
  Eur.\ Phys.\ J.\  C {\bf 71} (2011) 1722
  [arXiv:1106.2529 [hep-ph]].

\bibitem{mc7}
O.~Buchmueller, 
{\it et al.},
  Eur.\ Phys.\ J.\ C {\bf 72} (2012) 1878
  [arXiv:1110.3568 [hep-ph]].

\bibitem{mc75} 
O.~Buchmueller, 
{\it et al.},
Eur.\ Phys.\ J.\ C {\bf 72} (2012) 2020
 [arXiv:1112.3564 [hep-ph]].
  
\bibitem{mc8}
O.~Buchmueller, 
{\it et al.},
Eur.\ Phys.\ J.\ C {\bf 72} (2012) 2243
  [arXiv:1207.7315 [hep-ph]].
  
 \bibitem{newBNL} [The Muon g-2 Collaboration],
                 {\it Phys. Rev. Lett.} {\bf 92} (2004) 161802, 
                 hep-ex/0401008;
                 G.~Bennett et al.\ [The Muon g-2 Collaboration],
                  Phys.\ Rev. D {\bf 73} (2006) 072003
                  [arXiv:hep-ex/0602035].
                  
 \bibitem{fredl-gm2}
  M.~Benayoun, P.~David, L.~DelBuono and F.~Jegerlehner,
  Eur.\ Phys.\ J.\ C {\bf 73} (2013) 2453
  [arXiv:1210.7184 [hep-ph]].

                  
\bibitem{newerg-2}                  
T.~Blum, A.~Denig, I.~Logashenko, E.~de Rafael, B.~L.~Roberts, T.~Teubner and G.~Venanzoni,
  arXiv:1311.2198 [hep-ph].


\bibitem{ATLAS5}
ATLAS Collaboration, \\
{\tt https://cdsweb.cern.ch/record/1432199/} {\tt files/ATLAS-CONF-2012-033.pdf};
arXiv:1208.0949 [hep-ex].

\bibitem{CMS5}
S.~Chatrchyan {\it et al.}  [CMS Collaboration],
  arXiv:1207.1798 [hep-ex],
  arXiv:1207.1898 [hep-ex].

\bibitem{mcweb}
For more information and updates, please see {\tt http://cern.ch/mastercode/}.



\bibitem{post-LHC} 
 For a sampling of other post-LHC analyses, see:
T.~Li, J.~A.~Maxin, D.~V.~Nanopoulos and J.~W.~Walker,
  Europhys.\ Lett.\  {\bf 100}, 21001 (2012)
  [arXiv:1206.2633 [hep-ph]];
C.~Strege, G.~Bertone, F.~Feroz, M.~Fornasa, R.~Ruiz de Austri and R.~Trotta,
  JCAP {\bf 1304}, 013 (2013)
  [arXiv:1212.2636 [hep-ph]];
M.~E.~Cabrera, J.~A.~Casas and R.~R.~de Austri,
  JHEP {\bf 1307} (2013) 182
  [arXiv:1212.4821 [hep-ph]];
K.~Kowalska, L.~Roszkowski and E.~M.~Sessolo,
  JHEP {\bf 1306} (2013) 078
  [arXiv:1302.5956 [hep-ph]];
T.~Cohen and J.~G.~Wacker,
  JHEP {\bf 1309} (2013) 061
  [arXiv:1305.2914 [hep-ph]];
S.~Henrot-Versillé, Rém.~Lafaye, T.~Plehn, M.~Rauch, D.~Zerwas,
S.~ép.~Plaszczynski, B.~R.~éd'Orfeuil and M.~Spinelli, 
  arXiv:1309.6958 [hep-ph].


 \bibitem{Stefaniak}
P.~Bechtle, K.~Desch, H.~K.~Dreiner, M.~Hamer, M.~KrŠmer, B.~O'Leary, W.~Porod and X.~Prudent {\it et al.},
  arXiv:1310.3045 [hep-ph].


  
   \bibitem{125-other}
   H.~Baer, V.~Barger and A.~Mustafayev,
  Phys.\ Rev.\ D {\bf 85}, 075010 (2012)
  [arXiv:1112.3017 [hep-ph]];
   J.~L.~Feng, K.~T.~Matchev and D.~Sanford,
  Phys.\ Rev.\ D {\bf 85}, 075007 (2012)
  [arXiv:1112.3021 [hep-ph]];
  T.~Li, J.~A.~Maxin, D.~V.~Nanopoulos and J.~W.~Walker,
  Phys.\ Lett.\ B {\bf 710} (2012) 207
  [arXiv:1112.3024 [hep-ph]];
    S.~Heinemeyer, O.~Stal and G.~Weiglein,
  Phys.\ Lett.\ B {\bf 710}, 201 (2012)
  [arXiv:1112.3026 [hep-ph]];
A.~Arbey, M.~Battaglia, A.~Djouadi, F.~Mahmoudi and J.~Quevillon,
  Phys.\ Lett.\ B {\bf 708} (2012) 162
  [arXiv:1112.3028 [hep-ph]];
   P.~Draper, P.~Meade, M.~Reece and D.~Shih,
  Phys.\ Rev.\ D {\bf 85}, 095007 (2012)
  [arXiv:1112.3068 [hep-ph]];
S.~Akula, B.~Altunkaynak, D.~Feldman, P.~Nath and G.~Peim,
  Phys.\ Rev.\ D {\bf 85} (2012) 075001
  [arXiv:1112.3645 [hep-ph]];
M.~Kadastik, K.~Kannike, A.~Racioppi and M.~Raidal,
  JHEP {\bf 1205} (2012) 061
  [arXiv:1112.3647 [hep-ph]];
J.~Cao, Z.~Heng, D.~Li and J.~M.~Yang,
  Phys.\ Lett.\ B {\bf 710} (2012) 665
  [arXiv:1112.4391 [hep-ph]];
L.~Aparicio, D.~G.~Cerdeno and L.~E.~Ibanez,
  JHEP {\bf 1204}, 126 (2012)
  [arXiv:1202.0822 [hep-ph]];
  H.~Baer, V.~Barger and A.~Mustafayev,
  JHEP {\bf 1205} (2012) 091
  [arXiv:1202.4038 [hep-ph]];
C.~Balazs, A.~Buckley, D.~Carter, B.~Farmer and M.~White,
  arXiv:1205.1568 [hep-ph];
D.~Ghosh, M.~Guchait, S.~Raychaudhuri and D.~Sengupta,
  arXiv:1205.2283 [hep-ph].
 

  \bibitem{eo6}
   J.~Ellis and K.~A.~Olive,
  Eur.\ Phys.\ J.\ C {\bf 72}, 2005 (2012)
  [arXiv:1202.3262 [hep-ph]].
  
  \bibitem{elos}
   J.~Ellis, F.~Luo, K.~A.~Olive and P.~Sandick,
  Eur.\ Phys.\ J.\ C {\bf 73}, 2403 (2013)
  [arXiv:1212.4476 [hep-ph]].
  
   \bibitem{ehow+}
O.~Buchmueller, M.~J.~Dolan, J.~Ellis, T.~Hahn, S.~Heinemeyer, W.~Hollik, J.~Marrouche, K.~A.~Olive, H.~Rzehak, K.~de~Vries and G.~Weiglein,
CERN preprint CERN-PH-TH/2013-294.


\bibitem{moremuneg}  
  A.~Fowlie, M.~Kazana, K.~Kowalska, S.~Munir, L.~Roszkowski, E.~M.~Sessolo, S.~Trojanowski and Y.~-L.~S.~Tsai,
  Phys.\ Rev.\ D {\bf 86} (2012) 075010
  [arXiv:1206.0264 [hep-ph]];
  A.~Arbey, M.~Battaglia, A.~Djouadi and F.~Mahmoudi,
  Phys.\ Lett.\ B {\bf 720} (2013) 153
  [arXiv:1211.4004 [hep-ph]].
  
  \bibitem{0809.3437}
F.~Feroz, M.~P.~Hobson and M.~Bridges,
  Mon.\ Not.\ Roy.\ Astron.\ Soc.\  {\bf 398} (2009) 1601
  [arXiv:0809.3437 [astro-ph]];
  F.~Feroz, K.~Cranmer, M.~Hobson, R.~Ruiz de Austri and R.~Trotta,
  JHEP {\bf 1106} (2011) 042
  [arXiv:1101.3296 [hep-ph]].

 \bibitem{newFH} T.~Hahn, S.~Heinemeyer, W.~Hollik, H.~Rzehak and
                G.~Weiglein,
                [arXiv:1312.4937 [hep-ph]].
  
  \bibitem{Planck}
P.~A.~R.~Ade {\it et al.}  [Planck Collaboration],
  arXiv:1303.5076 [astro-ph.CO].
  

  
\bibitem{Svenetal}
  S.~Heinemeyer {\it et al.}, 
  JHEP {\bf 0608} (2006) 052
  [arXiv:hep-ph/0604147];
  S.~Heinemeyer, W.~Hollik, A.~M.~Weber and G.~Weiglein,
  JHEP {\bf 0804} (2008) 039
  [arXiv:0710.2972 [hep-ph]].
  
  \bibitem{Gfitter}
Gfitter Collaboration, \\
{\tt http://project-gfitter.web.cern.ch/} \\ {\tt project-gfitter/}.


\bibitem{Allanach:2001kg}
  B.~C.~Allanach,
  Comput.\ Phys.\ Commun.\  {\bf 143} (2002) 305
  [arXiv:hep-ph/0104145].

\bibitem{FeynHiggs}
 G.~Degrassi, S.~Heinemeyer, W.~Hollik, P.~Slavich and G.~Weiglein,
  Eur.\ Phys.\ J.\ C {\bf 28} (2003) 133
  [arXiv:hep-ph/0212020];
   S.~Heinemeyer, W.~Hollik and G.~Weiglein,
  Eur.\ Phys.\ J.\ C {\bf 9} (1999) 343
  [arXiv:hep-ph/9812472];
  S.~Heinemeyer, W.~Hollik and G.~Weiglein,
  Comput.\ Phys.\ Commun.\  {\bf 124} (2000) 76
  [arXiv:hep-ph/9812320];
   M.~Frank {\it et al.}, 
  JHEP {\bf 0702} (2007) 047
  [arXiv:hep-ph/0611326];
  T.~Hahn, S.~Heinemeyer, W.~Hollik, H.~Rzehak and G.~Weiglein,
  Comput.\ Phys.\ Commun.\  {\bf 180} (2009) 1426.
  see {\tt http://www.feynhiggs.de}~.



\bibitem{SuFla}
 G.~Isidori and P.~Paradisi,
  Phys.\ Lett.\ B {\bf 639} (2006) 499
  [arXiv:hep-ph/0605012];
  G.~Isidori, F.~Mescia, P.~Paradisi and D.~Temes,
  Phys.\ Rev.\  D {\bf 75} (2007) 115019
  [arXiv:hep-ph/0703035], and references therein.
  
\bibitem{SuperIso}
F.~Mahmoudi,
  Comput.\ Phys.\ Commun.\  {\bf 178} (2008) 745
  [arXiv:0710.2067 [hep-ph]]; 
  Comput.\ Phys.\ Commun.\  {\bf 180} (2009) 1579
  [arXiv:0808.3144 [hep-ph]];
  D.~Eriksson, F.~Mahmoudi and O.~Stal,
  JHEP {\bf 0811} (2008) 035
  [arXiv:0808.3551 [hep-ph]].





\bibitem{MicroMegas}
  G.~Belanger, F.~Boudjema, A.~Pukhov and A.~Semenov,
  Comput.\ Phys.\ Commun.\  {\bf 176} (2007) 367
  [arXiv:hep-ph/0607059];
  Comput.\ Phys.\ Commun.\  {\bf 149} (2002) 103
  [arXiv:hep-ph/0112278];
  Comput.\ Phys.\ Commun.\  {\bf 174} (2006) 577
  [arXiv:hep-ph/0405253].
  
\bibitem{SSARD}  Information about this code is available from K.~A.~Olive: it contains important contributions 
from T.~Falk, A.~Ferstl, G.~Ganis, F.~Luo, A.~Mustafayev, J.~McDonald, K.~A.~Olive, P.~Sandick, Y.~Santoso and M.~Srednicki.


\bibitem{SLHA}
P.~Skands {\it et al.},
  JHEP {\bf 0407} (2004) 036
  [arXiv:hep-ph/0311123];
  B.~Allanach {\it et al.},
  Comput.\ Phys.\ Commun.\  {\bf 180} (2009) 8
  [arXiv:0801.0045 [hep-ph]].
  
  \bibitem{MH-ATLAS} ATLAS Collaboration, 
                   ATLAS-CONF-2013-014, ATLAS-COM-CONF-2013-025.

\bibitem{MH-CMS} CMS Collaboration,
                 CMS-PAS-HIG-13-005.


\bibitem{SM2LRGE} H.~Arason et al.,
  Phys.\ Rev.\ D {\bf 46} (1992) 3945.

  \bibitem{bse}
M.~S.~Carena, H.~E.~Haber, S.~Heinemeyer, W.~Hollik, C.~E.~M.~Wagner and
G.~Weiglein, 
  Nucl.\ Phys.\ B {\bf 580} (2000) 29
  [hep-ph/0001002].
  

\bibitem{mhiggsFD3l} R.~Harlander, P.~Kant, L.~Mihaila and M.~Steinhauser,
                     Phys.\ Rev.\ Lett.\ {\bf 100} (2008) 191602
                     [Phys.\ Rev.\ Lett.\ {\bf 101} (2008) 039901]
                     [arXiv:0803.0672 [hep-ph]];
                     JHEP {\bf 1008} (2010) 104
                     [arXiv:1005.5709 [hep-ph]].

  \bibitem{FKPS}
J.~L.~Feng, P.~Kant, S.~Profumo and D.~Sanford,
  Phys.\ Rev.\ Lett.\  {\bf 111} (2013) 131802
  [arXiv:1306.2318 [hep-ph]].


\bibitem{BsmmATLAS}
 G.~Aad {\it et al.}  [ATLAS Collaboration],
 Phys.\ Lett.\ B {\bf 713} (2012) 387
 [arXiv:1204.0735 [hep-ex]].

\bibitem{BsmmCDF}
 T.~Aaltonen {\it et al.}  [CDF Collaboration],
 Phys.\ Rev.\ Lett.\  {\bf 107} (2011) 191801
  [Publisher-note {\bf 107} (2011) 239903]
 [arXiv:1107.2304 [hep-ex]].

\bibitem{BsmmD0}
 V.~M.~Abazov {\it et al.}  [D0 Collaboration],
 Phys.\ Lett.\ B {\bf 693} (2010) 539
 [arXiv:1006.3469 [hep-ex]].


\bibitem{Bobeth:2013uxa}
  C.~Bobeth, M.~Gorbahn, T.~Hermann, M.~Misiak, E.~Stamou and M.~Steinhauser,
  arXiv:1311.0903 [hep-ph];
T.~Hermann, M.~Misiak and M.~Steinhauser,
  arXiv:1311.1347 [hep-ph];
  C.~Bobeth, M.~Gorbahn and E.~Stamou,
  arXiv:1311.1348 [hep-ph].
  
\bibitem{BsmmSM}
 A.~J.~Buras, J.~Girrbach, D.~Guadagnoli and G.~Isidori,
 Eur.\ Phys.\ J.\ C {\bf 72} (2012) 2172
 [arXiv:1208.0934 [hep-ph]];
  K.~De Bruyn {\it et al.},
  Phys.\ Rev.\ Lett.\  {\bf 109} (2012) 041801
  [arXiv:1204.1737 [hep-ph]].



\bibitem{BsmmComb}
 R.Aaij {\it et al.}  [LHCb and CMS Collaborations],
LHCb-CONF-2013-012, CMS PAS BPH-13-007.


\bibitem{MFV}
  R.~S.~Chivukula and H.~Georgi,
  Phys.\ Lett.\ B {\bf 188} (1987) 99;
  G.~D'Ambrosio {\it et al.}
  Nucl.\ Phys.\ B {\bf 645} (2002) 155
  [hep-ph/0207036].
  
\bibitem{RmmMFV}
  A.~J.~Buras,
  Phys.\ Lett.\ B {\bf 566} (2003) 115
  [hep-ph/0303060];
  G.~Isidori and D.~M.~Straub,
  Eur.\ Phys.\ J.\ C {\bf 72} (2012) 2103
  [arXiv:1202.0464 [hep-ph]].

\bibitem{Fabrizio}
 Fabrizio Palla, CERN seminar, Aug. 6th, 2013.

\bibitem{Justine}
 Justine Serrano, CERN seminar, Aug. 6th, 2013.
 
\bibitem{Junk}
T.~Junk, {\tt http://www-cdf.fnal.gov/$\sim$trj/} {\tt mclimit/mclimit\_csm.pdf}, CDF/DOC/STATISTICS/PUBLIC/8128 (2007).

\bibitem{Shifman:1978zn}
  M.~A.~Shifman, A.~I.~Vainshtein and V.~I.~Zakharov,
  Phys.\ Lett.\  B {\bf 78}, 443 (1978).

\bibitem{Vainshtein:1980ea}
  A.~I.~Vainshtein, V.~I.~Zakharov and M.~A.~Shifman,
  Sov.\ Phys.\ Usp.\  {\bf 23}, 429 (1980)
  [Usp.\ Fiz.\ Nauk {\bf 131}, 537 (1980)].
  
  \bibitem{EFO}
  J.~R.~Ellis, A.~Ferstl and K.~A.~Olive,
 Phys.\ Lett.\ B {\bf 481}, 304 (2000)
 [hep-ph/0001005].

\bibitem{eos}
J.~R.~Ellis, K.~A.~Olive and C.~Savage,
  Phys.\ Rev.\ D {\bf 77}, 065026 (2008)
  [arXiv:0801.3656 [hep-ph]].
  
  \bibitem{FP}
  P.~Draper, J.~Feng, P.~Kant, S.~Profumo and D.~Sanford,
  Phys.\ Rev.\ D {\bf 88}, 015025 (2013)
  [arXiv:1304.1159 [hep-ph]].
  
 

\bibitem{CMSSM-SM-like-Higgs} J.~R.~Ellis, S.~Heinemeyer, K.~A.~Olive
  and G.~Weiglein, 
  Phys.\ Lett.\  B {\bf 515} (2001) 348
  [arXiv:hep-ph/0105061];
  S.~Ambrosanio, A.~Dedes, S.~Heinemeyer, S.~Su and G.~Weiglein,
  Nucl.\ Phys.\  B {\bf 624} (2002) 3
  [arXiv:hep-ph/0106255].


 \bibitem{Arbey:2012bp}
B.~C.~Allanach, C.~G.~Lester and A.~M.~Weber,
  JHEP {\bf 0612} (2006) 065
  [hep-ph/0609295];
  F.~Feroz, B.~C.~Allanach, M.~Hobson, S.~S.~AbdusSalam, R.~Trotta and A.~M.~Weber,
  JHEP {\bf 0810} (2008) 064
  [arXiv:0807.4512 [hep-ph]];

  
\bibitem{ATLAS-HL-LHC}
ATLAS Collaboration,
  arXiv:1307.7292 [hep-ex].






  





\end{thebibliography}
\end{document}